\documentclass[aps,prd,10pt,twocolumn,preprintnumbers,superscriptaddress,nofootinbib,letterpaper,longbibliography]{revtex4-2}

\usepackage{hyperref}
\usepackage{graphicx}
\usepackage{amsfonts,amsmath,amssymb,bm}
\usepackage{xcolor} 
\usepackage{tikz}
\usepackage[compat=1.1.0]{tikz-feynman} 
\usetikzlibrary{decorations.markings}
\tikzfeynmanset{
  double_arrow/.style={
    double distance=1.5pt,
    line width=0.6pt,
    decoration={
      markings,
      mark=at position 0.7 with {\arrow[scale=0.8]{latex}},
    },
    postaction={decorate},
  }
}

\usepackage{slashed}
\usepackage{xspace}
\usepackage{cleveref}
\usepackage{orcidlink}
\usepackage{braket} 
\usepackage{booktabs}

\hypersetup{
  colorlinks  = true,
  urlcolor    = blue,
  linkcolor   = blue,
  citecolor   = red
}

\definecolor{mustard}{RGB}{185, 135, 15} 

\newcommand{\nucleon}{\mathcal{N}}
\newcommand{\newcleon}{N}

\def\Irvine{Department of Physics and Astronomy, University of California, Irvine, CA 92697}
\def\IPPP{Institute of Particle Physics Phenomenology, Durham University, Durham, UK, DH1 3LF}
\def\TUD{Institute of Nuclear and Particle Physics, TUD Dresden University of Technology, D-01062 Dresden, Germany}
\def\MSU{Department of Physics and Astronomy, Michigan State University, East Lansing, MI 48824}

\graphicspath{{figures/}{figures_appendix/}}

\newcommand{\chiEFT}{$\chi$EFT\xspace}
\newcommand{\half}{\frac{1}{2}}

\begin{document}

\title{Phenomenological Chiral Effective Field Theory for Neutrino Event Generators}

\author{Zara Graham-Jones\,\orcidlink{0009-0009-0411-8966}}
\email{zara.m.graham-jones@durham.ac.uk}
\affiliation{\IPPP}

\author{Mareen Hoppe\,\orcidlink{0009-0003-0429-2792}}
\email{mareen.hoppe@tu-dresden.de}
\affiliation{\TUD}

\author{Joshua Isaacson\,\orcidlink{0000-0001-6164-1707}}
\email{isaacs21@msu.edu}
\affiliation{\MSU}

\author{Shirley Weishi Li\,\orcidlink{0000-0002-2157-8982}}
\email{shirley.li@uci.edu}
\affiliation{\Irvine}

\author{Mi\v sa Toman\,\orcidlink{0009-0003-2647-1281}}
\email{mtoman@uci.edu}
\affiliation{\Irvine}

\date{October 29, 2026}

\begin{abstract}
Single-pion production is a crucial interaction channel for accelerator-based neutrino experiments, which rely on neutrino event generators for its cross sections. However, no model implemented in current generators provides a consistent and systematically improvable description of the resonance region. In this work, we develop a phenomenological chiral effective field theory model of pion production, built from nucleons, pions, and the $\Delta(1232)$ with consistent, spin 3/2, gauge-invariant couplings, and we implement it in a fully automated pipeline from the Lagrangian to events using FeynRules, UFO 2.0, and \textsc{Sherpa}. We validate our results with pion photoproduction data and compare against neutrino bubble chamber, MINERvA, T2K, MicroBooNE, MiniBooNE, and electron-carbon scattering data. The framework can be systematically improved by extending the Lagrangian with higher-order operators and additional degrees of freedom, and by refitting couplings and form factors as new data become available. Combined with nuclear model interfaces, our approach opens a path to resonance-region predictions that meet the needs of next-generation experiments.
\end{abstract}

\preprint{UCI-HEP-TR-2026-15, MSUHEP-26-017, MCNET-26-035}

\maketitle


\section{Introduction}
\label{sec:intro}

The discovery of neutrino oscillations has opened a new avenue for investigating the Standard Model and physics beyond it. Accelerator-based neutrino oscillation experiments offer particularly exciting opportunities for such studies. As these experiments transition from a statistics-limited to a systematics-limited regime~\cite{MicroBooNE:2015bmn,Hyper-Kamiokande:2018ofw,DUNE:2020ypp}, understanding their systematic uncertainties has become essential. One crucial source of these uncertainties is the modeling of neutrino-nucleus scattering processes~\cite{NOvA:2021nfi,T2K:2023mcm,Ankowski:2015kya,Coyle:2022bwa,Coyle:2025xjk}, whose predictions typically disagree with data at the tens-of-percent level~\cite{Ankowski:2020qbe}. Next-generation experiments aim at percent-level precision, and meeting that goal requires significant investment in the theoretical underpinnings of these calculations.

Among the relevant interaction channels, single-pion production is especially important~\cite{Alvarez-Ruso:2014bla}. At the energies of current and next-generation beams, resonance production contributes a large fraction of the total event rate~\cite{DUNE:2020ypp,Ankowski:2020qbe}. Pions that are misreconstructed or missed distort the inferred neutrino energy, and neutral-current $\pi^0$ production is a leading background to electron-neutrino appearance measurements~\cite{MiniBooNE:2009dxl,MicroBooNE:2022zhr}. Therefore, accurate single-pion cross sections are needed both for signal and for background predictions in oscillation analyses.

Predicting these cross sections from first principles is currently not possible. The relevant hadronic invariant masses lie in the resonance region, where QCD is non-perturbative and the degrees of freedom are hadrons rather than quarks. The $\Delta(1232)$ gives the largest contribution to the cross section, but the non-resonant amplitudes required by chiral symmetry are also sizable, and the two interfere. The vector part of the interaction is constrained by electron and photon scattering data, while the axial part rests on limited bubble chamber measurements from the 1980s~\cite{Radecky:1981fn,Kitagaki:1986ct}. Lattice QCD calculations of the relevant transition amplitudes are advancing but do not yet constrain this region~\cite{Alexandrou:2006mc,Alexandrou:2010uk,Kronfeld:2019nfb}. 

In practice, single-pion production is therefore described by phenomenological models of varying theoretical inputs. These models are then implemented in neutrino event generators. The current state-of-the-art generators include \textsc{GENIE}~\cite{GENIE:2021npt,Andreopoulos:2009rq}, \textsc{NuWro}~\cite{Golan:2012wx,Golan:2012rfa}, \textsc{NEUT}~\cite{Hayato:2009zz,Hayato:2021heg,Abe:2024avs}, \textsc{GiBUU}~\cite{Buss:2011mx}, and \textsc{Achilles}~\cite{Isaacson:2022cwh,Isaacson:2025cnk}. All of these generators consist of three major components: the modeling of the nuclear initial state, the calculation of the hard scattering matrix element, and the final-state interactions (FSI). 

To compute the hard scattering matrix element for single-pion production, each generator takes a different approach. \textsc{GENIE} and \textsc{NEUT} rely on the Rein-Sehgal and Berger-Sehgal models~\cite{Rein:1980wg,Berger:2007rq}, which compute transition amplitudes in a relativistic harmonic-oscillator quark model, with the non-resonant background added incoherently. By default, \textsc{NuWro} describes the $\Delta$ with form factors fit to the ANL and BNL data and takes the non-resonant contribution as a fraction of the deep-inelastic cross section, added without interference~\cite{Golan:2012wx}. \textsc{GiBUU} describes resonance production with form factors fit to electroproduction data and obtains the background by subtracting the resonance contribution from data~\cite{Buss:2011mx,Mosel:2019vhx}. \textsc{Achilles}~\cite{Isaacson:2025cnk} adopts the dynamical coupled-channels (DCC) model~\cite{Kamano:2013iva,Nakamura:2015rta}, which provides a unitary and accurate description of $\pi N$ and $\gamma N$ data, but extending it to new channels or currents requires refitting. Furthermore, the DCC model is distributed as tables of precomputed amplitudes, preventing a systematic study of the underlying theory uncertainties.
 
What is missing is a consistent approach, in which resonant and non-resonant amplitudes are derived from a single Lagrangian and interfere coherently, constraints such as the conserved vector current (CVC) and the partially conserved axial current (PCAC) are satisfied automatically, and the same couplings and form factors relate photon, electron, and neutrino probes. Such a framework would also provide a method to quantify the theory uncertainties and a defined path for systematic improvement. None of the above generator implementations offer this.

Chiral effective field theory (\chiEFT) provides exactly this structure. As the low-energy effective field theory of QCD, it organizes the interactions of pions, nucleons, and the $\Delta(1232)$ in a systematic expansion, with chiral symmetry built-in and with electroweak currents obtained by gauging the same Lagrangian~\cite{scherer2003introduction}. To describe the data in the resonance region, we supplement the leading-order interactions with phenomenological ingredients, namely hadronic form factors, following Refs.~\cite{Fernandez-Ramirez:2005nuq,Hernandez:2007qq}. This takes the calculation beyond strict power counting, which is why we refer to the model as phenomenological chiral effective field theory. Even with these additions, all amplitudes are still derived from a single Lagrangian and interfere coherently, the symmetry constraints remain automatically satisfied, and the same couplings and form factors describe photon, electron, and neutrino scattering.

This strategy is not entirely new. Phenomenological \chiEFT models for neutrino-induced pion production were first proposed in Ref.~\cite{Hernandez:2007qq} and further developed in Refs.~\cite{Hernandez:2013jka,Alvarez-Ruso:2015eva,Gonzalez-Jimenez:2016qqq}, and a covariant \chiEFT calculation was presented in Ref.~\cite{Yao:2018pzc}. Of these, the model in Ref.~\cite{Gonzalez-Jimenez:2016qqq} has recently been implemented in \textsc{NuWro}~\cite{Yan:2024kkg}, and its amplitudes, like all the others, are derived and coded by hand. Our contribution is twofold. First, we formulate the model for the full electroweak sector using the gauge-invariant $N\Delta$ transition couplings, previously applied to $\pi N$ scattering and photoproduction~\cite{Pascalutsa:2000kd,Fernandez-Ramirez:2005nuq} and, for the $\Delta$ decay vertex alone, to neutrino-induced pion production~\cite{Praet:2008yn}. In contrast to Refs.~\cite{Hernandez:2007qq,Hernandez:2013jka,Alvarez-Ruso:2015eva,Gonzalez-Jimenez:2016qqq}, the axial $N \to \Delta$ transition strength is fixed by the strong coupling rather than fit to neutrino data, and the $W$ and $Z$ boson currents carry a consistent set of transition form factors obtained from CVC and PCAC arguments. Second, we automate the entire chain from Lagrangian to events. FeynRules~\cite{Alloul:2013bka} derives all interaction vertices, the UFO~2.0 format~\cite{Degrande:2011ua,Darme:2023jdn} carries the form factors attached to individual vertices, and \textsc{Sherpa}~\cite{Sherpa:2024mfk} computes and integrates the complete tree-level amplitude for any requested initial and final state through its matrix element generator \textsc{Comix}~\cite{Gleisberg:2008fv,Hoppe2026}. The UFO format provides a standardized set of Python files defining the particles, couplings, and vertices within a given model, and is supported by multiple event generators, including \textsc{MadGraph}~\cite{Alwall:2014hca} and \textsc{Whizard}~\cite{Kilian:2007gr}, so the model is not tied to our choice of \textsc{Sherpa}. Extending the calculations beyond free nucleon interactions, the nuclear initial and final state handling can be accomplished through an extension to the interface between \textsc{Sherpa} and \textsc{Achilles}~\cite{Isaacson:2021xty}. No amplitude is derived by hand, and computing a channel requires only specifying its initial and final states. 

We validate our model and the automated pipeline with \textsc{Sherpa} in the two settings where free-nucleon data exist. For the vector current, we compare against total cross sections for all four pion photoproduction channels; for the full electroweak interaction, we compare against the reanalyzed ANL and BNL bubble chamber measurements of charged-current single-pion production~\cite{Radecky:1981fn,Kitagaki:1986ct,Wilkinson:2014yfa,Rodrigues:2016xjj}. Finally, we apply the entire framework to nuclear targets using the \textsc{Achilles}+\textsc{Sherpa} interface. The scope of this work is deliberately limited. We restrict ourselves to the $\Delta(1232)$ region and to single-pion final states, omitting vector mesons and higher resonances, which we leave to future work. 

The paper is organized as follows. In Sec.~\ref{sec:chipt}, we introduce the two-flavor \chiEFT ingredients and their electroweak gauging. Section~\ref{sec:delta} introduces the $\Delta(1232)$ and its strong and electroweak couplings, and Sec.~\ref{sec:formfactors} specifies the phenomenological form factors. In Sec.~\ref{sec:free}, we present the free-nucleon calculation and its validation against photoproduction and neutrino data. Section~\ref{sec:bound} embeds the model in nuclei, and compares the full chain to single-pion data on carbon and argon. We conclude in Sec.~\ref{sec:conclusions}. 


\section{Chiral Effective Field Theory}
\label{sec:chipt}

In this section, we summarize the key elements of \chiEFT needed for this work: chiral symmetry breaking, the pion and nucleon Lagrangians, and the coupling to electroweak bosons. Our discussion follows Ref.~\cite{scherer2003introduction}.

\chiEFT describes the light pseudoscalar mesons as Goldstone bosons associated with the spontaneously broken chiral flavor symmetry of massless QCD. The relevant global symmetry group of $N$ massless quarks at the classical level is
\begin{align}
 G_{\textrm{classical}} & = U(N)_L\otimes U(N)_R\nonumber\\
 & \cong SU(N)_L\otimes SU(N)_R \otimes U(1)_V\otimes U(1)_A\,.
\end{align}
The axial $U(1)_A$ symmetry is broken by the Adler-Bell-Jackiw anomaly~\cite{Adler:1969gk,Bell:1969ts}, so it does not exist at the quantum level. The actual symmetry group is
\begin{align}
 G & =SU(N)_L\otimes SU(N)_R \otimes U(1)_V\,.
\end{align}
This symmetry is broken in two ways. First, the quark condensate $\braket{0|\overline{q}_iq_j|0} = v\delta_{ij}$ spontaneously breaks $G$ to its vector subgroup $H < G$, with $L = R$,
\begin{align}
 H & = SU(N)_V\otimes U(1)_V\,.
\end{align}
According to the Goldstone theorem~\cite{Goldstone:1962es}, the number of Goldstone bosons arising from the spontaneous symmetry breaking, $n$, is equal to the number of broken generators of $G$,
\begin{align}
 n \equiv \dim G - \dim H = N^2 - 1\,,
\end{align}
and all Goldstone bosons carry the quantum numbers of the corresponding broken generators. Because the broken generators are axial, the Goldstone bosons are pseudoscalar particles. For $N = 2$ ($u$ and $d$ quarks), the Goldstone bosons are the pion triplet $\pi^+, \pi^0, \pi^-$. Including the $s$ quark ($N = 3$) gives the meson octet consisting of $K^0, \overline{K}^0, K^+, K^-, \pi^+, \pi^0, \pi^-, \eta$. The second source of chiral symmetry breaking is the explicit breaking by the quark masses. If all quark masses are equal, the symmetry group of the theory is still $H$. Unequal quark masses, $m_u\neq m_d\neq m_s$, break the symmetry down to $U(1)_V$.

The rest of the paper focuses on the case of two degenerate quarks, $m_u = m_d$. Accordingly, we also take the proton and the neutron to be degenerate and use the physical pion masses, see Table~\ref{tab:inputs}. The extension to the $s$ quark is left for future work.


\subsection{Goldstone boson Lagrangian}

The general framework for writing the interactions of Goldstone bosons is the CCWZ construction~\cite{Coleman:1969sm,Callan:1969sn}. The pions enter through the exponential of the broken generators, and the lowest-order Lagrangian reads
\begin{align}
\label{eq:pion-kinetic-term}
 \mathcal{L} & \supset \frac{f_\pi^2}{4}\mathrm{Tr}\left(\partial_\mu U^\dagger \partial^\mu U\right) + \frac{f_\pi^2}{4}\mathrm{Tr}\left(U^\dagger \chi + U \chi^\dagger\right)\,.
\end{align}
The matrix $U$ is related to the pion fields $\phi^a$\,, $a=\{1, 2, 3\}$ by
\begin{align}
 U & = \exp\left(i\frac{\phi^a\tau^a}{f_\pi}\right), \quad\text{with}\quad \tau^a\phi^a = \begin{pmatrix}
 \pi^0 & \sqrt{2}\pi^+\\
 \sqrt{2}\pi^- & -\pi^0
 \end{pmatrix}\,,
 \label{eq:U-definition}
\end{align}
where $\tau^a$ are the Pauli matrices in isospin space, and $f_\pi \simeq 93$~MeV is the pion decay constant\footnote{Throughout, we use the theoretical convention of $f_\pi\simeq93$~MeV, as in Ref.~\cite{Peskin:1995ev}. Sources such as PDG~\cite{ParticleDataGroup:2026mpi} normalize the pion state differently and adopt the value $f_\pi' = \sqrt{2}f_\pi\simeq130\,\rm MeV\,$ for the pion decay constant.}. The matrix $\chi$ encodes the quark masses. In general, $\chi = 2B_0\,\mathrm{diag}(m_u, m_d)$, where $B_0$ is a constant proportional to the quark condensate. At leading order, the pion masses depend only on the trace of the quark mass matrix, $m_\pi^2 = B_0\left(m_u + m_d\right)$, so we can parametrize
\begin{equation}
\chi = m_\pi^2\, 1\!\!1_{2\times2}
\end{equation}
directly in terms of the physical pion mass.

Under a chiral transformation $$(L, R) \in SU(2)_L\otimes SU(2)_R\,,$$ the pion matrix transforms as
\begin{align}\label{eq:U-transformation-law}
 U' = R\, U L^\dagger\,,
\end{align}
where we follow the convention of Ref.~\cite{Scherer:2012xha}. The kinetic term in Eq.~\eqref{eq:pion-kinetic-term} is invariant under this transformation, while the $\chi$ term is invariant only for $L = R$, carrying the explicit breaking of the axial symmetry by the quark masses.


\subsection{Including nucleons}

Nucleons form an $SU(2)$ doublet $\nucleon_{L/R} = (p, n)^T_{L/R}$, transforming as $\nucleon_L' = L\nucleon_L$ and $\nucleon_R' = R\nucleon_R$. The kinetic term,
\begin{align}
 i\overline{\nucleon}_L\slashed{\partial} \nucleon_L
 + i\overline{\nucleon}_R\slashed{\partial} \nucleon_R\,,
\end{align}
is manifestly invariant under $G$, but the mass term $-m_N\left(\overline{\nucleon}_L \nucleon_R + \overline{\nucleon}_R \nucleon_L\right)$ is not. An invariant mass term is obtained by dressing the naive expression with the pion matrix $U$, which absorbs the transformation properties of nucleons~\cite{Srednicki:2007qs},
\begin{align}
 -m_N\left(\overline{\nucleon}_L U^\dagger \nucleon_R
 + \overline{\nucleon}_R U \nucleon_L\right)\,.
 \label{eq:nuc-mass-term}
\end{align}
At zeroth order in $\phi^a /f_\pi$, this reduces to the usual mass term. However, at higher orders the $\nucleon$ fields are not mass eigenstates. The redefinition that diagonalizes the mass term is
\begin{align}
    \newcleon_R & = u^\dagger\nucleon_R\,,\qquad\newcleon_L = u\nucleon_L\,,\quad U = u^2\,.
    \label{eq:nucleon-redefinition}
\end{align}
The field $u = \sqrt{U}$ dresses the nucleon fields, and every pion coupling in this work is built from it.

At lowest order, the only additional interaction term that respects parity, charge conjugation, and $G$ is
\begin{align}
 i c\,\left(\overline{\nucleon}_R\gamma^\mu
 U\partial_\mu U^\dagger\nucleon_R
 + \overline{\nucleon}_L\gamma^\mu U^\dagger\partial_\mu U\nucleon_L\right)\,,
\end{align}
where the real coefficient $c$ is not fixed by symmetry. The kinetic term, the mass term, and this interaction term constitute the lowest-order pion-nucleon Lagrangian.

In terms of the new fields, the Lagrangian reads
\begin{align}
\label{eq:nucleon-lagrangian}
 \mathcal{L} \supset \overline{\newcleon}\left(i\slashed{\partial}
 - m_N\right)\newcleon
 + \overline{\newcleon}\gamma^\mu\left(v_\mu + g_A\gamma^5 a_\mu\right)\newcleon\,,
\end{align}
with
\begin{align}
 v_\mu & = \frac{i}{2}\left(u^\dagger\partial_\mu u
 + u\partial_\mu u^\dagger\right)\,,\nonumber\\
 a_\mu & = \frac{i}{2}\left(u^\dagger\partial_\mu u
 - u\partial_\mu u^\dagger\right)\,.
 \label{eq:v-a-defs}
\end{align}
The vector coupling in Eq.~\eqref{eq:nucleon-lagrangian} is fixed entirely by the field redefinition, a consequence of chiral symmetry. The axial coupling $g_A = 1 - 2c$ inherits the free coefficient and is therefore a free parameter of the theory, which we fix to its measured value, see Table~\ref{tab:inputs}.


\subsection{Electroweak interactions}
\label{sec:gauging}

The interactions constructed so far follow from QCD alone. To include the electroweak interactions, we gauge the global symmetry group $G$ by introducing the gauge fields
\begin{alignat}{3}
 l_\mu & = l_\mu^a T^a + b_\mu\,, &\qquad l'_\mu &= L\bigl(l_\mu &&+ i\partial_\mu\bigr)L^\dagger\,,\nonumber\\
 r_\mu & = r_\mu^a T^a + b_\mu\,, &\qquad r'_\mu &= R\bigl(r_\mu &&+ i\partial_\mu\bigr)R^\dagger\,,
\end{alignat}
where $l_\mu^a$ and $r_\mu^a$ gauge $SU(2)_L$ and $SU(2)_R$, $T^a = \tau^a/2$ are the $SU(2)$ generators and $b_\mu$ gauges $U(1)_V$. The individual components of the gauge field are then given by
\begin{align}
\label{eq:chiral-symmetry-generator-normalization}
    l_\mu^a & = \textrm{Tr}\left(l_\mu\tau^a\right)\,,\quad r_\mu^a = \textrm{Tr}\left(r_\mu\tau^a\right)\,,\nonumber\\
    b_\mu & = \frac{1}{2}\textrm{Tr}\left(l_\mu\right) = \frac{1}{2}\textrm{Tr}\left(r_\mu\right)\,.
\end{align}
The covariant derivatives of the pion and nucleon fields are
\begin{align}
 D_{\mu} U & = \partial_\mu U - i r_\mu U + i U l_\mu\,,\nonumber\\
 D_\mu \nucleon_L & = \partial_\mu \nucleon_L - i l_\mu \nucleon_L\,,\nonumber\\
 D_\mu \nucleon_R & = \partial_\mu \nucleon_R - i r_\mu \nucleon_R\,,
\end{align}
with the form dictated by the transformation law in Eq.~\eqref{eq:U-transformation-law}. The fields $l_\mu$ and $r_\mu$ are identified with the Standard Model gauge bosons by matching these covariant derivatives to the Standard Model ones acting on the quark doublet. The derivation is given in Appendix~\ref{app:gauge-matching}. In terms of the physical gauge boson fields, the result is
\begin{align}
    l_\mu & = e\,Q\,A_\mu
    + \frac{g_2}{\cos\theta_W}\left(T^3 - \sin^2\theta_W\, Q\right)Z_\mu
    \nonumber\\
    & \quad + \frac{g_2}{\sqrt{2}}V_{ud}\left(W^\dagger_\mu T^+
    + W_\mu T^-\right)\,,\nonumber\\
    r_\mu & = e\,Q\,A_\mu
    - \frac{g_2}{\cos\theta_W}\sin^2\theta_W\, Q\, Z_\mu\,,
    \label{eq:lr-physical}
\end{align}
where $T^\pm = T^1 \pm iT^2$, $W_\mu = \left(W^1_\mu + i W^2_\mu\right)/\sqrt{2}$ is the physical charged $W^-$ boson field and $g_2 = e/\sin\theta_W$. The parameter $V_{ud}$ is the CKM matrix element $V_{ud}=\cos \theta_c$ with the Cabibbo angle $\theta_c$ and the matrix $Q = T^3 + \frac{1}{2}1\!\!1$ is the electric charge matrix of the fields the sources act on. Note that $r_\mu$ contains no $W$ boson, since the charged current is purely left-handed, as it must be. The photon term is the same in $l_\mu$ and $r_\mu$, reflecting the purely vector coupling of electromagnetism. Equation~\eqref{eq:lr-physical} is the form used throughout the remainder of this work. Gauging the Lagrangian of Eq.~\eqref{eq:nucleon-lagrangian} by $\partial_\mu \to D_\mu$ gives the interaction terms
\begin{align}
    \mathcal{L}_{\pi N gauge} & \supset \overline{\newcleon}\gamma^\mu\left(v_\mu +\frac{1}{2}\left(\Tilde{r}_\mu + \Tilde{l}_\mu\right)\right)\newcleon\nonumber\\
    &\qquad+g_A\overline{\newcleon}\gamma^\mu\gamma^5\left(a_\mu +\frac{1}{2}\left(\Tilde{r}_\mu - \Tilde{l}_\mu\right)\right)\newcleon
    \label{eq:pion-nucleon-gauge-lagrangian}
\end{align}
with the dressed gauge fields
\begin{equation}
    \Tilde{l}_\mu = ul_\mu u^\dagger\,,\qquad \Tilde{r}_\mu = u^\dagger r_\mu u\,.
\end{equation}
Because pions are pseudoscalars, the parity transformation acts as
\begin{align}
    (\phi^a)' = -\phi^a\,,\quad u' = u^\dagger\,,\quad\tilde{r}_\mu' = \tilde{l}_\mu\,,\quad\tilde{l}_\mu' = \tilde{r}_\mu\,.
\end{align}
Therefore, the combinations
\begin{align}
    v_\mu + \frac{1}{2}\left(\tilde{r}_\mu +\tilde{l}_\mu\right)\,,\nonumber\\
    a_\mu + \frac{1}{2}\left(\tilde{r}_\mu -\tilde{l}_\mu\right)\nonumber\,,
\end{align}
transform as a vector and axial vector, respectively.


\subsection{Chiral connection and vielbein}

The Lagrangian of Eq.~\eqref{eq:pion-nucleon-gauge-lagrangian} contains the pion and gauge fields in two combinations, a vector one, $v_\mu + \frac{1}{2}(\Tilde{r}_\mu + \Tilde{l}_\mu)$, and an axial one, $a_\mu + \frac{1}{2}(\Tilde{r}_\mu - \Tilde{l}_\mu)$. For the nucleon this is all that is needed. For the $\Delta$ of Sec.~\ref{sec:delta}, however, writing its couplings directly in terms of $u$, $l_\mu$, and $r_\mu$ becomes cumbersome. It is simpler to give the two combinations names, the chiral connection $\Gamma_\mu$ and the chiral vielbein $u_\mu$, and to work out how they transform. Every coupling of the $\Delta$ to pions and gauge bosons in Sec.~\ref{sec:delta} is then written in terms of $\Gamma_\mu$ and $u_\mu$, and the same two objects are what the literature uses~\cite{Scherer:2012xha}.

The physical nucleon field $\newcleon$ of Eq.~\eqref{eq:nucleon-redefinition} is built from $\nucleon$, which transforms with $L$ or $R$, and from $u = \sqrt{U}$, which depends on the pion fields. Under a chiral transformation, it therefore transforms as $\newcleon' = K\newcleon$ with a matrix $K = K(L, R, U)$ that depends on the pion fields and is neither $L$ nor $R$. For an ordinary isospin rotation, $L = R$, it reduces to $K = L = R$. Since $\newcleon_R = u^\dagger\nucleon_R$ and $\newcleon_L = u\nucleon_L$ must transform with the same $K$, the field $u$ itself transforms as
\begin{align}
    u' & = K u L^\dagger = R u K^\dagger\,.
\end{align}
Since $K$ depends on $x$ through the $U(x)$ field, gauging the symmetry introduces the chiral covariant derivative of the nucleon field,
\begin{align}
    D_\mu \newcleon & = \partial_\mu \newcleon + \Gamma_\mu \newcleon\,,
\end{align}
with an anti-Hermitian chiral connection
\begin{align}
    \Gamma_\mu & = -\Gamma_\mu^\dagger = -i\left(v_\mu + \frac{1}{2}\left(\Tilde{r}_\mu + \Tilde{l}_\mu\right)\right)\nonumber\\
    & = \frac{1}{2}\left(u^\dagger\left(\partial_\mu - i r_\mu\right)u + u\left(\partial_\mu - i l_\mu\right)u^\dagger\right)\,.
    \label{eq:chiral-connection}
\end{align}
The chiral connection contains an even number of pion fields, and it transforms inhomogeneously, like a gauge field,
\begin{align}
    \Gamma'_\mu & = K\left(\Gamma_\mu +  \partial_\mu\right)K^\dagger\,.\label{eq:chiral-connection-transformation-law}
\end{align}

The other remaining structure is the hermitian chiral vielbein
\begin{align}
    u_\mu & = u_\mu^\dagger = 2a_\mu + \Tilde{r}_\mu - \Tilde{l}_\mu\, \nonumber\\
    & = i\left(u^\dagger\left(\partial_\mu - i r_\mu\right)u
    - u\left(\partial_\mu - i l_\mu\right)u^\dagger\right)\,.
    \label{eq:chiral-vielbein}
\end{align}
It contains an odd number of pion fields and transforms homogeneously
\begin{align}
    u_\mu' & = K u_\mu K^\dagger\,.\label{eq:chiral-vielbein-transformation-law}
\end{align}

The chiral connection and vielbein can be used to construct chiral Lagrangians owing to their well-defined transformation properties. In terms of these objects, the pion-nucleon Lagrangian of Eq.~\eqref{eq:pion-nucleon-gauge-lagrangian} takes the compact form
\begin{align}
    \mathcal{L}_{\pi N gauge} & = \overline{\newcleon}\left(i\slashed{D} - m_N\right)\newcleon+\frac{1}{2}g_A\overline{\newcleon}\gamma^\mu\gamma^5u_\mu\newcleon\,.
    \label{eq:Lagrangian-pion-nucleon-gauge}
\end{align}
%


\section{The \texorpdfstring{$\Delta(1232)$}{Delta(1232)} Baryons}
\label{sec:delta}

The $\Delta(1232)$ is the lightest baryon resonance, and it makes the largest contribution to single-pion production. Describing it requires two ingredients beyond the framework of Sec.~\ref{sec:chipt}. First, the $\Delta$ has spin 3/2, and its covariant description contains unphysical components that must be removed. Second, the $\Delta$ has isospin 3/2, and it must be embedded in the $SU(2)$ structure of the chiral Lagrangian. In this section, we construct the free field, its isospin structure, and its couplings to pions and nucleons as well as to the electroweak bosons.


\subsection{The free Rarita-Schwinger field}

Delta baryons are described by the usual covariant Rarita-Schwinger formalism for spin-3/2 particles~\cite{Rarita:1941mf}\footnote{An alternative is based on S.~Weinberg's formulation of theories with arbitrary spin using $(j,0)\oplus(0,j)$ Lorentz representations~\cite{Weinberg:1964cn}, which contain only the physical degrees of freedom and require no additional constraints. Interactions have been constructed in this framework for general spin-3/2 particles~\cite{DelgadoAcosta:2015ypa}, dark matter particles~\cite{Criado:2020jkp}, and for the $\Delta$ resonance itself~\cite{Criado:2021gcb} in the $(3/2,0)\oplus(0,3/2)$ representation with applications to $\pi N$ scattering and pion photoproduction~\cite{Mart:2019jtb}. These formulations are less developed for the electroweak sector.}, where they are represented by the vector-spinor field, $\Psi^\mu_s$, a 16-component wave function containing both a spinor index~$s=\{1, 2, 3, 4\}$, which is usually dropped, and a vector index~$\mu=\{0,1,2,3\}$. A spin-3/2 particle has 4 spin states, doubled to 8 by the antiparticle, whereas the field~$\Psi^\mu_s$ contains 16 components. The 8 unphysical degrees of freedom have to be removed by imposing constraints. The most general free Lagrangian whose equations of motion impose the required constraints is given by~\cite{Moldauer:1956zz}: 
\begin{equation}
    \mathcal{L}_{\frac{3}{2}} = \overline{\Psi}_{\mu}\Lambda^{\mu\nu}(A)\Psi_{\nu}\,,
    \label{eq:Delta-Lagrangian}
\end{equation}
with $\overline{\Psi}_{\mu} = \Psi_{\mu}^{\dagger}\gamma^{0}$, and
\begin{equation}
\begin{aligned}
\Lambda^{\mu \nu}(A)= & -\left[\left(i \slashed{\partial}-m_{\Delta}\right) g^{\mu \nu}+i A\left(\gamma^\mu \partial^\nu+\gamma^\nu \partial^\mu\right)\right. \\
& \left.+\frac{i}{2}\left(3 A^2+2 A+1\right) \gamma^\mu \slashed{\partial} \gamma^\nu \right.\\
&\left.+m_{\Delta}\left(3 A^2+3 A+1\right) \gamma^\mu \gamma^\nu\right]\,,
\end{aligned}
\end{equation}
where $A\neq -\frac{1}{2}$ is an arbitrary real parameter, usually called the oﬀ-shell parameter, and $m_\Delta$ is the mass of the $\Delta$ baryon. The parameter $A$ appears because the field contains spurious spin-1/2 components, $\gamma^\nu\Psi_\nu$. It is shown~\cite{Nath:1971wp} that Eq.~\eqref{eq:Delta-Lagrangian} is invariant under so-called point transformations:
\begin{equation}
    \Psi_\mu(x) \rightarrow \Psi_\mu(x) + a \gamma_\mu \gamma_\nu \Psi^\nu(x), \,\quad A \rightarrow \frac{A-2a}{1+4a}\,,
    \label{eq:point-transformation}
\end{equation} 
ensuring that physical quantities are independent of $A$. The same holds in the interacting theory once consistent couplings are used, as discussed in Sec.~\ref{sec:delta-couplings}. The constraints that eliminate the unphysical degrees of freedom follow from the Euler-Lagrange equations,
\begin{equation}
    \gamma^{\mu} \Psi_{\mu} = 0,\, \quad \partial^\mu \Psi_{\mu} = 0\,.
    \label{eq:equations-of-motion}
\end{equation}

We set $A = -1$ throughout, which brings the propagator to its simplest form,
\begin{align}
    S_F^{\mu\nu}(p) = & \frac{1}{p^2-m_{\Delta}^2
    + i\varepsilon}P_{3/2}^{\mu\nu}(p),
    \label{eq:delta-propagator}
\end{align}
where
\begin{align}
    P_{3/2}^{\mu\nu}(p) & =  -\left(\slashed{p}+m_{\Delta}\right)\Big[g^{\mu\nu}-\frac{1}{3} \gamma^\mu \gamma^\nu\nonumber\\
    & +\frac{1}{3 m_{\Delta}}\left(p^\mu \gamma^\nu-\gamma^\mu p^\nu\right)
    -\frac{2}{3 m_{\Delta}^2}\, p^\mu p^\nu\Big]\nonumber\\
    & = \sum_{s}u^\mu_{s}(p)\overline{u}^{\nu}_s(p)\,.
    \label{eq:polarization-sum-3/2}
\end{align}
is the spin $3/2$ particle polarization sum~\cite{Christensen:2013aua}.

The propagator for general $A$ carries an additional term proportional to $(1+A)$ and is given in Appendix~\ref{app:delta-propagator}. Because $\Delta$ is unstable, we use the complex-mass scheme~\cite{Denner:1999gp} in which $m_\Delta$ is replaced by $\sqrt{m_\Delta^2-im_\Delta\Gamma_\Delta}$ and the $i\varepsilon$ term is no longer needed. There are different prescriptions for handling the mass and the width. In this work, we use a fixed width $\Gamma_\Delta = 102$~MeV, together with the pole mass $m_\Delta = 1.209$~GeV~\cite{Fernandez-Ramirez:2005nuq,ParticleDataGroup:2026mpi}. One could instead use an energy-dependent width, which changes the cross section near and above the resonance peak, and we discuss this choice in Sec.~\ref{sec:free} and Appendix~\ref{app:lineshape}. Additionally, neither choice gives the amplitude the phases required by Watson's theorem~\cite{Watson:1952ji}, which is also discussed there.


\subsection{Isospin structure}
\label{sec:delta-isospin}

Including $\Delta$ baryons in the $\chi$PT formulation requires their description in terms of isospin. Delta baryons have isospin $I=\frac{3}{2}$ with four states: $\Delta^{++}, \Delta^+, \Delta^0, \Delta^-$, which can be collected into a 4-component vector. Since $\Delta$ baryons couple to pions (isovector) and nucleons (isospinor), we re-express these 4 $\Delta$ states in a tensor-product form which makes the coupling structure manifest at the cost of some redundancy.

An $I=\frac{3}{2}$ representation can be constructed from the tensor product of $I=1$ and $I=\frac{1}{2}$ representations, i.e. the $\Delta$ baryons are represented by an isovector-isospinor vector-spinor wave function $\Psi_{\mu, s, a, r}$, with $a=\{1,2,3\}$ and $r=\{1,2\}$ denoting the isovector and the isospinor index, respectively. This product decomposes as $1\otimes\frac{1}{2} = \frac{3}{2}\oplus\frac{1}{2}$; it has 6 components, the 4 physical $\Delta$ states plus a spurious isospin-1/2 doublet. The projection operators
\begin{align}
    \xi_{ab}^{\frac{3}{2}} = \delta^{ab}-\frac{1}{3} \tau^a \tau^b\,
\end{align} 
project out the isospin-1/2 components, leaving the physical isospin-3/2 combinations. The isospin-1/2 contamination is the trace $\tau^a\Psi_a$, in analogy with the spin-1/2 components $\gamma^\mu\Psi_\mu$ of the previous subsection, and the projected field satisfies $\tau^a \xi^{\frac{3}{2}}_{ab}\Psi_b = 0$. In all interactions of this work, we use $\Psi_a$ to denote the projected field. Applying the projectors to the isovector-isospinor vector-spinor wave function leads to the following isospin doublets, with coefficients given by the Clebsch-Gordan decomposition of $1\otimes\frac{1}{2}$:
\begin{align}
    & \Psi_1 = \xi_{1 ; b}^{\frac{3}{2}} \Psi_{\mu, s, b, r}=\frac{1}{\sqrt{2}}\binom{\frac{1}{\sqrt{3}} \Delta^0-\Delta^{++}}{\Delta^{-}-\frac{1}{\sqrt{3}} \Delta^{+}}, \nonumber\\
    & \Psi_2 = \xi_{2 ;b}^{\frac{3}{2}} \Psi_{\mu, s, b, r}=-\frac{i}{\sqrt{2}}\binom{\frac{1}{\sqrt{3}} \Delta^0+\Delta^{++}}{\Delta^{-}+\frac{1}{\sqrt{3}} \Delta^{+}}, \nonumber\\
    & \Psi_3 =  \xi_{3 ; b}^{\frac{3}{2}} \Psi_{\mu, s, b, r}=\sqrt{\frac{2}{3}}\binom{\Delta^{+}}{\Delta^0}\,.
    \label{eq:delta-doublets}
\end{align}
On the left-hand sides, the vector and spinor indices are suppressed, and the isospinor index $r$ labels the two entries of each doublet. We note that this convention has the opposite sign of the one adopted by Ref.~\cite{Fernandez-Ramirez:2005nuq}.


\subsection{Couplings to pions and nucleons}
\label{sec:delta-couplings}

The $\Delta$ couples to pions in two ways. A pion can attach to two $\Delta$ fields, giving the diagonal $\pi\Delta\Delta$ vertices, or it can convert a nucleon into a $\Delta$, giving the $\pi N\Delta$ transition vertex. Both couplings are built from the two objects through which pions enter the Lagrangian: the chiral connection of Eq.~\eqref{eq:chiral-connection} and the chiral vielbein of Eq.~\eqref{eq:chiral-vielbein}.

The chiral connection enters by replacing the derivatives in Eq.~\eqref{eq:Delta-Lagrangian} with the $\Delta$ chiral covariant derivative,
\begin{align}
    \left(D_\mu\Psi\right)^{a, r}_\nu & = D_{\mu}^{aa', rr'}\Psi_\nu^{a', r'}\nonumber\\
    D_{\mu}^{aa', rr'} & = \partial_\mu\delta^{aa'}\delta^{rr'} - 2 i \varepsilon^{aa'c}\,\Gamma_\mu^c\delta^{rr'} + \delta^{aa'}\Gamma_\mu^{r r'}\,.
    \label{eq:delta_cov}
\end{align}
This generates the vector couplings of the $\Delta$ to two pions and to the gauge bosons. Because the $\Delta$ transforms as $1\otimes\frac{1}{2}$, its covariant derivative acts on both indices: in the adjoint representation on the isovector index $a$, through the isospin Levi-Civita symbol $\varepsilon^{aa'c}$, and in the fundamental representation on the isospinor index $r$. Both are built from the nucleon chiral connection of Eq.~\eqref{eq:chiral-connection},
\begin{align}
    \Gamma^{rr'}_\mu \equiv \left(\Gamma_\mu\right)^{rr'}\,,\quad
    \Gamma^a_\mu = \frac{1}{2}\mathrm{Tr}\left(\Gamma_\mu\tau^a\right)\,.
\end{align}
Note the difference in normalization compared to Eq.~\eqref{eq:chiral-symmetry-generator-normalization}.

The chiral vielbein transforms homogeneously so it can enter interaction terms directly. We write its isovector components as
\begin{align}
    u^a_\mu \equiv \frac{1}{2}\mathrm{Tr}\left(u_\mu\tau^a\right)= -\partial_\mu\phi^a/f_\pi + \frac{1}{2}\left(r_\mu^a - l_\mu^a\right) + \dots
\end{align}
At lowest order, each factor of $u_\mu$ carries one pion or one gauge boson field, so it generates the single-pion couplings and the axial couplings to the gauge bosons. For the diagonal couplings, the vielbein can attach to two $\Delta$ fields in three ways. These vertices do not contribute to single-pion production off a nucleon at tree level, and we give the explicit Lagrangian in Appendix~\ref{app:delta-propagator} for completeness.

For the $\pi N\Delta$ transition coupling, the natural choice is to attach one factor of the vielbein between a $\Delta$ and a nucleon field,
\begin{align}
    g\,\overline{\Psi}^a_\mu\left(g^{\mu\nu}
    + \Tilde{z}\gamma^\mu\gamma^\nu\right)u^a_\nu\, N + \mathrm{h.c.}\,, \quad \Tilde{z} = \frac{1}{2}(3A + 1)\,,
    \label{eq:conventional-coupling}
\end{align}
which is the conventional coupling used in much of the literature, consistent with Eqs.~\eqref{eq:point-transformation} and~\eqref{eq:equations-of-motion}~\cite{Nath:1971wp}. However, this coupling is pathological. The interacting theory propagates the unphysical spin-1/2 components, leading to acausal modes~\cite{Pascalutsa:1998pw,Pascalutsa:1999zz}, known as the Johnson-Sudarshan and Velo-Zwanziger problems~\cite{Johnson:1960vt,Velo:1969bt}. The cure is an interaction that cannot excite the unphysical components in the first place. Since they enter $\Psi_\mu$ the way a pure-gauge piece $\partial_\mu\epsilon$ enters a gauge field, an interaction invariant under the spin-3/2 gauge transformation $\Psi_\mu \to \Psi_\mu + \partial_\mu\epsilon$ does not couple to them~\cite{Pascalutsa:1998pw,Pascalutsa:1999zz}. Such interactions are called gauge invariant (GI)~\cite{Pascalutsa:2006up}, where the terminology refers to this transformation and not to the electroweak gauge symmetry. In analogy with electromagnetism, invariance is achieved by building interactions from the field-strength tensor
\begin{align}
    \Psi_{\mu\nu}^a & = \partial_\mu\Psi_\nu^a - \partial_\nu\Psi^a_\mu\,,
\end{align}
which decouples the spurious components identically. The index $a$ again labels the isospin doublets of Eq.~\eqref{eq:delta-doublets}. Following Refs.~\cite{Pascalutsa:2000kd,Pascalutsa:2006up,Mart:2019jtb}, we use the GI pion-nucleon-delta Lagrangian
\begin{align}
    \mathcal{L}_{\pi N\Delta} & = -\frac{h_D}{m_\Delta}\,
    \varepsilon^{\mu\nu\rho\sigma}\,
    \big(\partial_\rho\overline{\Psi}^a_{\sigma}\big)\gamma_\mu\gamma_5\, u^a_\nu\, N + \mathrm{h.c.}\,,
    \label{eq:LpiNDelta-GI}
\end{align}
where $\varepsilon^{\mu\nu\rho\sigma}$ is the totally antisymmetric tensor with $\varepsilon^{0123} = -1$. The leading term of $u^a_\nu$ is $-\partial_\nu\phi^a/f_\pi$, which gives the $\pi N\Delta$ vertex. The next term is the axial gauge field $\frac{1}{2}\left(r^a_\nu - l^a_\nu\right)$, which contains the $W$ and $Z$ but not the photon and gives the axial $N\to\Delta$ transition induced by the weak bosons. Because $\varepsilon^{\mu\nu\rho\sigma}$ is antisymmetric, only the field strength enters, $\varepsilon^{\mu\nu\rho\sigma}\,\partial_\rho\overline{\Psi}^a_\sigma = \tfrac12\,\varepsilon^{\mu\nu\rho\sigma}\,\overline{\Psi}^a_{\rho\sigma}$ which is what makes the coupling gauge invariant. For an on-shell $\Delta$ this coupling reduces to the conventional coupling of Eq.~\eqref{eq:conventional-coupling} with $g = h_D$. We also note that the $\Delta$ field derivative of Eq.~\eqref{eq:LpiNDelta-GI} and all future $\Delta$ field derivatives should be replaced with the $\Delta$ covariant derivative Eq.~\eqref{eq:delta_cov}. Such a replacement would lead to contributions to $2\pi$ production, which are left to future work.

The partial width that follows from Eq.~\eqref{eq:LpiNDelta-GI} is
\begin{align}
    \Gamma\left(\Delta\to N\pi\right) = \frac{h_D^2}{12\pi f_\pi^2}\,\frac{k^3}{m_\Delta}\left(E_N+m_N\right)\,,
    \label{eq:Delta-width}
\end{align}
where $k$ and $E_N$ are the nucleon breakup momentum and energy in the $\Delta$ rest frame. For the same Lagrangian, Eq.~(B.7) of Ref.~\cite{Fernandez-Ramirez:2005nuq} is larger by a factor of four.

The coupling $h_D$ can be estimated in the quark model or fixed from experiment, and the two do not agree. With $g_A = 1.2754$, spin-flavor SU(6) symmetry of the quark model gives $h_D = \frac35\sqrt2\,g_A \simeq 1.08$, which Ref.~\cite{Scherer:2012xha} presents as a rough estimate of a coupling that must ultimately be fitted. It predicts $\Gamma(\Delta\to N\pi) \simeq 49$~MeV, about half the observed width. We instead fix $h_D = 1.557$ from the pole width $\Gamma(\Delta\to N\pi) = 101.4$~MeV.\footnote{Care is required when comparing values of this coupling across the literature. For example, the pseudovector coupling $(f_{\pi N\Delta}/m_\pi)\,\overline{\Psi}_\mu \partial^\mu\pi\, N$ of Ref.~\cite{Praet:2008yn}, with $f_{\pi N\Delta} = 2.21$ fixed from $\Gamma = 120$~MeV at $m_\Delta = 1.232$~GeV, corresponds to $h_D = (f_\pi/m_\pi)\,f_{\pi N\Delta} = 1.47$.} The conventional coupling is not included in our Lagrangian; the GI coupling replaces it entirely.


\subsection{Electroweak couplings}
\label{sec:delta-ew}

We organize the electroweak couplings of the $\Delta$ by their origin. Most of them were already generated by the gauging of the previous sections, with no new operators and no new parameters. The vector $N \to \Delta$ transitions are the missing piece. The gauging cannot produce them, and they enter through one new operator, whose two strengths are the only new parameters of this section, fixed by photon data and extended to the $W$ and $Z$ by the conserved vector current relation (CVC).

The Lagrangians in Sec.~\ref{sec:delta-couplings} already contain the gauge bosons through $\Gamma_\mu$ and $u_\mu$. Reading off their terms with one gauge boson gives three kinds of vertices. First, the covariant derivative of Eq.~\eqref{eq:delta_cov} in Eq.~\eqref{eq:Delta-Lagrangian} generates the couplings of the photon, $W$, and $Z$ to the $\Delta$ alone. Second, the $\pi N\Delta$ operator of Eq.~\eqref{eq:LpiNDelta-GI} contains axial $N \to \Delta$ vertices with one $W$ or $Z$, since the gauge-field part of the vielbein of Eq.~\eqref{eq:chiral-vielbein} is the axial combination of $r_\mu$ and $l_\mu$. Keeping only that part, the operator reads,
\begin{equation}
 \mathcal{L}^A_{N\Delta gauge} = -\frac{h_D}{m_\Delta}\,\varepsilon^{\mu\nu\rho\sigma}
 \big(\partial_\rho\overline{\Psi}^a_{\sigma}\big)\gamma_\mu\gamma_5\, \frac{1}{2}\left(r_\nu^a - l_\nu^a\right)\, N + \mathrm{h.c.}\,,
 \label{eq:LAND}
\end{equation}
with the same coupling $h_D$ as the $\pi N\Delta$ vertex and with the axial source
\begin{equation}
 r_\mu^a - l_\mu^a = -g_2\times
 \begin{cases}
  V_{ud}\, W^a_\mu\,, & a = 1,2\,,\\[2pt]
  Z_\mu/\cos\theta_W\,, & a = 3\,.
 \end{cases}
 \label{eq:axial-source}
\end{equation}
Equation~\eqref{eq:LAND} introduces no new parameter and is the off-diagonal analogue of the Goldberger--Treiman relation~\cite{Goldberger:1958tr}. For the nucleon, the $g_A$ term of Eq.~\eqref{eq:nucleon-lagrangian} gives both the $\pi NN$ vertex and the axial coupling to the $W$, so the two strengths are tied, which is the Goldberger--Treiman relation~\cite{Goldberger:1958tr} (see Appendix~\ref{app:cvc,pcac,ppd}). Here the same mechanism ties the axial $N \to \Delta$ transition to the strong coupling $h_D$. Finally, the same operator generates four-point contact vertices, in which the gauge boson and a pion attach to the $N\Delta$ transition at a single point, as required by electroweak gauge invariance. These vertices do not enter single-pion production at tree level, and we do not discuss them further.

We now turn to the new interactions. Historically, the best studied one is the electromagnetic excitation $p \overset{\gamma}{\to} \Delta^+$~\cite{Pascalutsa:2006up,Burkert:2025coj}. It dominates pion photoproduction in the resonance region, which is also the data we validate against in Sec.~\ref{sec:free}. The gauging cannot produce this coupling, since the photon is absent from $u_\mu$, entering $l_\mu$ and $r_\mu$ equally and canceling in their axial combination. Isospin dictates the form of the operator. Since the nucleon carries isospin $\tfrac{1}{2}$ and the $\Delta$ carries isospin $\tfrac{3}{2}$, an isosinglet cannot excite this transition, and only the isotriplet part of a gauge boson couples to it. The photon contains both pieces, an isosinglet and an isotriplet component, as can be seen from the charge matrix $Q = T^3 + \tfrac{B}{2}1\!\!1$ in Eq.~\eqref{eq:lr-physical}. An operator coupling the photon to this transition must therefore single out the triplet's third direction. Note that, unlike the operators of the strong Lagrangian, whose isospin indices are fully summed, this operator attaches an independent strength to one fixed direction in isospin space, which is why it is typically called an isospin-breaking term. Following Ref.~\cite{Fernandez-Ramirez:2005nuq}, we use
\begin{align}
    \mathcal{L}_{\gamma N\Delta} & = \kappa\overline{N}\left(iG_1
    \Tilde{F}_{\mu\nu} + G_2\gamma_5 F_{\mu\nu}\right)\partial^\mu\Psi^{3\,\nu}
    + \mathrm{h.c.}\,,
    \label{eq:gammaNDelta}
\end{align}
where $\kappa = \frac{3e}{2m_N\left(m_N + m_\Delta\right)}$, $\Tilde{F}_{\mu\nu} = \frac{1}{2}\varepsilon_{\mu\nu\rho\sigma}F^{\rho\sigma}$ is the dual field strength, and $\Psi^{3\,\nu}$ is the third isovector component, i.e. one doublet of Eq.~\eqref{eq:delta-doublets}. The couplings $G_1, G_2$ are set to $G_1 = 3.07$ and $G_2 = 1.22$.\footnote{We have neglected the Coulomb term in Eq.~\eqref{eq:gammaNDelta}, which we tested to have less than a 0.5\% impact on electron scattering results. Inclusion will be left to future work.} The ratio $G_2/G_1 = 0.398$ is taken from the fit of Ref.~\cite{Fernandez-Ramirez:2005nuq} to photoproduction data, and the overall scale is set by the radiative width $\Gamma(\Delta \to N\gamma) = 0.572$~MeV at the pole mass obtained by scaling the $\Delta$ width by the PDG branching ratio~\cite{ParticleDataGroup:2026mpi}. For the coupling of Eq.~\eqref{eq:gammaNDelta} this width reads
\begin{align}
    \Gamma\left(\Delta^+\to\gamma p\right) 
    & = \frac{e^2 (m_\Delta-m_N)^4 (m_\Delta+m_N)^2}{64 \pi  m_\Delta^3 m_N^2}\nonumber\\
   & \quad\times\left(\frac{1}{\xi} G_1^2+G_1 G_2+\xi G_2^2\right)\,,
   \label{eq:gammaNDelta-width}
\end{align}
with $\xi = \frac{m_\Delta-m_N}{m_\Delta+m_N}$. Note that photoproduction fixes only the products $h_D G_i$. The fitted set 1 of Ref.~\cite{Fernandez-Ramirez:2005nuq} gives $h\,G_1 = 4.63$ and $h\,G_2 = 1.84$, against $4.78$ and $1.90$ here, so the width-derived couplings agree with their fit to within about three percent.

Finally, we turn to the vector transitions of $W$ and $Z$. CVC relates them to the electromagnetic transition. CVC is the statement~\cite{Feynman:1958ty} that the vector part of the charged weak current, $\bar u\gamma^\mu d$, and the isovector part of the electromagnetic current, $J^\mu_{\mathrm{EM},\,I=1} = \tfrac{1}{2}\left(\bar u\gamma^\mu u - \bar d\gamma^\mu d\right)$, are components of one conserved isotriplet $V^{a\,\mu} = \bar q\gamma^\mu T^a q$, with $q = (u, d)^T$, 
\begin{align}
    V^{1\,\mu} + i\, V^{2\,\mu} = \bar u\gamma^\mu d\,,
    \qquad
    V^{3\,\mu} = J^\mu_{\mathrm{EM},\,I=1}\,.
    \label{eq:cvc-triplet}
\end{align}
The weak vector transitions therefore follow from Eq.~\eqref{eq:gammaNDelta}. The $W^\pm$ couple to $\bar u\gamma^\mu d$ and $\bar d\gamma^\mu u$, the two charged components of the triplet, while the photon and the $Z$ both couple to its third component, $V^{3\,\mu}$. To obtain the corresponding Lagrangian, we replace the photon field strength by that of the isovector source,
\begin{align}
    V^a_\mu \equiv\frac{1}{2e}\left(r_\mu^a + l_\mu^a\right) =
    \begin{cases}
        \displaystyle\frac{V_{ud}}{2\sin\theta_W} \, W^a_\mu\,, & a = 1,2\,, \\[1em]
        \displaystyle A_\mu + \cot\left(2\theta_W\right) Z_\mu\,, & a = 3\,.
    \end{cases}
    \label{eq:isovector-source}
\end{align}
Its field strength is $V^a_{\mu\nu} = \partial_\mu V^a_\nu - \partial_\nu V^a_\mu$, $\Tilde{V}^{\,a}_{\mu\nu}$ is its dual, and the physical charged fields are again $W^\pm_\mu = (W^1_\mu \mp i\,W^2_\mu)/\sqrt{2}$, so that Eq.~\eqref{eq:gammaNDelta} becomes
\begin{align}
    \mathcal{L}_{N\Delta gauge} & = \kappa \overline{N}\left(iG_1
    \Tilde{V}^{\,a}_{\mu\nu} + G_2\gamma_5 V^a_{\mu\nu}\right)\partial^\mu\Psi^{a\,\nu}
    + \mathrm{h.c.}
    \label{eq:VNDelta}
\end{align}
Note that the normalization and signs of Eq.~\eqref{eq:isovector-source} are fixed, since its third component must reproduce the photon operator of Eq.~\eqref{eq:gammaNDelta}, and the other two components follow by isospin. In Eq.~\eqref{eq:VNDelta} all isospin indices are again summed, so the full operator has the same invariant structure as the couplings inherited from the gauging, and the isospin-breaking label really attaches to its photon-only form, Eq.~\eqref{eq:gammaNDelta}. What distinguishes the operator is that its strengths $G_1$ and $G_2$ are new parameters rather than inherited ones. The $Q^2$ dependence assigned to this transition is discussed in Sec.~\ref{sec:formfactors}.


\section{Form Factors}
\label{sec:formfactors}

The vertices of the preceding two sections are point-like, as dictated by the lowest-order Lagrangian. The composite structure of the hadrons, which in strict \chiEFT would enter through higher-order operators, is incorporated instead through empirical form factors, and this is the defining phenomenological step of our approach. Each form factor is attached to an individual vertex and depends only on that vertex's own kinematics, which is what the UFO format supports and what the automated pipeline propagates to every amplitude. 

In general, form factors are defined for on-shell hadrons. In this work, we assume that on-shell form factors apply to off-shell hadrons as well. This allows the CVC, PCAC, and pion-pole-dominance relations in Appendix~\ref{app:cvc,pcac,ppd} to tie the form factors to one another and keeps the fitted input to a minimum. The explicit parameterizations and parameter values are collected in Appendix~\ref{app:FFs}.

As an example, consider first the coupling of a gauge boson to the nucleon. At lowest order, it is a single Lorentz structure with a constant coupling, $\gamma^\mu$ for the photon and $\gamma^\mu(1 - g_A\gamma^5)$ for the $W$. For a composite nucleon, the current can contain every structure that a spin-1 boson coupling to two spin-1/2 particles allows, and the most general form is
\begin{align}
    \braket{N|J^\mu_N\!\left(0\right)|N} & \propto \bar{u}\left[\Gamma_V^\mu\!\left(Q^2\right)
    - \Gamma_A^\mu\!\left(Q^2\right)\right]u\,,
\end{align}
with~\cite{scherer2003introduction,Mukhopadhyay:1998iw}
\begin{align}
    \Gamma_V^\mu & = \mathcal{F}_1 \gamma^\mu
    + \mathcal{F}_2\, \frac{i\sigma^{\mu\nu}q_\nu}{2m_N}
    + \mathcal{F}_S\, \frac{q^\mu}{2m_N}\,,\nonumber\\
    \Gamma_A^\mu & = \mathcal{F}_A \gamma^\mu\gamma^5
    + \mathcal{F}_T\, \frac{i\sigma^{\mu\nu}q_\nu}{2m_N}\gamma^5
    + \mathcal{F}_P\, \frac{q^\mu}{2m_N}\gamma^5\,,
    \label{eq:general-current}
\end{align}
where $q$ is the momentum transfer, $Q^2 = -q^2$, and each form factor $\mathcal{F}_i(Q^2)$ is a real, dimensionless function. At $Q^2 = 0$, the current must reduce to that of a point particle with the nucleon's charge, anomalous magnetic moment, and axial coupling, which fixes the normalizations $\mathcal{F}_i(0)$, and the $Q^2$ dependence describes the departure from point-like behavior. The structures in $\Gamma_V^\mu$ come from the vector part of the current and the structures in $\Gamma_A^\mu$ from the axial part, and we refer to the functions multiplying them as the vector and the axial form factors. The photon has only the former. $\mathcal{F}_1$ and $\mathcal{F}_2$ are the Dirac and Pauli form factors, which encode the charge and magnetization distributions of the nucleon and $\mathcal{F}_A$ is the axial form factor, probed only by the weak interaction. 

The same construction applies to every vertex at which a gauge boson couples to the hadrons of this model, with the Lorentz structures appropriate to the particles involved. These are the $\gamma NN$, $WNN$, and $ZNN$ vertices just discussed, the $\gamma\pi\pi$, $W\pi\pi$, and $Z\pi\pi$ vertices, the $N \to \Delta$ transition vertices of Sec.~\ref{sec:delta-ew}, and the four-point contact vertices in which a boson and a pion attach to the nucleon line at one point. Each structure at each of these vertices carries a form factor, and the question is how many of them are independent. Symmetry fixes most of it.

The vector form factors are fixed by CVC. CVC~(see Appendix~\ref{app:cvc,pcac,ppd}) requires $q_\mu \Gamma_V^\mu = 0$, which ensures that the vector current has no component along the momentum transfer $q$, known as \textit{transversality}. This forces $\mathcal{F}_S = 0$ for on-shell nucleons and, more usefully, ties the weak vector form factors to the electromagnetic ones measured in electron scattering. We take the Dirac and Pauli form factors of the proton and the neutron from the Kelly parametrization~\cite{Kelly:2004hm}. The $W$ and $Z$ vector form factors are isospin combinations of these. The charged current is purely isovector and carries $F_{1,2}^V = F_{1,2}^p - F_{1,2}^n$, and the neutral current carries the corresponding combination given in Appendix~\ref{app:FFs}. The Ward identity further requires the same $F_1^V$ on the $\gamma\pi\pi$, $W\pi\pi$, and $Z\pi\pi$ vertices and on the vector part of the contact term. For the $N \to \Delta$ transition, CVC relates the $W$ and $Z$ vector transitions to the photon one of Eq.~\eqref{eq:VNDelta}, and $G_1$ and $G_2$ are multiplied by a common dipole that equals unity at $Q^2 = 0$.

The axial form factors, $\mathcal{F}_A$ and $\mathcal{F}_P$, are tied to one another by PCAC. The axial current is not conserved, but its divergence is proportional to the pion field, $\partial_\mu A^{a\,\mu} = f_\pi m_\pi^2\,\pi^a$, so that it is conserved in the chiral limit (see Appendix~\ref{app:cvc,pcac,ppd}). Physically, the axial current can convert into a pion, which the nucleon then absorbs, and this process relates $F_P$ to $F_A$. At $Q^2 = 0$ the same connection is the Goldberger--Treiman relation between $g_A$ and $g_{\pi NN}$. Finally, $\mathcal{F}_T$ disappears for all bosons and nucleons in the on-shell limit with $m_n=m_p$, so what remains to be measured is therefore $F_A$ alone. Electron scattering does not reach it, so its normalization is the axial coupling $g_A$ of Sec.~\ref{sec:chipt} and its $Q^2$ dependence is taken from the $z$-expansion fit to deuterium data of Ref.~\cite{Meyer:2016oeg}, and the $Z$ axial form factor follows from $F_A$ by isospin. The axial part of the contact term is not tied to $F_A$ by any such relation, and there we use the $\rho$-dominance form~\cite{Sakurai:1960ju}. PCAC constrains the $N \to \Delta$ transition in the same way, with the $\Delta$ treated as on shell. The leading-order Lagrangian generates a single axial transition form factor, $C^A_5$, whose normalization $C^A_5(0) = \sqrt{2/3}\,h_D$ is fixed by the strong coupling through the off-diagonal Goldberger--Treiman relation of Sec.~\ref{sec:delta-ew}, and whose $Q^2$ dependence we take from the fit of Ref.~\cite{Lalakulich:2005cs} with $M_{A\Delta} = 1.05$~GeV. The pion pole of the transition, $C^A_6$, follows from PCAC as $F_P$ does, and the remaining Adler form factors $C^A_3$ and $C^A_4$ vanish at this order.

The pion is on shell in every diagram except the pion-in-flight term, where the Ward--Takahashi identity would be violated by a form factor on the $\pi NN$ vertex. In the minimal-substitution prescription of Ref.~\cite{Ohta:1989ji}, the identity is restored by an additional contact current, by which this form factor is canceled exactly for on-shell external hadrons. The $\pi NN$ vertex is therefore kept point-like, with the leading-order Goldberger--Treiman coupling $g_{\pi NN} = g_A m_N/f_\pi$, about 4\% below its value at the pion pole (Appendix~\ref{app:FFs}), and $h_D$ is fixed from the $\Delta$ width. The $\Delta$ leg of the strong vertex, however, is off shell, and the vertex grows with its virtuality. We therefore multiply the $\pi N\Delta$ vertex, and the contact vertices that gauging it generates, by a regulating function of the $\Delta$ invariant mass, $F_\Delta(p_\Delta^2)$ of Eq.~\eqref{eq:FDeltaLam}, which is unity on shell and damps the resonance away from its pole. Reference~\cite{Fernandez-Ramirez:2005nuq} instead applies form factors at the amplitude level~\cite{Davidson:2001rk}, one function multiplying each Born or resonance amplitude as a whole, together with extra terms that restore current conservation. We do not need them, because every vertex of Secs.~\ref{sec:chipt} and~\ref{sec:delta} is gauge invariant by construction, and the UFO format in any case supports only vertex-level form factors.

What remains as fitted input is therefore small. The electromagnetic form factors $F_1$ and $F_2$ of the proton and the neutron come from electron scattering, the $Q^2$ dependence of $F_A$ from deuterium, the shape of $C^A_5$ and the two dipole masses from the fits quoted above, and the cutoff $\Lambda$ of the $\Delta$ regulator from Ref.~\cite{Fernandez-Ramirez:2005nuq}. Every normalization is fixed by $g_A$, by the $\Delta$ widths, or by the photon-point couplings, and the inputs are summarized in Table~\ref{tab:inputs}.


\section{Scattering on Free Nucleons}
\label{sec:free}

\begin{figure*}
    \centering
    \includegraphics[width=0.9\linewidth]{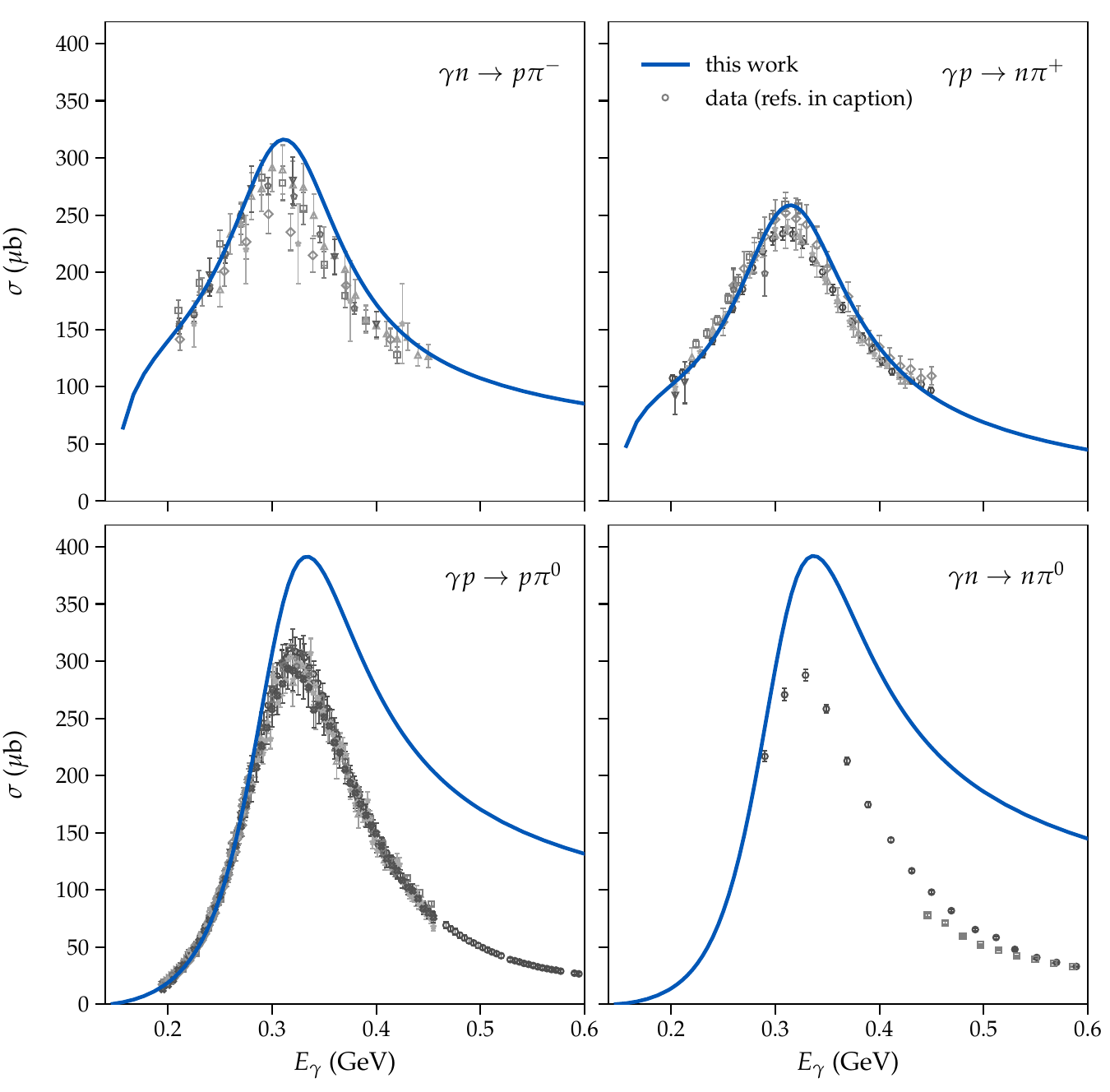}
    \caption{Total cross sections for single-pion photoproduction on the nucleon in the $\Delta(1232)$ region. The four panels show the four charge channels: $\gamma n \to p\pi^-$ (top left), $\gamma p \to n\pi^+$ (top right), $\gamma p \to p\pi^0$ (bottom left), and $\gamma n \to n\pi^0$ (bottom right). The solid blue curves show the result of this work, obtained from the automated pipeline in \textsc{Sherpa} with vertex-level form factors. Data for $\gamma n \to p\pi^-$ are from Refs.~\cite{White:1960ukk,Aachen-Bonn-Hamburg-Heidelberg-Muenchen:1973kaj,Rossi:1973wf,VonHoltey:1974br,Fujii:1976jg,Bagheri:1987kf,Wang:1992thesis}; for $\gamma p \to n\pi^+$ from Refs.~\cite{Knapp:1959zz,Fischer:1972mt,Fujii:1976jg,Fissum:1996fi,MacCormick:1996jz,Blanpied:2001ae,GDH:2004ydy}; for $\gamma p \to p\pi^0$ from Refs.~\cite{Vasilkov:1960jetp,Govorkov:1968sjnp,Hitzeroth:1969du,Fischer:1971zp245,Fuchs:1996ja,Haerter:1996thesis,MacCormick:1996jz,GDH:2004ydy,CB-ELSA:2004sqg,Schumann:2010js,A2:2012lnr,A2:2015mhs}; and for $\gamma n \to n\pi^0$ from Refs.~\cite{A2:2018jcd,A2:2019yud}. Adding vector-meson exchange and an energy-dependent $\Delta$ width with the corresponding Breit--Wigner parameters accounts for most of the discrepancy with the data, as Fig.~\ref{fig:pi0p-variants} shows for $\gamma p \to p\pi^0$. For details see Appendix~\ref{app:lineshape}.}
    \label{fig:photoproduction}
\end{figure*}

\begin{figure}
    \centering
    \includegraphics[width=\columnwidth]{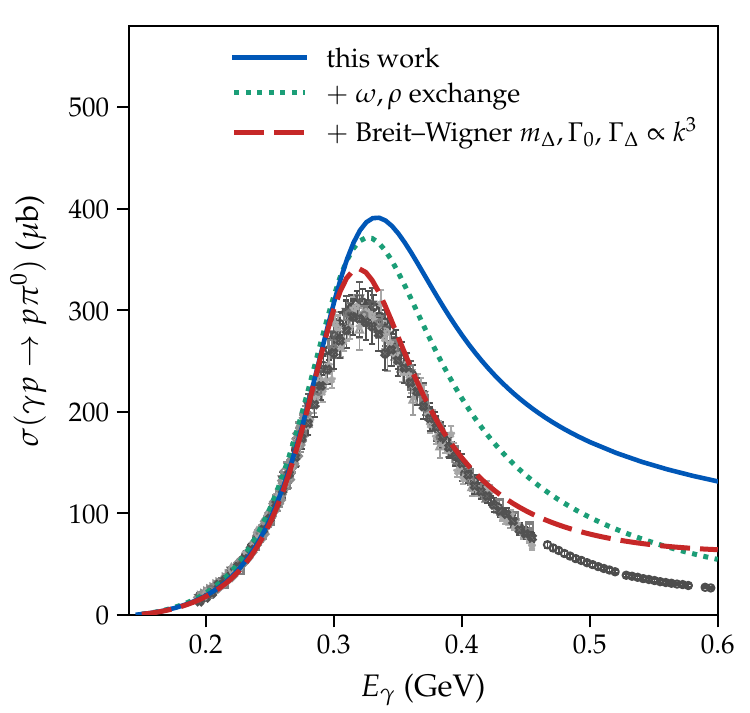}
    \caption{The $\gamma p \to p\pi^0$ channel of Fig.~\ref{fig:photoproduction} with the modifications discussed in the text, applied one after another. Solid blue: this work. Dotted green: with $t$-channel $\omega$ and $\rho$ exchange added with the couplings of Ref.~\cite{Fernandez-Ramirez:2005nuq}. Dashed red: adding the energy-dependent $\Delta$ width $\Gamma(W) = \Gamma_0\,(k/k_0)^3\,(m_\Delta/W)$ and the Breit--Wigner parameters $m_\Delta = 1.232$~GeV, $\Gamma_0 = 117.7$~MeV, with $h_D$, $G_1$, and $G_2$ re-derived from the Breit--Wigner widths $\Gamma(\Delta \to N\pi) = 117$~MeV and $\Gamma(\Delta \to N\gamma) = 0.66$~MeV and ratio $G_2/G_1$ from Ref.~\cite{Fernandez-Ramirez:2005nuq}. Open markers are the data of Fig.~\ref{fig:photoproduction}. The other three channels are shown in Appendix~\ref{app:lineshape}.}
    \label{fig:pi0p-variants}
\end{figure}

\begin{figure}
    \centering
    \includegraphics[width=\columnwidth]{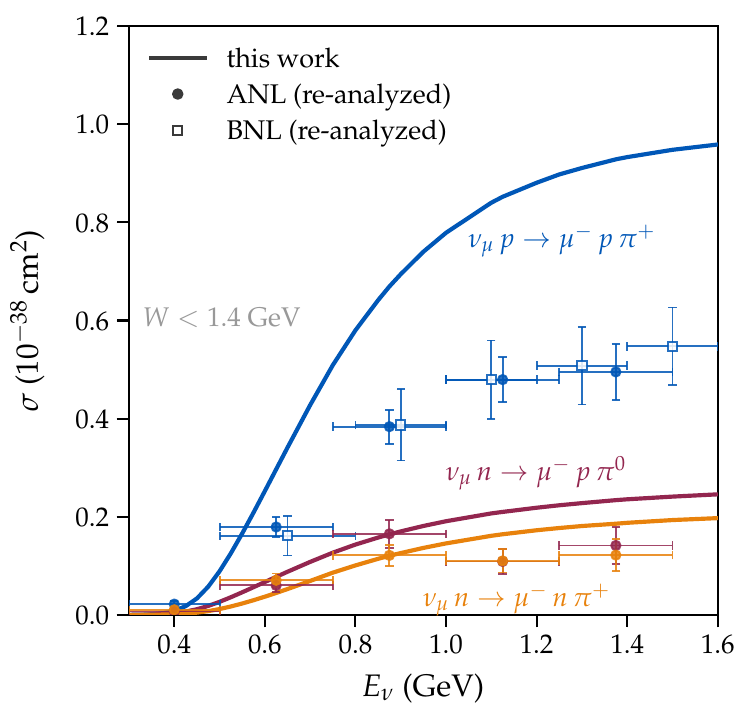}
    \caption{Total cross sections for charged-current single-pion production on free nucleons, computed with the automated pipeline in \textsc{Sherpa} and compared to data. A cut on the hadronic invariant mass, $W < 1.4$~GeV, is applied to both the calculation and the data. Data are from the ANL~\cite{Radecky:1981fn} and BNL~\cite{Kitagaki:1986ct} deuterium bubble-chamber experiments, with the normalizations of the re-analysis of Refs.~\cite{Wilkinson:2014yfa,Rodrigues:2016xjj}. The excess over the data is a property of the tree-level $\Delta$ line shape. An energy-dependent width with the corresponding Breit--Wigner parameters brings all three channels into broad agreement with the data, see Appendix~\ref{app:lineshape}.}
    \label{fig:nuN}
\end{figure}

So far, we have described the analytic ingredients of the model. A major technical component of this work is to turn them into event-generator predictions automatically, without any process-specific intervention. The Lagrangian, with its nucleon, pion, and $\Delta$ sectors and their couplings to the electroweak bosons, is implemented in \textsc{FeynRules}, which derives the complete set of interaction vertices and exports them in the UFO format. Form-factor support in UFO~2.0 lets each vertex carry its own $Q^2$-dependent form factors, specified once, at the level of the Lagrangian.  \textsc{Sherpa} reads the resulting model, computes the complete tree-level matrix element for a requested initial and final state with \textsc{Comix}, and integrates over phase space. Computing a new channel therefore requires no additional amplitude derivation. Once the external states are named, every diagram the Lagrangian permits, resonant, non-resonant, and their interference, is included coherently and automatically. The workflow, the model files, and an example run card are given in Appendix~\ref{app:howtouse}.

We test the pipeline in the two settings where free-nucleon, or nearly free-nucleon, measurements exist. Pion photoproduction isolates the vector current, and charged-current neutrino scattering on deuterium probes the full electroweak structure. The inputs used in the model can be found in Table~\ref{tab:inputs}. Both are discussed below.

\begin{table*}[htbp]
\centering
\caption{Inputs of the model. The upper block is the chiral Lagrangian and the
$\Delta$; the middle block names the form-factor prescriptions; the lower block provides the electroweak parameters.}
\label{tab:inputs}
\renewcommand{\arraystretch}{1.2}
\begin{tabular}{@{}lp{0.42\textwidth}l@{}}
\toprule
Quantity & Value or prescription & Source \\
\midrule
\multicolumn{3}{@{}l}{\emph{Chiral Lagrangian}} \\
$f_\pi$             & $93$~MeV                   & \\
$g_A = F_A(0)$               & $1.2754$                    &~\cite{ParticleDataGroup:2026mpi} \\
$m_{\pi^0}$, $m_{\pi^\pm}$ & $134.98$, $139.57$~MeV &~\cite{ParticleDataGroup:2026mpi} \\
$m_N$               & $938.272$~MeV              &~\cite{ParticleDataGroup:2026mpi} \\
$m_\Delta$, $\Gamma_\Delta$ & $1209$~MeV, $102$~MeV (pole values) &~\cite{ParticleDataGroup:2026mpi} \\
$\Gamma(\Delta\to N\pi)$    & $101.43$~MeV; fixes $h_D$ &~\cite{ParticleDataGroup:2026mpi} \\
$\Gamma(\Delta\to N\gamma)$ & $0.5722$~MeV; fixes the scale of $G_1,G_2$
                            &~\cite{ParticleDataGroup:2026mpi} \\
$G_2/G_1$                   & $0.398$; gives E2/M1 $=-2.45\%$
                            &~\cite{Fernandez-Ramirez:2005nuq} \\
\midrule
\multicolumn{3}{@{}l}{\emph{Form factors}} \\
Nucleon vector $F_{1,2}$ & Kelly parametrization &~\cite{Kelly:2004hm} \\
Nucleon axial $F_A$      & deuterium $z$-expansion, $F_A(Q^2)=g_A\tfrac{\widetilde{F}_A(Q^2)}{\widetilde{F}_A(0)}$
                         &~\cite{Meyer:2016oeg} \\
Induced pseudoscalar $F_P$ & PCAC \& PPD fix this, not a free parameter & \\
$N\to\Delta$ vector        & CVC dipole, $M_V^2 = 0.71$~GeV$^2$ &~\cite{Lalakulich:2005cs}\\
$N\to\Delta$ axial $C_5^A$ &  Paschos–Lalakulich form, $M_{A\, \Delta}=1.05$~GeV&~\cite{Lalakulich:2005cs}\\
$\pi N\Delta$~off-shell $\Delta$ & $\Lambda^4/[\Lambda^4 + (p_\Delta^2 - m_\Delta^2)^2]$, $\Lambda = 1.121$~GeV &~\cite{Fernandez-Ramirez:2005nuq} \\
Axial contact              & $\rho$ dominance, $m_\rho = 775.8$~MeV &~\cite{Sakurai:1960ju}\\
\midrule
\multicolumn{3}{@{}l}{\emph{Electroweak}} \\
$1/\alpha$        & $137.036$, the $Q^2 = 0$ value
                  &~\cite{ParticleDataGroup:2026mpi} \\
$G_F$             & $1.16637\times10^{-5}$~GeV$^{-2}$
                  &~\cite{ParticleDataGroup:2026mpi} \\
$\sin^2\theta_W$  & $0.23863$ at $Q^2\to0$, for the NC charges
                  &~\cite{ParticleDataGroup:2026mpi} \\
$\theta_c$        & $0.2277$ &~\cite{ParticleDataGroup:2026mpi} \\
\bottomrule
\end{tabular}
\end{table*}


\subsection{Pion photoproduction}
\label{sec:photoprod}

Pion photoproduction in the $\Delta(1232)$ region is the cleanest consistency check of the hadronic model. The photon couples only to the vector current, the four charge channels are related by isospin, and total cross sections are measured at the few-percent level in three of the four channels, most precisely at MAMI, ELSA, and LEGS~\cite{A2:2015mhs,CB-ELSA:2004sqg,Blanpied:2001ae}. Figure~\ref{fig:photoproduction} shows the total cross sections for $\gamma n \to p\pi^-$, $\gamma p \to n\pi^+$, $\gamma p \to p\pi^0$, and $\gamma n \to n\pi^0$ as functions of the photon energy, computed with the automated pipeline and with every parameter fixed as described in Secs.~\ref{sec:delta} and~\ref{sec:formfactors}. Nothing here has been adjusted to these data.

The charged-pion channels are described from threshold through the resonance peak to within about ten percent, in both normalization and shape. Since no parameter has been adjusted to these data, this agreement tests both the physics content of the model, the $\Delta$ with couplings fixed by its widths on top of the chiral Born terms, and its implementation in the automated pipeline. 

There are two visible discrepancies in the neutral-pion channel $\gamma p \to p\pi^0$. First, the peak is about 30\% too high and sits about 15~MeV above the data. Second, above the resonance the calculation falls off too slowly compared to the data, by a factor of three at $E_\gamma = 0.5$~GeV. The $\gamma n \to n\pi^0$ channel shows the same two discrepancies, with the peak about 40\% too high and the tail a factor of three too high at 0.5~GeV.

Both discrepancies are understood, and Fig.~\ref{fig:pi0p-variants} shows their resolution. There, we add the ingredients that are not yet in our automated pipeline one at a time to a semi-analytic calculation that reproduces the pipeline result, and show how the calculation changes. The solid blue curve is the pipeline result of Fig.~\ref{fig:photoproduction}. Next, we add dynamical degrees of freedom beyond the $\Delta$, the $\omega$ and $\rho$ mesons, and the dotted green curve shows that they damp the tail. The $\omega$, whose $\omega\pi\gamma$ coupling involves the neutral pion only, dominates the effect and interferes destructively with the Born terms, while the charged channels are nearly untouched. Lastly, the dashed red curve shows the effect of changing the $\Delta$ description from a constant width at the pole parameters to an energy-dependent width at the Breit--Wigner parameters. With a constant width in the propagator the resonance is too strong above its pole. A width that grows with the pion momentum, together with the Breit--Wigner parameters that such a width defines, brings the peak to within about ten percent of the data. Appendix~\ref{app:lineshape} traces the mechanism and shows the same curves for the other three channels. Above $E_\gamma \simeq 0.5$~GeV the red curve overshoots the data because the energy-dependent width is derived from a threshold expansion, which away from the resonance pole produces an excessively large resonant contribution~\cite{ParticleDataGroup:2025Resonances}.


\subsection{Charged-current single-pion production}
\label{sec:cc1pi}

Neutrino scattering probes the axial current, which photoproduction cannot reach, and it is the process the generator is built for. Free-nucleon neutrino data are scarce. The only measurements at these energies are the ANL and BNL deuterium bubble-chamber experiments of the 1980s, which cover the three charged-current single-pion channels, $\nu_\mu p \to \mu^- p\pi^+$, $\nu_\mu n \to \mu^- p\pi^0$, and $\nu_\mu n \to \mu^- n\pi^+$~\cite{Radecky:1981fn,Kitagaki:1986ct}. Figure~\ref{fig:nuN} compares the pipeline prediction with these data. Note that the ANL and BNL cross sections as originally published differ in normalization by 30--40\%, a long-standing puzzle traced to the neutrino flux prediction each experiment used. References~\cite{Wilkinson:2014yfa,Rodrigues:2016xjj} re-analyzed the data with a method in which the flux cancels, and the two data sets then agree. We therefore show the re-analyzed data. A cut on the hadronic invariant mass, $W < 1.4$~GeV, is applied identically to the data and to the calculation. It restricts the comparison to the $\Delta$ region where the theory applies.

The channel $\nu_\mu p \to \mu^- p\pi^+$ proceeds through the $\Delta^{++}$ alone and is pure isospin $3/2$, which makes it the weak analogue of $\gamma p \to p\pi^0$. The two $\nu_\mu n$ channels contain isospin-$1/2$ amplitudes and non-resonant contributions in comparable measure. The calculation is about 80\% above the data in the pure isospin-$3/2$ channel, about 50\% above in the $p\pi^0$ channel, and about 15\% above in the $n\pi^+$ channel. This is the same effect seen at the $\gamma p \to p\pi^0$ peak, now integrated over the resonance. With a constant width the $\Delta$ carries too much strength above its pole. Appendix~\ref{app:lineshape} shows that the energy-dependent width with Breit--Wigner parameters brings all three channels into broad agreement with the data, with the couplings re-derived from the Breit--Wigner widths and no other changes.

We emphasize that no parameter is tuned to these data. The axial $N\to\Delta$ coupling $C_5^A(0)$ is the Goldberger--Treiman value implied by the $\pi N \Delta$ coupling and the $\Delta$ width, $C_5^A(0) = \sqrt{2/3}\, h_D = 1.27$, and not a fit. Reference~\cite{Alvarez-Ruso:2015eva} fitted $C_5^A(0)$ to the re-analyzed $\nu_\mu p \to \mu^- p\pi^+$ data below $1.1$~GeV with the Breit--Wigner $\Delta$ parameters and an energy-dependent width and found $1.14 \pm 0.07$. With the same $\Delta$ parameters our Goldberger--Treiman value is $1.18$, consistent with their fit.
This value includes the Olsson phases~\cite{Olsson:1974sw} that impose Watson's theorem~\cite{Watson:1952ji} on the $\Delta$ amplitude, which our calculation does not. A detailed discussion of the differences is given in Appendix~\ref{app:lineshape}.


\section{Scattering on Nuclei}
\label{sec:bound}

Up to this point, we have been considering interactions of pions, photons, and leptons with individual nucleons. However, in the case of electron and neutrino scattering, the interaction target consists of complex nuclei to increase the overall interaction rate. In order to compare to experimental data, the calculations need to include the nuclear initial state and final state interactions (FSI). In this work, we will consider the spectral function approach~\cite{Benhar:1994hw} for describing the initial state of the nucleons within a nucleus and the \textsc{Achilles} cascade model for FSI~\cite{Isaacson:2020wlx,Isaacson:2025cnk}. Details of these two components are given in the following sections. Furthermore, we will show comparisons using an extended \textsc{Achilles}+\textsc{Sherpa} pipeline to MINERvA, MicroBooNE, MiniBooNE, T2K, and electron-carbon scattering measurements for validation of the entire framework.


\subsection{Spectral function}
\label{subsec: spectral funcs}

Following the \textsc{Achilles} framework laid out in Refs.~\cite{Isaacson:2022cwh} and~\cite{Isaacson:2025cnk}, the hadronic final state can be approximated by the factorized expression
\begin{equation}
    \left\lvert \Psi_f\right\rangle = \left\lvert X\right\rangle \otimes \left\lvert \Psi_f^{A-1}\right\rangle\,,
\end{equation}
where $\left\lvert X\right\rangle$ can be expressed as a plane wave $\left\lvert N \right\rangle$ describing the propagation of a single final state nucleon in the case of quasielastic scattering or as $\left\lvert N \pi\right\rangle$ describing the propagation of a final state nucleon and pion in the case of resonance production.
Similar to the final state factorization, we can factorize the initial state using the plane wave impulse approximation and keep only the one-body current contributions as
\begin{equation}
    \left\langle \Psi_0\right\rvert \approx \left\langle k\right\rvert \otimes \left\langle \Psi_f^{A-1}\right\rvert\,,
\end{equation}
where $k$ represents the hole-momentum of the removed nucleon.
Projecting this against the true nuclear ground state, we obtain the spectral function given as
\begin{equation}
    \begin{aligned}
        S(\textbf{k}, E') = \sum_{f_{A-1}} \Bigl\lvert \left\langle \Psi_0 \right\rvert \left. k \right\rangle &\otimes \lvert \Psi_f^{A-1}\bigr\rangle\Bigr\rvert^2 \\  &\times\mskip\medmuskip \delta\left(E' + E_0^A - E_f^{A-1}\right)\,,
    \end{aligned}
\end{equation}
where the sum runs through the possible final states of the $A-1$ spectator nucleons, $E'$ is the excitation energy, $E_0^A$ is the nuclear ground state energy, and $E_f^{A-1}$ is the final nuclear state energy.

The spectral function is generally expressed as a mean-field component and a correlation component. The mean-field component describes the low momentum region, while the correlation component describes the high momentum region in which the $A-1$ state is unbound, with at least one spectator in the continuum.
In this work, we will use the carbon spectral function obtained using correlated-basis function theory~\cite{Benhar:1994hw}, which is the standard in \textsc{Achilles} for carbon.
The argon spectral function used is the one extracted in Ref.~\cite{Nikolakopoulos:2024mjj} which used Thomas Jefferson Lab (JLab) data on ${}^{40}$Ar and ${}^{48}$Ti~\cite{Murphy:2019wed,JeffersonLabHallA:2022cit,JeffersonLabHallA:2022ljj} for the proton and neutron spectral function respectively.

When using the spectral function approach, the removal energy results in an off-shell initial state nucleon. This is traditionally handled by the de Forest prescription, in which the off-shellness of the nucleon is shifted to the gauge boson and then modifying the current to restore gauge invariance~\cite{DeForest:1983ahx}. However, the choice of restoring the gauge invariance is itself gauge dependent as demonstrated in Ref.~\cite{Kelly:1997zza}. We therefore directly use off-shell currents as defined in Ref.~\cite{Isaacson:2025lyx}.


\subsection{Final state interactions}

The modeling of final state interactions is handled through the interface to \textsc{Achilles}, and is modeled through the impact-parameter based cascade introduced in Ref.~\cite{Isaacson:2020wlx} with the extensions to pions from Ref.~\cite{Isaacson:2025cnk}. Here, we summarize the key components of the cascade.

First, a random spatial configuration of protons and neutrons are created. In the case of the ${}^{12}$C nucleus, these are obtained from Green's function Monte Carlo calculations~\cite{Carlson:2014vla}, and thus retain correlation effects. While for argon, we use just mean-field configurations due to the inability to handle the more sophisticated calculations for heavier nuclei.

After setting up the initial nuclear configuration, a struck nucleon is selected from the configuration randomly and used to seed the hard interaction. The outgoing particles from the hard interaction are then propagated from that starting location. Interactions with background nucleons are determined using either the \emph{cylinder} or \emph{Gaussian} impact parameter probability models which depend on the interaction cross section between the propagating particle and the background nucleon~\cite{Isaacson:2020wlx}. Pauli blocking is applied to nucleons using a local Fermi gas model, while pions are never Pauli blocked. Resonances are handled either as propagating particles or are decayed instantly, providing either on-the-fly in-medium corrections to the lifetime through Pauli blocking or interference with non-resonant contributions, respectively. Finally, particles that reach the radius of the nucleus are removed from the list of propagating particles and added to the list of outgoing particles.


\subsection{Comparisons to Data}
\label{sec:nuclear-data}

In this subsection, we compare the full simulation chain, from the Lagrangian to the final-state particles leaving the nucleus, with neutrino single-pion measurements on carbon- and argon-based targets and electron-carbon scattering. We generate events with \textsc{Achilles} using the \textsc{Sherpa} matrix elements of the model exactly as validated in Sec.~\ref{sec:free}, with spectral functions for the initial state and the cascade for the final state, and without two-body currents. The predictions are binned, selected, and normalized against the published data with \textsc{NUISANCE}~\cite{Stowell:2016jfr}, which applies the same final-state selection and kinematic cuts in each measurement to our event samples. We focus on measurements that isolate the $\Delta$, namely the charged-current $\pi^0$ measurement of MINERvA~\cite{MINERvA:2017okh}, the charged-current $\pi^+$ measurements of T2K on hydrocarbon~\cite{T2K:2019yqu,T2K:2021naz}, the neutral-current $\pi^0$ measurement of MicroBooNE, and electron-carbon inclusive cross sections. In Appendix~\ref{app:nuclear}, we also show the $\pi^+$ and $\pi^0$ measurements of MiniBooNE~\cite{MiniBooNE:2010eis,MiniBooNE:2010cxl} and the $\pi^\pm$ measurement of MINERvA~\cite{MINERvA:2014ogb,MINERvA:2016sfc}. All of the neutrino measurements are unfolded, flux-averaged cross sections, so the comparison is made in terms of true final-state quantities and inherits the unfolding model of each experiment. The run card of Appendix~\ref{app:howtouse} is the one used here.

\begin{figure*}
    \centering
    \includegraphics[width=0.9\linewidth]{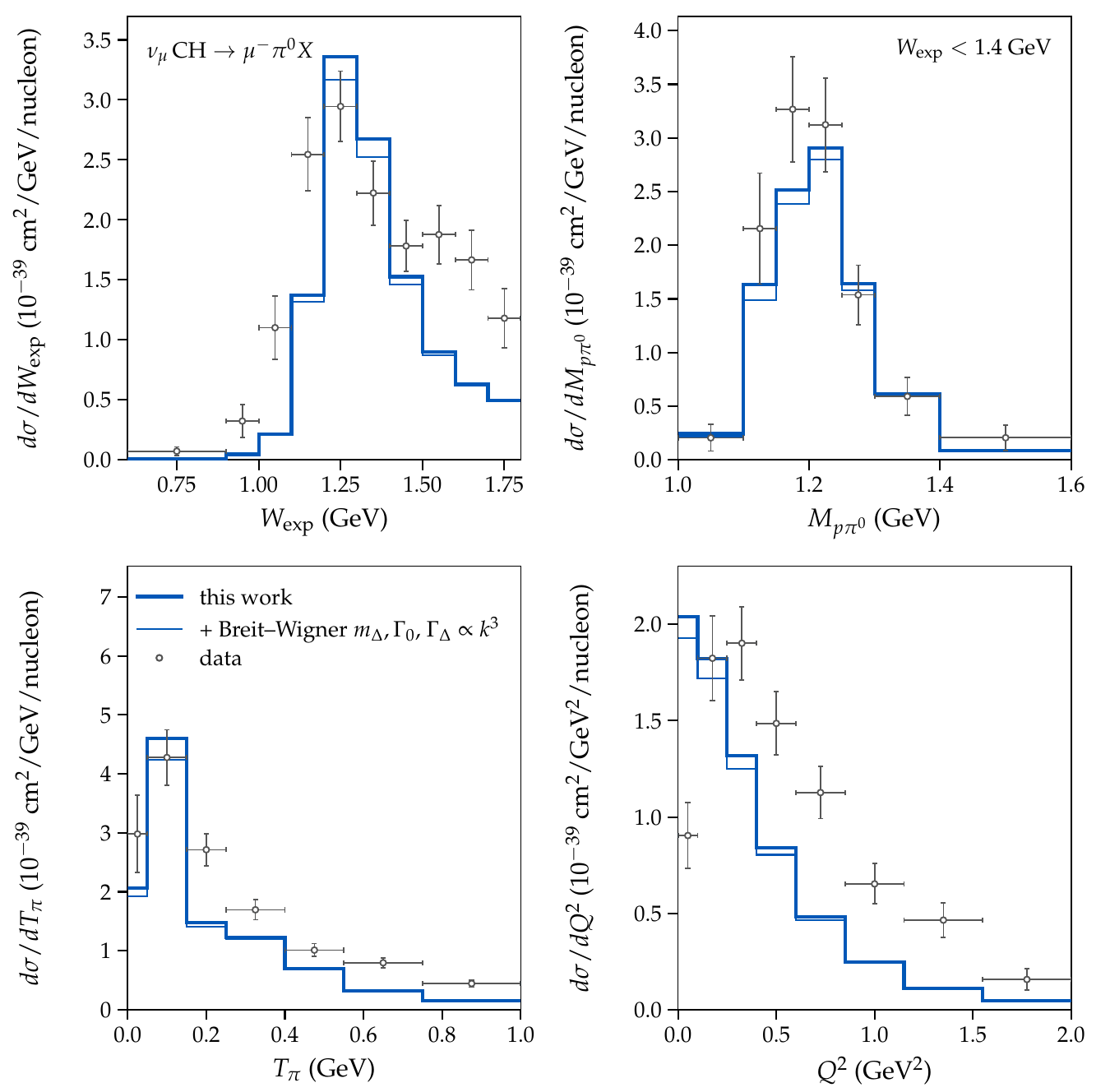}
    \caption{Charged-current $\pi^0$ production on hydrocarbon measured by MINERvA~\cite{MINERvA:2017okh}, $1.5 < E_\nu < 20$~GeV, compared to the full chain of this work. Top left: the hadronic invariant mass $W_{\rm exp}$ reconstructed from the muon, with the experiment's cut $W_{\rm exp} < 1.8$~GeV used in all plots unless noted otherwise. Top right: the $p\pi^0$ invariant mass for events with $W_{\rm exp} < 1.4$~GeV and a proton above $100$~MeV kinetic energy. Bottom: the $\pi^0$ kinetic energy and $Q^2$. Thick lines: this work. Thin lines: the Breit--Wigner variant of Appendix~\ref{app:lineshape}. The $\Delta$ appears at the right mass and with the right strength, while the data above $W_{\rm exp} \simeq 1.4$~GeV, about $40\%$ of the rate at this beam energy, come from higher resonances and non-resonant production that the leading-order model does not contain.}
    \label{fig:minerva-pi0}
\end{figure*}

\begin{figure*}
    \centering
    \includegraphics[width=0.9\linewidth]{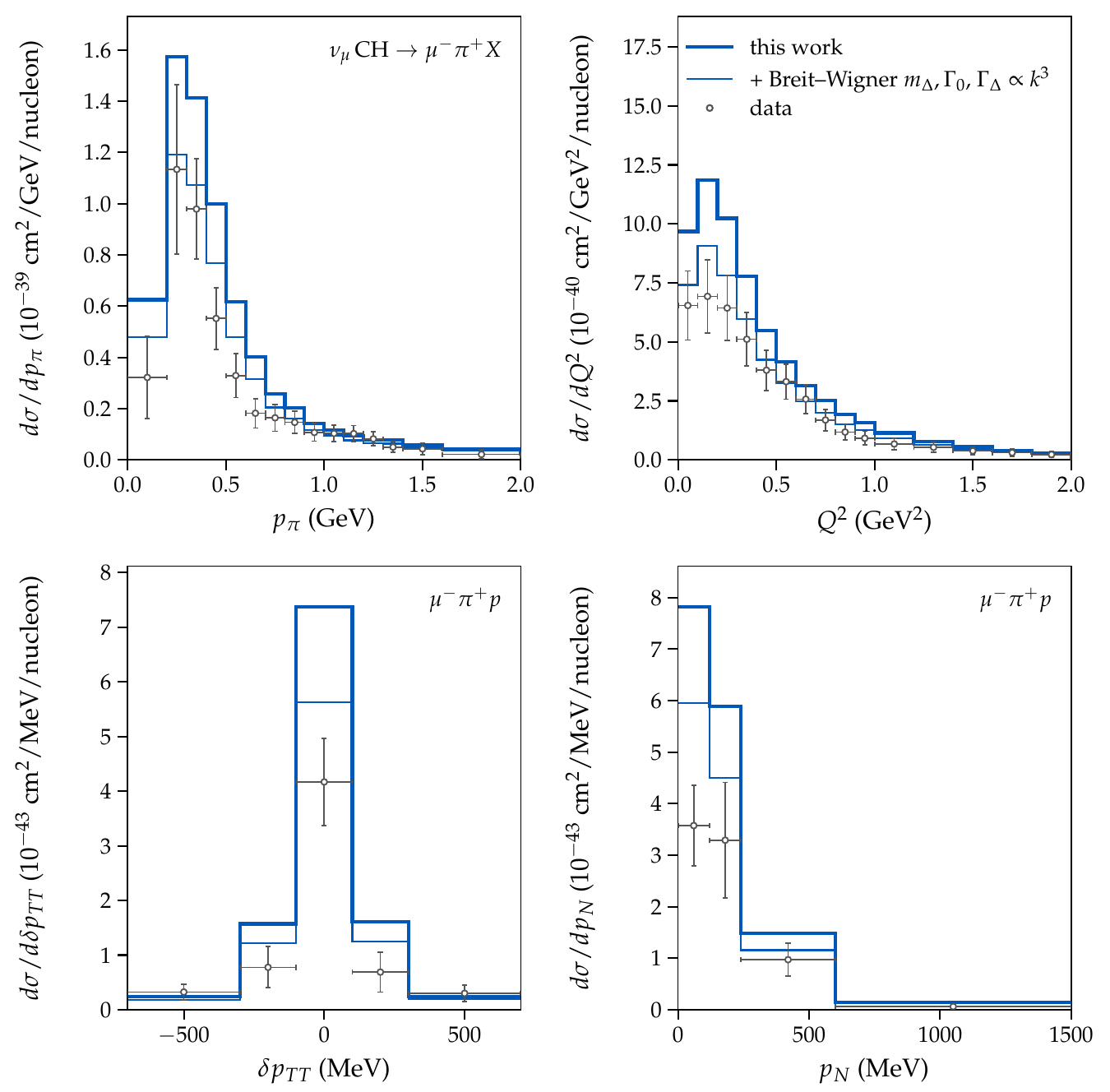}
    \caption{Charged-current $\pi^+$ production on hydrocarbon measured by T2K, compared to the full chain of this work. Top: the pion momentum and $Q^2$ for events with one $\pi^+$, $\cos\theta_\mu > 0.2$, $p_\mu > 200$~MeV, and $\cos\theta_\pi > 0.2$~\cite{T2K:2019yqu}, where the $Q^2$ sample requires $p_\pi > 200$~MeV in addition. Bottom: the double-transverse momentum imbalance $\delta p_{TT}$ and the inferred initial nucleon momentum $p_N$ for events with a $\pi^+$ and a proton~\cite{T2K:2021naz}. Thick lines: this work. Thin lines: the Breit--Wigner variant of Appendix~\ref{app:lineshape}. At the T2K energy the $\Delta$ dominates, and the pipeline sits above the data by the same amount as on free nucleons (Fig.~\ref{fig:nuN}), while the variant significantly improves the comparison to the data in normalization and in the $Q^2$ dependence.
    }
    \label{fig:t2k-pip}
\end{figure*}

Figure~\ref{fig:minerva-pi0} shows the MINERvA $\pi^0$ measurement. For each distribution, we show our nominal result as a thick blue line and, as a thin blue line, an estimate of the effect of the ingredients the pipeline is missing. We have already seen in Sec.~\ref{sec:free} and Appendix~\ref{app:lineshape} that the leading-order calculation overshoots the free-nucleon data through the resonance, and that the energy-dependent width with the Breit--Wigner parameters accounts for most of the excess. However, running widths are currently not implemented within the \textsc{Sherpa} event generator. To estimate the impact this would have on the results, we calculate analytically the ratio between the fixed width and running width cross sections for free nucleons (see Appendix~\ref{app:lineshape} for additional details). This ratio is then applied event-by-event during the analysis step to estimate the impact including the running width would have on our results, and is shown with the thin line. This reweighting is applied to the entire event weight, which is a reasonable approximation since the Delta contribution is dominant.

The top panels show the distributions of $W_{\rm exp}$ and $M_{p\pi^0}$, which are the same quantity measured on the leptonic and on the hadronic side of the event. $W_{\rm exp}$ is inferred from the muon through $W_{\rm exp}^2 = m_N^2 + 2m_N(E_\nu - E_\mu) - Q^2$ assuming the struck nucleon was at rest. $W_{\rm exp}$ therefore represents the mass of the whole hadronic system, smeared by the Fermi motion.  $M_{p\pi^0}$ is the invariant mass of the observed proton and pion, and it represents $m_\Delta$ itself, for the events in which the proton leaves the nucleus with more than $100$~MeV of kinetic energy. For a free nucleon at rest the two coincide. In $W_{\rm exp}$ the model peaks where the data peak and with the same height, and similarly in the $M_{p\pi^0}$ comparison, where the experiment requires $W_{\rm exp} < 1.4$~GeV to isolate the $\Delta^+$ resonance. Above $W_{\rm exp} \simeq 1.4$~GeV, contributions from other resonances like the $N^\ast(1440)$ become dominant, explaining the lack of strength in our model. The inclusion of these higher resonances is left to future work. The same missing tail lowers the model at large $T_\pi$ and at large $Q^2$ in the bottom panels. Under the $\Delta$ peak the model is about $15\%$ below the data, and the Breit--Wigner variant about $20\%$ below. Reference~\cite{MINERvA:2017okh} finds the same deficit in the $\Delta$ region for GENIE with final-state interactions, so it is not specific to this work, and it is intriguing that the deficit appears in a model whose $\Delta$ is tuned to the bubble-chamber data as much as in ours, which is not. The most likely explanation of the deficit is in-medium modification of $\Delta$ properties~\cite{MINERvA:2017okh,MINERvA:2019kfr}. We do not pursue this modification in this work.

Figure~\ref{fig:t2k-pip} shows the T2K $\pi^+$ measurements, taken with a flux that peaks at $0.6$~GeV, where the $\Delta$ is the only resonance that contributes. The top panels show the measurement with one pion and any number of nucleons, and the bottom panels the measurement with one pion and one proton. The pipeline sits about $50\%$ above the data for the first measurement, and $40$--$80\%$ above for the second, which is the excess on free nucleons carried into the nucleus. The Breit--Wigner variant significantly improves the description of the data. It reproduces the pion-momentum and the $Q^2$ distributions to within $5$--$10\%$ in normalization and follows the $Q^2$ shape from $0$ to $2$~GeV$^2$, so that the axial form factor of the $N \to \Delta$ transition is tested at the same time as the line shape. The two variables in the bottom panels are built from the momenta of all three final-state particles~\cite{T2K:2021naz}. The double-transverse momentum imbalance $\delta p_{TT}$ is the sum of the pion and proton momentum components along the axis perpendicular to both the neutrino and the muon directions, which vanishes for scattering on a free nucleon at rest, and $p_N$ is the initial nucleon momentum inferred from the transverse and longitudinal momentum imbalance of the three particles, assuming a carbon nucleus at rest. In both, the central bin is populated by the hydrogen in the target and by carbon events without final-state interactions, while the tails are populated by the Fermi motion and by the cascade, and the model reproduces both.

\begin{figure*}
    \centering
    \includegraphics[width=\linewidth]{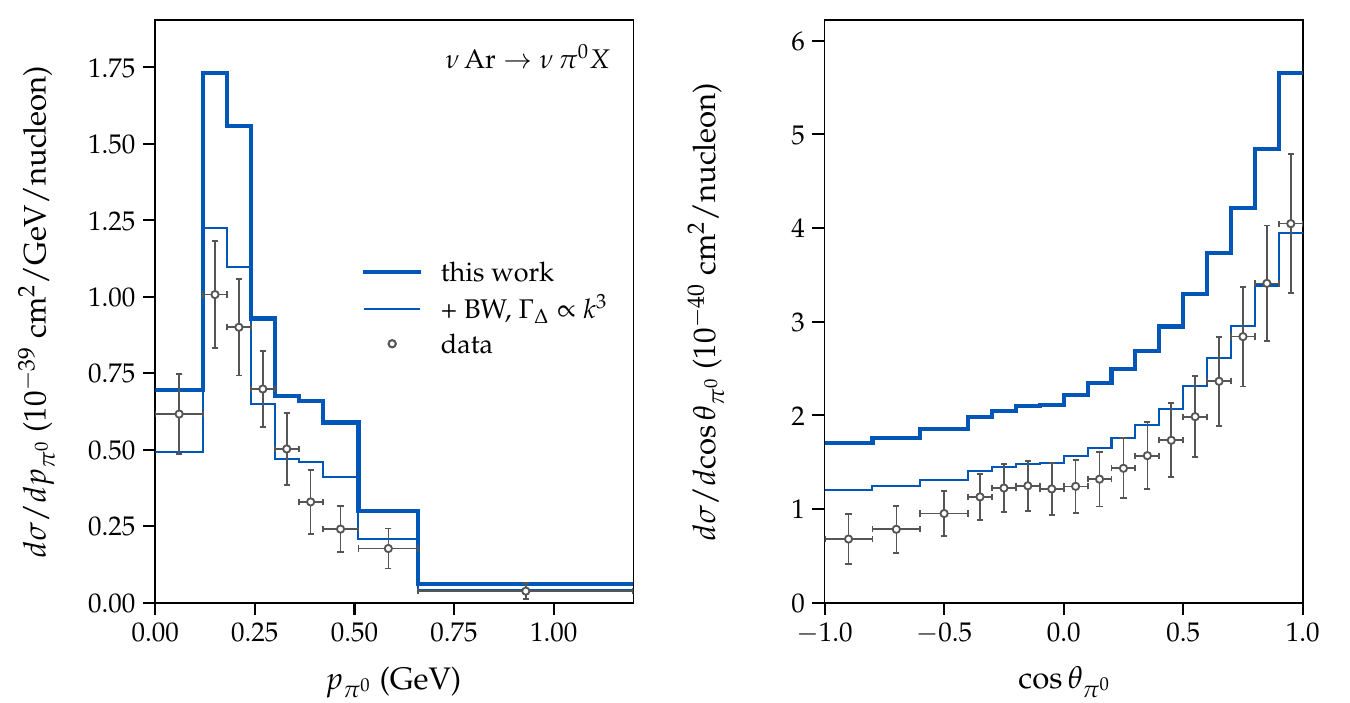}
    \caption{Neutral-current $\pi^0$ production on argon measured by MicroBooNE in the Booster Neutrino Beam~\cite{MicroBooNE:2024pdj}, compared to this work. The signal is exactly one $\pi^0$ with momentum below $1.2$~GeV and any number of nucleons, summed over all neutrino flavors in the beam. The data are unfolded with the Wiener--SVD method, so the predictions are multiplied by the additional smearing matrix $A_C$ provided by the data release. Left: the $\pi^0$ momentum. Right: the cosine of the $\pi^0$ scattering angle with respect to the beam. Thick lines: this work. Thin lines: the Breit--Wigner variant of Appendix~\ref{app:lineshape}, with the running width computed for the neutral-current channels. Coherent nucleus-$\pi^0$ production is not included in the calculation.}
    \label{fig:microboone}
\end{figure*}

Figure~\ref{fig:microboone} shows the neutral-current $\pi^0$ measurement of MicroBooNE on argon~\cite{MicroBooNE:2024pdj}, which the model describes with no additional input, since the $Z$ couplings follow from the same Lagrangian. The left panel shows the $\pi^0$ momentum and the right panel the $\pi^0$ scattering angle.
As for the charged-current channels, the pipeline sits $60$--$70\%$ above the data in the integrated rate, and the Breit--Wigner variant, with the running width computed for the neutral-current channels as in Appendix~\ref{app:lineshape}, reduces the excess to $10$--$20\%$. The remaining excess sits at $p_{\pi^0} \simeq 0.36$--$0.51$~GeV and at backward angles. The agreement at forward angles has to be read with care, since the calculation does not include coherent nucleus-$\pi^0$ production off the argon nucleus as a whole. These events are part of the signal and are strongly forward peaked~\cite{Rein:1982pf,MiniBooNE:2008mmr,MINERvA:2014ani}, so adding them would raise the most forward bins above the data.

\begin{figure*}
    \centering
    \includegraphics[width=\linewidth]{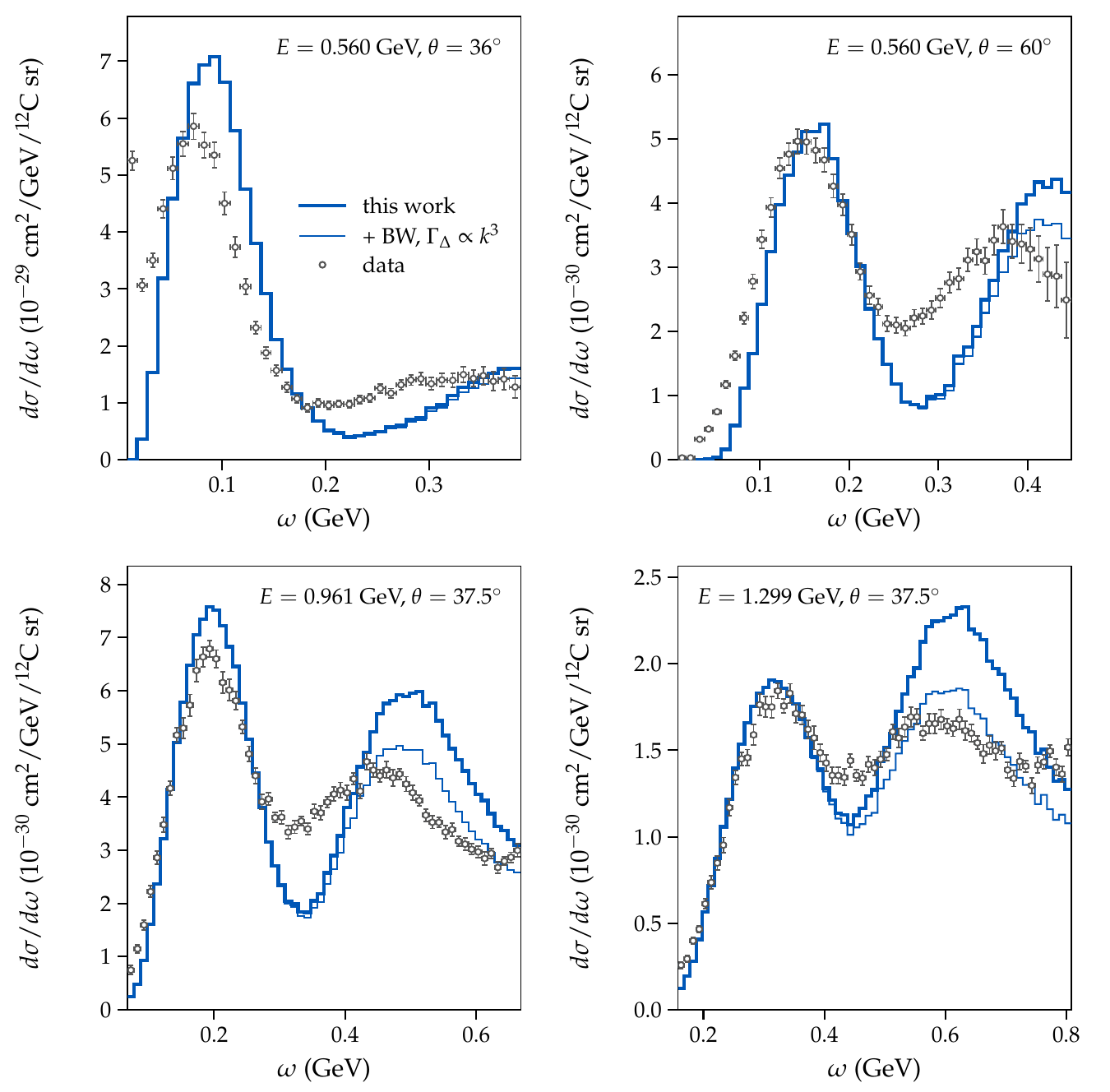}
    \caption{Electron-carbon scattering data measured by various experiments compared to this work. All the plots show the prediction from our model in the thick blue line, with the estimated impact of running width in the thin blue line. All plots show the electron beam energy ($E$) along with the angle for the outgoing electron ($\theta$). The data comes from Refs.~\cite{Barreau:1983ht,Sealock:1989nx}.}
    \label{fig:electron}
\end{figure*}

Figure~\ref{fig:electron} shows comparisons to electron scattering data taken at different beam energies and outgoing electron angles. The plot shows the double differential cross section versus the energy transfer ($\omega = E - E'$).
The first peak is the quasielastic peak and is modeled by nucleon-nucleon-photon interactions, while the second peak contains the contributions from the $\Delta$ resonance. The missing strength in the theory prediction in the dip region between these two peaks arises from the missing meson-exchange currents, and the missing strength at higher $\omega$ will be filled in with higher resonances and deep-inelastic scattering calculations. Furthermore, these calculations only include FSI through the intranuclear cascade, which does not allow for the modification of the lepton kinematics. These corrections are typically included through the use of a folding function~\cite{Benhar:2006hh}, and result in shifting the peak locations and broadening the peaks. However, including it directly in our calculations would result in double counting the FSI physics. The FSI corrections along with including the running width effects and vector mesons should help to resolve the extra strength and shifts in the peak position. A detailed investigation of this is left to future work.

The comparisons with MiniBooNE and with the MINERvA $\pi^\pm$ measurement in Appendix~\ref{app:nuclear} show that the nuclear data themselves limit how precisely any model can be tested. Relative to the Breit--Wigner variant that describes T2K, the MiniBooNE $\pi^+$ data are about $20\%$ higher, the MiniBooNE $\pi^0$ data about a factor of two higher, and the MINERvA $\pi^\pm$ data with $W < 1.4$~GeV about $10\%$ higher. These are the known normalization tensions among the nuclear single-pion datasets~\cite{Betancourt:2018bpu}. They bound what any single model can achieve on nuclear targets.

\section{Conclusions}
\label{sec:conclusions}

In this work we present a model of single-pion production on nucleons and nuclei, based on chiral effective field theory. The model has explicit pions, nucleons, and $\Delta(1232)$ baryons. Chiral symmetry fixes the form of the interaction vertices, and empirical form factors describe the composite structure of the hadrons. We implement the Lagrangian in \textsc{FeynRules}, which derives the vertices and exports them in the UFO format together with their form factors. \textsc{Sherpa} then computes the complete tree-level matrix elements for any requested final state, and \textsc{Achilles} adds the nuclear initial state and the final-state cascade. None of these steps is specific to a process. Every diagram the Lagrangian permits, resonant and non-resonant with their interference, is included automatically. The model files and run cards are public, see Appendix~\ref{app:howtouse}.

We pay particular attention to theoretical consistency, especially for the hadronic form factors. We use CVC to relate the weak vector form factors to the ones measured in electron scattering, and PCAC to relate the pseudoscalar form factors to the axial ones. We fix the the $\pi N\Delta$ and $\gamma N\Delta$ couplings to the $\Delta$ widths and photoproduction. What remains to be taken from neutrino data is the $Q^2$ dependence of the axial form factors, for which we use published fits to data. In particular, the axial $N \to \Delta$ coupling $C_5^A(0) = 1.27$ is a prediction of the model.

We compare the model with pion photoproduction on the nucleon, with neutrino scattering on deuterium, with neutrino scattering for charged current on carbon and neutral current on argon, and with electron scattering on nuclei. In photoproduction, the charged-pion channels are described from threshold through the resonance to within about ten percent. The two discrepancies in the neutral-pion channels are understood. The tail above the resonance is damped by $\omega$ and $\rho$ exchange, and the peak is brought to the data by an energy-dependent $\Delta$ width with the corresponding Breit--Wigner parameters. The excess over the deuterium data and over the T2K measurements is the same line-shape effect, and the energy-dependent width reduces it. The T2K pion-momentum and $Q^2$ distributions are then significantly improved in normalization and in shape. On MINERvA, the $\Delta$ appears at the right mass and with the right strength. There is a deficit under the peak, which GENIE shares and which points to in-medium modification of the $\Delta$, and the strength above $W \simeq 1.4$~GeV comes from higher resonances that our model does not contain yet. The neutral-current $\pi^0$ measurement of MicroBooNE on argon follows the same pattern as the charged-current data. The excess is reduced from $60$--$70\%$ to $10$--$20\%$ by the energy-dependent width, and the forward angles still lack coherent nucleus-$\pi^0$ production. In addition, the normalization tensions among the nuclear single-pion datasets limit how precisely any model can be tested on nuclear targets.

The model is not yet complete, and we plan to develop it further. The energy-dependent width is not yet in the generator and should be implemented there.
These improvements should be implemented together with the phases that Watson's theorem requires for the interference of the $\Delta$ with the non-resonant amplitudes.
The Lagrangian can be extended with the $\rho$ and $\omega$ mesons, with the resonances above the $\Delta$, and with the strange sector. Because the pipeline is automated, each of these extensions only requires adding the corresponding operators to the Lagrangian.
On the nuclear side, two-body currents, in-medium $\Delta$ properties, and coherent nucleus-pion production are the missing ingredients.


\begin{acknowledgments}
We thank George Fleming, Raúl González-Jiménez, Stefan Höche, Alessandro Lovato, Alexis Nikolakopoulos, Noah Steinberg, Tim Tait, Adrian Thompson, and Michael Wagman for discussions. The work of J.I. was supported by the U.S. Department of Energy, Office of Science, Office of Advanced Scientific Computing Research, Scientific Discovery through Advanced Computing (SciDAC-5) program, grant “NeuCol”. The work of S.W.L. was supported by the National Science Foundation under Grant No. PHY-2544442. M.H. acknowledges financial support from a doctoral fellowship of the Friedrich-Ebert-Stiftung during the final steps of this work. The work of M. T. is supported in part by U.S. National Science Foundation Grant PHY-2514888 and Simons Investigator Award No. 376204. The work of Z.GJ is supported by the Science and Technology Research Council. 

Part of our numerical checks and validating that the Lagrangian was correctly translated to FeynRules were carried out with the help of Claude Code. All physical arguments, calculations, and conclusions were developed and verified by
the authors, who take full responsibility for them.
\end{acknowledgments}


\appendix
\setcounter{figure}{0}
\renewcommand{\thefigure}{A\arabic{figure}}
\setcounter{table}{0}
\renewcommand{\thetable}{A\arabic{table}}
\renewcommand{\theHfigure}{A\arabic{figure}}   
\renewcommand{\theHtable}{A\arabic{table}}


\section{Matching to the Electroweak Bosons}
\label{app:gauge-matching}

In this appendix, we derive Eq.~\eqref{eq:lr-physical}. The chiral gauge fields $l_\mu$ and $r_\mu$ are identified with the Standard Model gauge bosons by requiring that the chiral covariant derivatives reproduce the Standard Model ones acting on the light-quark doublet. On the chiral side, the quark doublet couples through $D_\mu q_L = \partial_\mu q_L - il_\mu q_L$ and $D_\mu q_R = \partial_\mu q_R - ir_\mu q_R$, with $l_\mu = l^a_\mu T^a + b_\mu B$ and $r_\mu = r^a_\mu T^a + b_\mu B\,.$ The isoscalar field $b_\mu$ couples to baryon number: a field of baryon number $B$ enters its covariant derivative through $-iB\,b_\mu$. The nucleon and $\Delta$ fields of Sec.~\ref{sec:gauging} have $B = 1$ and the quark doublet has $B = 1/3$, so the isoscalar entries below are $\frac{1}{3}b_\mu\,.$ Turning to the Standard Model side, the hypercharge assignments are
\begin{align}
    Y(Q_L) = \begin{pmatrix}
        1/6 & 0\\
        0 & 1/6
    \end{pmatrix}\,,
    \qquad
    Y(Q_R) = \begin{pmatrix}
        2/3 & 0\\
        0 & -1/3
    \end{pmatrix}\,,
\end{align}
and the Standard Model covariant derivatives are
\begin{align}
    \mathcal{D}_\mu q_L & = \partial_\mu q_L - ig_1 B_\mu Yq_L
    - ig_2 W^{a}_\mu T^a q_L\,,\nonumber\\
    \mathcal{D}_\mu q_R & = \partial_\mu q_R - ig_1 B_\mu Y q_R\,,
\end{align}
where $B_\mu$ is the $U(1)_Y$ boson and $W^a_\mu$ are the $SU(2)_L$ gauge bosons. In the Standard Model, the $SU(2)_L$ partner of $u_L$ is not the mass eigenstate $d_L$ but the combination $d'_L = V_{ud}\, d_L + V_{us}\, s_L + V_{ub}\, b_L$. In the matching below, this amounts to replacing the physical vector boson field $W_\mu=W_\mu^1 +iW^2_\mu$ and its adjoint by $ V_{ud}\,W^{(\dagger)}_\mu$ in the off-diagonal entries. The photon and $Z$ couplings carry no mixing factors because the neutral currents are flavor diagonal. Equating the two covariant derivatives on the left-handed fields gives
\begin{align}
    g_1\begin{pmatrix}
        \frac{1}{6}B_\mu & 0\\
        0 & \frac{1}{6}B_\mu
    \end{pmatrix} + \frac{g_2}{2}\begin{pmatrix}
        W^3_\mu & \sqrt{2} V_{ud} W^\dagger_\mu\\
        \sqrt{2} V_{ud}W_\mu & - W^3_\mu
    \end{pmatrix} & =\nonumber\\
    \begin{pmatrix}
        \frac{1}{2} l^3_\mu + \frac{1}{3} b_\mu& \frac{1}{2}\left(l^1_\mu - i l^2_\mu\right)\\
        \frac{1}{2}\left(l^1_\mu + i l^2_\mu\right) & - \frac{1}{2} l^3_\mu + \frac{1}{3} b_\mu
    \end{pmatrix}\,,
\end{align}
and on the right-handed fields
\begin{align}
    g_1\begin{pmatrix}
        \frac{2}{3}B_\mu & 0\\
        0 & -\frac{1}{3}B_\mu
    \end{pmatrix} = \begin{pmatrix}
        \frac{1}{2} r^3_\mu + \frac{1}{3} b_\mu& \frac{1}{2}\left(r^1_\mu - i r^2_\mu\right)\\
        \frac{1}{2}\left(r^1_\mu + i r^2_\mu\right) & - \frac{1}{2} r^3_\mu + \frac{1}{3} b_\mu
    \end{pmatrix}\,.
\end{align}
This gives us the following relations, 
\begin{align}
    b_\mu & = \frac{g_1}{2}B_\mu\,,\nonumber\\
    l_\mu^3 & = g_2 W_\mu^3\,,\nonumber\\
    l_\mu^1 + i l_\mu^2 & = \sqrt{2}\,g_2 V_{ud} W_\mu\,,\nonumber\\
    l_\mu^1 - i l_\mu^2 & = \sqrt{2}\,g_2 V_{ud} W^\dagger_\mu\,,\nonumber\\
    r_\mu^3 & = g_1 B_\mu\,,\nonumber\\
    r^1_\mu & = r^2_\mu = 0\,,
\end{align}
such that the chiral gauge fields take the form
\begin{align}
    l_\mu & = \frac{g_1}{2}B_\mu 1\!\!1
    + \frac{g_2}{2}\begin{pmatrix}
        W^3_\mu & \sqrt{2}V_{ud} W^\dagger_\mu\\
        \sqrt{2}V_{ud} W_\mu & -W^3_\mu
    \end{pmatrix}\,,\nonumber\\
    r_\mu & = g_1 B_\mu\begin{pmatrix}
        1 & 0\\
        0 & 0
    \end{pmatrix} = g_1 B_\mu\, Q\,.
    \label{eq:lr-BW}
\end{align}
Here, $Q = T^3 + \frac{1}{2}1\!\!1$ is the $B = 1$ case of the charge matrix $Q = T^3 + \frac{B}{2}1\!\!1$ in Eq.~\eqref{eq:lr-physical}, and it gives the physical charges of the proton, the neutron, and the $\Delta$ baryons.

Finally, we rotate to the physical fields,
\begin{align}
    W^3_\mu & = \cos\theta_W\, Z_\mu + \sin\theta_W\, A_\mu\,,\nonumber\\
    B_\mu & = -\sin\theta_W\, Z_\mu + \cos\theta_W\, A_\mu\,,
\end{align}
with
\begin{align}
    e = \frac{g_1 g_2}{\sqrt{g_1^2 + g_2^2}}\,,
    \qquad \tan\theta_W = \frac{g_1}{g_2}\,.
\end{align}
Note that $e = g_2\sin\theta_W = g_1\cos\theta_W$. Substituting into Eq.~\eqref{eq:lr-BW} reproduces Eq.~\eqref{eq:lr-physical} with $B = 1$.


\section{The Spin-3/2 Propagator and the \texorpdfstring{$\pi\Delta$}{Pion-Delta} Lagrangian}
\label{app:delta-propagator}

In this appendix, we collect the propagator following from Eq.~\eqref{eq:Delta-Lagrangian} for general off-shell parameter $A$, together with the equivalent form used in the literature, and explain why the choice among them does not affect our results.

For general $A \neq -\tfrac{1}{2}$, the propagator reads~\cite{Benmerrouche:1989uc}
\begin{equation}
    \begin{aligned}
    S_F^{\mu\nu}(p)= & -\frac{\slashed p+m_{\Delta}}{p^2-m_{\Delta}^2+i\varepsilon}\Big[g^{\mu\nu}-\frac{1}{3} \gamma^\mu \gamma^\nu\\
    &+\frac{1}{3 m_{\Delta}}\left(p^\mu \gamma^\nu-\gamma^\mu p^\nu\right)-\frac{2}{3 m_{\Delta}^2}\, p^\mu p^\nu\Big] \\
    & +\frac{1}{3 m_{\Delta}^2} \frac{1+A}{1+2 A}\Big\{\left[\frac{A}{1+2 A} m_{\Delta}-\frac{1+A}{2(1+2 A)} \slashed p\right] \gamma^\mu \gamma^\nu\\
    &-\gamma^\mu p^\nu-\frac{A}{1+2 A}\, p^\mu \gamma^\nu\Big\}\,.
    \end{aligned}
\end{equation}
The second bracket is proportional to $(1+A)$ and vanishes for $A = -1$, which reduces the propagator to the form quoted in Eq.~\eqref{eq:delta-propagator}. The pole part carries the spin-3/2 projector, while the remaining terms propagate the spurious spin-1/2 components; these terms are non-pole contributions and amount to contact interactions.

Another form appears in the literature~\cite{Pascalutsa:1998pw}: 
\begin{align}
    S^{\alpha\beta}_{\Delta_1}(p) = &- \frac{1}{\slashed{p} - m_\Delta}
    \Big[ g^{\alpha\beta} - \frac{1}{3}\gamma^\alpha\gamma^\beta
    \nonumber\\
    & - \frac{1}{3m_\Delta^2}\big(\slashed{p}\gamma^\alpha p^\beta
    + p^\alpha\gamma^\beta \slashed{p}\big)\Big]\,.
    \label{eq:RS_delta_prop}
\end{align}
This is the Rarita-Schwinger form, written here with the overall sign adjusted to our conventions so that both forms share the same pole residue. 

The choice among these forms does not affect our results, for two separate reasons. The interactions of Sec.~\ref{sec:delta-couplings} are built from the tensor $\Psi_{\mu\nu}$, so every vertex attached to the $\Delta$ line is transverse in the $\Delta$ momentum. Every term by which the two forms differ carries an explicit factor of $p^\mu$ or $p^\nu$ on a free index, so between transverse vertices both forms yield identical amplitudes. The off-shell parameter $A$ is a freedom of a different kind. Changing $A$ at fixed couplings shifts the amplitude by a contact term with no pole, which the point transformation of Eq.~\eqref{eq:point-transformation} compensates by transforming the couplings~\cite{Pascalutsa:1998pw,Pascalutsa:2000kd}. We therefore fix $A = -1$ and use Eq.~\eqref{eq:delta-propagator} in the complex-mass scheme, as discussed in Sec.~\ref{sec:delta}.

The diagonal couplings of the $\Delta$ to pions, summarized in Sec.~\ref{sec:delta-couplings}, are collected here. Putting together the kinetic term of Eq.~\eqref{eq:Delta-Lagrangian}, with its derivatives promoted to covariant ones, and the three attachments of the vielbein~$u_\mu$ ($g_{1D}$, $g_{2D}$, $g_{3D}$), the lowest-order $\pi\Delta$ Lagrangian reads~\cite{Scherer:2012xha}
\begin{align}
    \mathcal{L}_{\pi \Delta} = \bar{\Psi}_{\mu} \xi^{\frac{3}{2}} \Lambda^{\mu \nu}_{\pi \Delta} \xi^{\frac{3}{2}} \Psi_\nu\,,
\end{align}
with
\begin{equation}
    \begin{aligned}
    \Lambda_{\pi \Delta}^{\mu \nu}= & -\Big[\left.\left(i \slashed{D}-m_{\Delta}\right) g^{\mu \nu}+i A\left(\gamma^\mu D^\nu+\gamma^\nu D^\mu\right)\right. \\
    & +\frac{i}{2}\left(3 A^2+2 A+1\right) \gamma^\mu \slashed{D} \gamma^\nu \\
    &+m_{\Delta}\left(3 A^2+3 A+1\right) \gamma^\mu \gamma^\nu \\
    & +\frac{g_{1D}}{2} \slashed{u}\gamma_5 g^{\mu \nu}+\frac{g_{2D}}{2}\left(\gamma^\mu u^\nu+u^\mu \gamma^\nu\right) \gamma_5 \\
    &+\frac{g_{3D}}{2} \gamma^\mu \slashed{u}\gamma_5 \gamma^\nu \Big]\,.
    \end{aligned}
\end{equation}
Consistency under the point transformations of Eq.~\eqref{eq:point-transformation} fixes two of the three couplings,
\begin{align}
    g_{2D} = A\, g_{1D}\,, \qquad
    g_{3D} = -\tfrac{1}{2}\left(1 + 2A + 3A^2\right) g_{1D}\,,
\end{align}
where $g_{1D}$ is a free parameter. $g_{1D}$ is only present when calculating multi-pion final states, a task left to future work.


\section{Vertex Structures and Their Explicit Form Factors}
\label{app:FFs}

This appendix collects the explicit vertices and form factors the model implements. To fix notation, we first reduce the scattering amplitude to a product of lepton and hadron currents, following  Ref.~\cite{halzenmartinQUARKSLEPTONSINTRODUCTORY1984}. We assume that the amplitude factorizes into a lepton current and a hadron current, linked by a single vector boson. Corrections require the exchange of a second boson and are suppressed by an additional power of the weak $W$ and $Z$ couplings, and of $\alpha$ for the photon. Under this assumption, the hadronic current, which can include pions emitted from the main nucleon line and hadronic resonances, remains isolated from the lepton current.

For a single-boson exchange, a $2 \to 2$ scattering amplitude can be written as
\begin{equation}
    T_{fi} = -i \int d^4 x d^4 x' \, \bra{f} \bar{\psi}_{\ell_f} \Gamma_\ell^\mu \psi_{\ell_i} D_{\mu \nu}(x-x') \bar{\psi}_{N_f} \Gamma_N^\nu \psi_{N_i} \ket{i}\,.
\end{equation}
We denote by $J^\mu(x) = \bar{u} \Gamma^\mu u\, e^{i(P_f-P_i)\cdot x}$ the plane-wave matrix element of a current in position space and by $J^\mu(Q^2) = \bar{u} \Gamma^\mu u$ its momentum-space form, so that
\begin{equation}
    T_{fi} = -i \int d^4 x d^4 x' \, j^{\mu}_{\ell}(x) D_{\mu \nu}(x-x') J^{\nu}_{N}(x')\,.
\end{equation}
The subscripts $\ell$ and $N$ label the lepton and hadron currents.

For a neutrino scattering off a nucleon via a $W$-boson, with incoming and final neutrino momenta $p_i, p_f$ and nucleon momenta $P_i,P_f$, the plane-wave currents are
\begin{align}
    j^{\mu}_{W,\ell}(x) = \bar{u}_\ell(p_f) \, \bigg[ \frac{g_2}{2\sqrt{2}}  \gamma^{\mu} (1-\gamma^5) \bigg] \, u_\nu(p_i) \,  e^{i(p_f-p_i)\cdot x}\,, 
\end{align}
\begin{align}
    J^{\mu}_{W,N}(x) = \bar{u}_N(P_f) \ \bigg[ \, \Gamma^\mu \, \bigg] \ u_N(P_i) \,  e^{i(P_f-P_i)\cdot x}\,.
\end{align}
Since the nucleon is a composite particle, the interaction vertex $\Gamma^\mu$ must be parametrized. For any form of the vertex, momentum conservation reduces the scattering amplitude $T_{fi}$ to a reduced amplitude $\mathcal{M}$,
\begin{align}
    T_{fi}= (2\pi)^4 \delta^{(4)}\left(\sum p_{out}- \sum p_{in}\right) \left(-i\mathcal{M} \right)\,. 
\end{align}
Writing the propagator through its Fourier transform and
substituting the plane-wave currents gives
\begin{align}
    T_{fi} &= -i \int d^4 x d^4 x' \frac{d^4 q}{(2\pi)^4} \bigg(  \bar{u}_\ell(p_f) \Gamma^\mu_{\ell} u_\nu(p_i) \,  e^{i(p_f-p_i)\cdot x} \nonumber\\
    & \cdot D_{\mu \nu}(q) e^{-iq\cdot (x-x')} \bar{u}_N(P_f) \Gamma^\nu_{N} u_N(P_i) \,  e^{i(P_f-P_i)\cdot x'} \bigg) \nonumber \\
    &=(2\pi)^4 \delta^{(4)}\!\left(p_f + P_f - p_i - P_i\right) \nonumber \\
    &\quad \times \left(-i\, j^{\mu}_{\ell}(Q^2)  D_{\mu \nu }(q) J^{\nu}_{N}(Q^2)\right)\,.
\end{align}
Comparing with the definition of $\mathcal{M}$ above gives
\begin{align}
    \mathcal{M} &= j^\mu_\ell (Q^2) D_{\mu \nu}(Q^2) J^\nu_N (Q^2)\,,
\end{align}
where $Q^2=-q^2=-(p_i-p_f)^2 = -(P_f-P_i)^2$, positive and Lorentz invariant for the processes considered here. The only unknown is the hadronic current $J_N^\nu$, whose explicit forms are collected in the following subsections. 

In the rest of this appendix, $u$ denotes a nucleon spinor and $u_\alpha$ a $\Delta$ spinor, with momentum labels shown only where needed.


\subsection{Nucleon currents}

At $Q^2 = 0$ the nucleon couples as a point particle, with the currents
\begin{align}
    \bra{N}J^\mu_{\gamma}(0)\ket{N} &= & Q_f \ e \quad & \bar{u} \; \gamma^{\mu} \; u &  \nonumber \\
    \bra{N}J_{W}^\mu (0)\ket{N} &= &\frac{g_2}{2\sqrt{2}}V_{ud} \quad & \bar u \; \gamma^{\mu} (1-g_A \gamma^5) \; u & \nonumber \\
    \bra{N}J_{Z}^\mu (0)\ket{N} &= &\frac{g_2}{2\cos(\theta_W)} 
    & \bar u \; \gamma^{\mu} (c_V - c_A\gamma^5) \; u\,. &
    \label{eq:pointlike-currents}
\end{align}
Here $c_V = T^3 - 2 Q_f \sin^2\theta_W$ and $c_A = g_A T^3$, where $T^3 = +\half$ for the proton and $-\half$ for the neutron, and $g_A$ is the axial constant. 

As discussed in Sec.~\ref{sec:formfactors}, the nucleon is composite, and each constant coupling in Eq.~\eqref{eq:pointlike-currents} becomes a form factor, a function of $Q^2$ that reduces to the point-like value at $Q^2 = 0$. The general decomposition of the current into form factors is given in Eq.~\eqref{eq:general-current}, and the $\mathcal{F}_i$ used below are those of that equation.

\begin{center}
\begin{tikzpicture}
    \begin{feynman}
        \vertex (i1) at (0, 1) {$\gamma\ (q)$};   
        \vertex (v1) at (0,0); 
        \vertex (o1) at (-1.5, -0.9) {$N(k_1)$};     
        \vertex (o2) at (1.5, -0.9) {$N(k_2)$};
        \diagram* {
            (i1) -- [photon] (v1) ,   
            (o1) -- [fermion] (v1) -- [fermion] (o2), 
            };
    \end{feynman}
    \fill (v1) circle (1.5pt);
\end{tikzpicture}
\end{center}

\begin{figure}[t]
    \centering
    \includegraphics[width=0.45\textwidth]{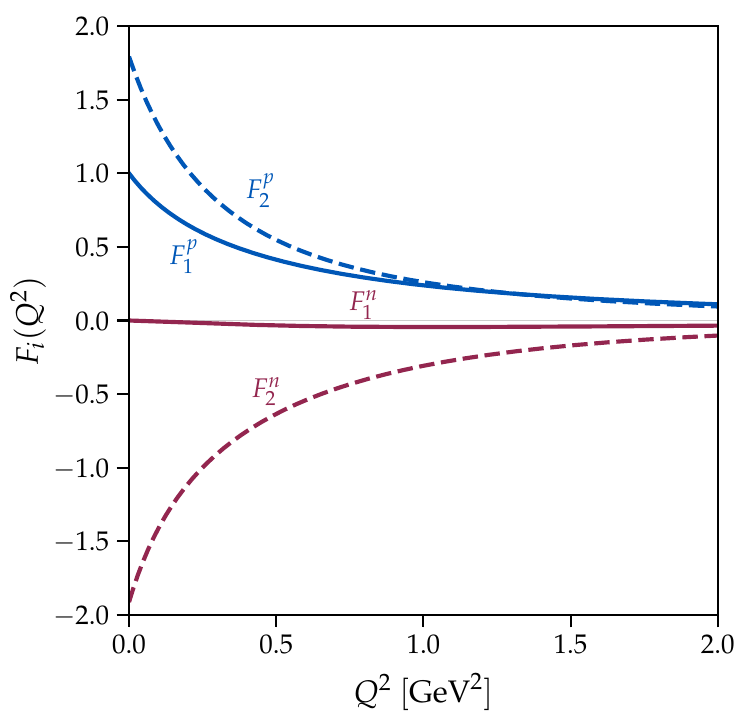}
    \caption{Electromagnetic form factors of the proton and neutron from the Kelly parametrization~\cite{Kelly:2004hm}.}
    \label{fig:EM_FF}
\end{figure}

For the photon, only the vector form factors enter, and we have $\mathcal{F}^{\gamma}_1 = F_1$ and $\mathcal{F}^{\gamma}_2 = F_2$. To cover the proton and neutron at once, we collect them into a matrix in isospin space,
\begin{equation}
    \hat F_{1,2}(Q^2) = \left( \begin{array}{cc}
        F_{1,2}^p(Q^2) & 0 \\
        0 & F_{1,2}^n(Q^2)
    \end{array} \right)\,,
\end{equation}
in terms of which the electromagnetic current is
\begin{multline}
    \bra{N(k_2)} J_{\mathrm{EM}}^\mu(0) \ket{N(k_1)} = \\ \ e\,\bar{u}(k_2) \; \left[ \hat{F}_1(Q^2)\gamma^\mu + \hat{F}_2(Q^2) \frac{i\sigma^{\mu\nu}q_\nu}{2m_N} \right]u(k_1) \,.
\end{multline}
We work with the Sachs combinations $G_E = F_1 - \tau F_2$ and $G_M = F_1 + F_2$, with $\tau = Q^2/4m_N^2$, in terms of which the elastic cross section separates into pure $G_E^2$ and $G_M^2$ terms, and take them from the Kelly parametrization~\cite{Kelly:2004hm},
\begin{equation}
    G(Q^2) = \frac{\sum_{k=0}^{\eta} a_k \tau^k}{1+ \sum_{k=1}^{\eta+2} b_k \tau^k}\,.
\end{equation}
The coefficients are listed in Table~\ref{tab:kelly}. Kelly gives the neutron electric form factor in the Galster form, $G_E^n = A_n \tau/(1 + B_n \tau)\, G_D$ with the dipole $G_D = (1 + Q^2/\Lambda^2)^{-2}$, $A_n = 1.7$, $B_n = 3.3$.
We use $\Lambda^2 = 0.7174~\mathrm{GeV}^2$, the default of \textsc{Achilles}~\cite{Isaacson:2022cwh}. The corresponding entries in Table~\ref{tab:kelly} follow from rewriting this form as a ratio of polynomials in $\tau$. The magnetic moments are $\mu_p = 2.793$ and $\mu_n = -1.913$ in units of the nuclear magneton. The resulting form factors are shown in Fig.~\ref{fig:EM_FF}. Note that if the neutron were a point-like neutral particle, its $F_1^n$ would vanish at all $Q^2$. The measured $F_1^n$ is nonzero, which is direct evidence that the neutron is composite.

\begin{table}[ht]
\caption{Coefficients of the Kelly parametrization for the nucleon electromagnetic form factors. The neutron electric entries are written in terms of the Galster parameters $A_n$, $B_n$, and $\Lambda$ defined in the text.}
\label{tab:kelly}
\centering
\renewcommand{\arraystretch}{1.2}
\begin{tabular}{lccccc}
\toprule
 & $a_0$ & $a_1$ & $b_1$ & $b_2$ & $b_3$ \\
\midrule
$G_E^p$ & $1$ & $-0.24$ & $10.98$ & $12.82$ & $21.97$ \\ 
$G_M^p/\mu_p$ & $1$ & $0.12$ & $10.97$ & $18.86$ & $6.55$ \\
\rule[-1.0em]{0pt}{2.8em}%
$G_E^n$ & $0$ & $A_n$ & $B_n + \dfrac{8 m_n^2}{\Lambda^2}$ & $\dfrac{16 m_n^4}{\Lambda^4} + \dfrac{8 B_n m_n^2}{\Lambda^2}$ & $\dfrac{16 B_n m_n^4}{\Lambda^4}$ \\ 
$G_M^n/\mu_n$ & $1$ & $2.33$ & $14.72$ & $24.2$ & $84.1$ \\
\bottomrule
\end{tabular}
\end{table}

\begin{center}
\begin{tikzpicture}
    \begin{feynman}
        \vertex (i1) at (0, 1) {$W\!/Z\ (q)$};   
        \vertex (v1) at (0,0); 
        \vertex (o1) at (-1.5, -0.9) {$N(k_1)$};     
        \vertex (o2) at (1.5, -0.9) {$N(k_2)$};
        \diagram* {
            (i1) -- [photon] (v1) ,    
            (o1) -- [fermion] (v1) -- [fermion] (o2), 
            };
    \end{feynman}
    \fill (v1) circle (1.5pt);
\end{tikzpicture}
\end{center}

For the $W$ and $Z$ bosons, the currents follow from the general nucleon decomposition of Eq.~\eqref{eq:general-current}~\cite{Isaacson:2021xty}, with the isospin combinations
\begin{align*}
    \mathcal{F}^W_1 &= F^p_1 - F^n_1 = F^V_1 & \quad \mathcal{F}^W_A &= F_A \\
    \mathcal{F}^{Z,p}_1 &= +\tfrac{1}{2} F^V_1 - 2\sin^2{\theta_W} \, F^p_1  & \quad \mathcal{F}^{Z,p}_A &= +\tfrac{1}{2} F_A \\
    \mathcal{F}^{Z,n}_1 &= -\tfrac{1}{2} F^V_1 - 2\sin^2{\theta_W} \, F^n_1  & \quad \mathcal{F}^{Z,n}_A &= -\tfrac{1}{2} F_A\,.
\end{align*}
The $\mathcal{F}_2$ combinations follow the same pattern as $\mathcal{F}_1$. The pseudoscalar form factor $\mathcal{F}_P$ follows from PCAC and pion-pole dominance, Eqs.~\eqref{eq:pcac} and~\eqref{eq:ppd} of Appendix~\ref{app:cvc,pcac,ppd}, which follows Ref.~\cite{alexandrouetalAxialNucleonNucleon2007},
\begin{equation}
    \mathcal{F}_P(Q^2) = \frac{4m_N^2}{Q^2+m_\pi^2}\mathcal{F}_A(Q^2).
    \label{eq:F_P}
\end{equation}
We take the $Q^2$ dependence of $F_A$ from the $z$-expansion fit to deuterium data of Ref.~\cite{Meyer:2016oeg},\footnote{We note here that Meyer et. al. uses a negative prescription for $\widetilde{g}_A$, so all coefficients $a_k$ have their sign flipped here to ensure $\widetilde{F}_A(0)>0$.}
\begin{align}
    \widetilde{F}_A(Q^2)=\sum_{k=0}^{8} a_k z^k \nonumber \, ,\\
    z = \frac{\sqrt{t_c+Q^2}-\sqrt{t_c-t_0}}{\sqrt{t_c+Q^2}+\sqrt{t_c-t_0}}\,.
    \label{eq:FAtilde form}
\end{align}
with the inputs listed in Table~\ref{tab:FA-zexp}.
The constant term $a_0$ is not fitted but fixed by the normalization $\widetilde{F}_A(0) = g_A$, and the implementation uses the ratio
\begin{equation}
    F_A(Q^2)=g_A \frac{\widetilde{F}_A(Q^2)}{\widetilde{F}_A(0)}
    \label{eq:FA form}
\end{equation}

\begin{table}[t]
\caption{Inputs of the axial form factor function $\widetilde{F}_A$, from the deuterium $z$-expansion fit of Ref.~\cite{Meyer:2016oeg}, with the $1\sigma$ uncertainties given for completeness.}
\label{tab:FA-zexp}
\renewcommand{\arraystretch}{1.2}
\begin{tabular}{lll}
\toprule
Quantity & Value & Note \\
\midrule
\multicolumn{3}{l}{\textit{Fixed parameters}} \\
$t_c$ & $0.17532~\text{GeV}^2$ & $m_\pi = 0.14~\text{GeV}$ is used in their fit, \\
      &                         & but we use our charged pion mass. \\
$t_0$ & $-0.28~\text{GeV}^2$ & $t_0^{\text{optimal}}(1~\text{GeV}^2)$ \\
\midrule
\multicolumn{3}{l}{\textit{Coefficients}} \\
$a_0$ & $\phantom{-}0.7580$ & from $\widetilde{F}_A(0)=g_A=1.2754$ \\
$a_1$ & $-2.30 \pm 0.13$ & free \\
$a_2$ & $\phantom{-}0.6\phantom{0} \pm 1.0$ & free \\
$a_3$ & $\phantom{-}3.8\phantom{0} \pm 2.5$ & free \\
$a_4$ & $-2.3\phantom{0} \pm 2.7\phantom{0}$ & free \\
$a_5$ & $-2.16$ & sum rule \\
$a_6$ & $\phantom{-}0.896$ & sum rule \\
$a_7$ & $\phantom{-}1.58$ & sum rule \\
$a_8$ & $-0.823$ & sum rule \\
\bottomrule
\end{tabular}
\end{table}

The full currents are then
\begin{multline}
    \bra{N(k_2)} J^\mu _{W^\pm}(0) \ket{N(k_1)} = \\
    g_{NNW} \; \bar{u}(k_2) \Bigg[ F^V_1(Q^2) \gamma^{\mu} + F^V_2(Q^2)  \frac{i\sigma^{\mu \nu}q_\nu}{2m_N}  \\
    -\Big(F_A(Q^2) \gamma^{\mu} + F_P(Q^2) \frac{q^\mu }{2m_N}\Big) \gamma^5 \Bigg] T^\pm \, u(k_1) \,, 
    \label{eq:nucleonW}
\end{multline}
\begin{multline}
    \bra{N(k_2)} J^\mu _{Z}(0) \ket{N(k_1)} = \\
    g_{NNZ} \; \bar{u}(k_2) \Bigg[ \bigg( F^V_1(Q^2)  \gamma^{\mu} + F^V_2(Q^2) \frac{i\sigma^{\mu \nu}q_\nu}{2m_N} \\
    \qquad \qquad -\Big(F_A(Q^2) \gamma^{\mu} + F_P(Q^2) \frac{q^\mu }{2m_N}\Big) \gamma^5 \bigg) T^3  \\
    -2\sin^2{\theta_W}\Big( \hat F_1(Q^2)  \gamma^{\mu} + \hat F_2(Q^2) \frac{i\sigma^{\mu \nu}q_\nu}{2m_N} \Big)\hat{1}  \ \Bigg] u(k_1)\,, 
\end{multline}
\begin{multline}
    g_{NNW}=\frac{g_2}{2\sqrt{2}}V_{ud} \ \, , \,  \quad g_{NNZ}=\frac{g_2}{2\cos{\theta_W}}\,. 
\end{multline}
The isospin matrices $T^a = \tau^a/2$, $a = 1,2,3$, are the $SU(2)$ isospin generators, and the charged combinations are $T^\pm = T^1 \pm i T^2$. 

Note that in the amplitude, the $q^\mu$ of the pseudoscalar $F_P$ term contracts with the lepton current, and the Dirac equation reduces its contribution to
\begin{eqnarray}
    \mathcal{M}_P &\sim& (m_{\ell_i} + m_{\ell_f} )\overline{u}_\ell(p_f) u_\ell(p_i) \overline{u}_N u_N\,,
\end{eqnarray}
which is suppressed relative to the other terms by the lepton mass. Off the nucleon mass shell, as in the pole diagrams of single-pion production, the induced pseudoscalar term is implemented in the pseudovector form $F_A(Q^2)\,q^\mu\slashed{q}\gamma^5/(Q^2+m_\pi^2)$, which the pion pole of the pseudovector $\pi NN$ coupling generates. Between on-shell spinors it reduces to the $F_P$ term of Eq.~\eqref{eq:nucleonW} through $\slashed q\gamma^5 \to 2m_N\gamma^5$, and off shell it keeps the axial current conserved in the chiral limit, in the same way as the pion pole of the $N \to \Delta$ transition in Eq.~\eqref{eq:C6Apole}.


\subsection{Pion nucleon current}

\begin{center}
\begin{tikzpicture}
 \begin{feynman}
 \vertex (i1) at (0, 1) {$\pi\ (q)$};
 \vertex (v1) at (0,0);
 \vertex (o1) at (-1.5, -0.9) {$N(k_1)$};
 \vertex (o2) at (1.5, -0.9) {$N(k_2)$};
 \diagram* {
 (i1) -- [scalar] (v1) , 
 (o1) -- [fermion] (v1) -- [fermion] (o2), 
 };
 \end{feynman}
 \fill (v1) circle (1.5pt);
\end{tikzpicture}
\end{center}

The strong pion-nucleon vertex follows from the axial coupling of Eq.~\eqref{eq:Lagrangian-pion-nucleon-gauge}. Note that the pion couples through the isospin Pauli matrices $\tau^a$, unlike the electroweak currents which couple through the generators $T^a=\frac{\tau^a}{2}$, and we get 
\begin{equation}
    \mathcal{L}_{\pi NN} 
    = -\frac{g_A}{2 f_\pi}\, \overline{N} \gamma^\mu \gamma^5 \tau^a N\, \partial_\mu \phi^a \,,
    \label{eq:LpiNN}
\end{equation}
This gives the current
\begin{multline}
     \bra{N(k_2)} J^a_{\pi NN}(0) \ket{N(k_1)} = \\
     i g_{\pi NN}(Q^2) \, \bar u (k_2) \frac{\left(\slashed{k}_2 - \slashed{k}_1 \right)}{2 m_N} \gamma^5 \tau^a u(k_1) \, .
    \label{eq:V-pi-N-N}
\end{multline}
where $a$ runs over $(1,2,3)$ and $g_{\pi NN}$ is the pion-nucleon form factor, 
\begin{equation}
    g_{\pi NN}(Q^2)= \frac{m_N}{f_\pi}F_A(Q^2) \, ,
    \label{eq:GT-function}
\end{equation}
with 
\begin{equation}
     g_{\pi NN}(0) =\frac{g_A m_N}{f_\pi}\, .
\end{equation} 
As discussed in Sec.~\ref{sec:formfactors}, we keep the vertex point-like with the Goldberger--Treiman coupling $g_{\pi NN}(0)=12.8674$, which lies $3.9\%$ below $g_{\pi NN}(-m_\pi^2)=13.3684$.

However, introducing a form factor on this vertex results in the pion-in-flight and the Kroll-Ruderman contributions violating the Ward-Takahashi identity by a term proportional to $1 - g_{\pi NN}(Q^2)/g_{\pi NN}(-m_\pi^2)$. Following the prescription of Ohta~\cite{Ohta:1989ji}, the Ward-Takahashi identity can be restored by an additional contact current proportional to the same factor [cf.\ Eq.~(4.16) of Ref.~\cite{Ohta:1989ji}], while the Kroll--Ruderman term retains the point-like coupling.
For on-shell external hadrons, the form factor on the pion-in-flight term is canceled exactly by this current, and the point-like amplitude is recovered. The vertex is therefore treated as point-like, with the Goldberger--Treiman coupling $g_{\pi NN}=\tfrac{g_A m_N}{f_\pi}=12.8674$. In extensions to multipion production, this relationship no longer holds, and the more general prescription from Haberzettl~\cite{Haberzettl:1997jg,Haberzettl:1998aqi} or amplitude level form factors like the Davidson-Workman form~\cite{Workman:1992hr,Davidson:2001rk} is required. In this prescription, the form factors depend on multiple variables which is currently not supported with the UFO-\textsc{Sherpa} interface and is thus left to future work.

With both nucleons on shell, Eq.~\eqref{eq:V-pi-N-N} reduces to the standard pseudoscalar pion-nucleon vertex,
\begin{equation}
    \bra{N(k_2)} J^a_{\pi NN}(0) \ket{N(k_1)} \underset{\text{on-shell}}{=} i g_{\pi NN}  \, \bar u (k_2) \, \gamma^5 \tau^a \, u(k_1) \,.
    \label{eq:V-pi-N-N onshell}
\end{equation}
Away from the nucleon mass shell the two Lorentz structures differ by contact terms, and we keep the form of Eq.~\eqref{eq:V-pi-N-N}, which follows directly from \chiEFT.

In the physical charge basis, expanding $\tau^a \phi^a$ gives the vertices the relative factors
\begin{equation}
     \Gamma_{\pi^0 pp}=-\Gamma_{\pi^0 nn}= \frac{1}{\sqrt{2}} \Gamma_{\pi^\pm N'N} = i g_{\pi NN} \, \frac{\left(\slashed{k}_2 - \slashed{k}_1 \right) }{2 m_N} \gamma^5 \, .
\end{equation}
This holds for incoming and outgoing pions alike.


\subsection{Pion electromagnetic currents}

\begin{center}
\begin{tikzpicture}
 \begin{feynman}
 \vertex (i1) at (0, 1) {$\gamma\!/W\!/Z\ (q)$};
 \vertex (v1) at (0,0);
 \vertex (o1) at (-1.5, -0.9) {$\pi(k_1)$};
 \vertex (o2) at (1.5, -0.9) {$\pi(k_2)$};
 \diagram* {
 (i1) -- [photon] (v1) , 
 (v1) -- [scalar] (o1), 
 (v1) -- [scalar] (o2),
 };
 \end{feynman}
 \fill (v1) circle (1.5pt);
\end{tikzpicture}
\end{center}

The gauge bosons couple to two pions through the covariant derivative of the pion kinetic term. This vertex enters single-pion production at tree level, when the boson produces a pion pair and one of the two pions is reabsorbed by the nucleon. Together with the contact term below, it is required for the gauge invariance of the nucleon-pole amplitudes. The coupling is pointlike, with the single Lorentz structure $(k_1+k_2)^\mu$. In physical charge eigenstates, the currents are
\begin{align}
 \bra{\pi^{\pm}(k_2)} J^\mu_{\gamma \pi \pi}(0) \ket{\pi^{\pm}(k_1)} &= \pm \, e \, F_1^V(Q^2) \, (k_1+k_2)^\mu \nonumber \\
 \bra{\pi^{0}(k_2)} J^\mu_{W^\pm \pi \pi}(0) \ket{\pi^{\mp}(k_1)} &= \frac{g_2}{2} V_{ud} \, F_1^V(Q^2) \, (k_1+k_2)^\mu \nonumber \\
 \bra{\pi^{\pm}(k_2)} J^\mu_{Z \pi \pi}(0) \ket{\pi^{\pm}(k_1)} &= \pm \frac{g_2}{2\cos\theta_W} \Big(1 - 2\sin^2\theta_W\Big) \nonumber \\
 &\qquad \times F_1^V(Q^2) \, (k_1+k_2)^\mu\,,
\end{align}
with the remaining $W^\pm$ channels related by isospin. Four features of these currents deserve comment. At $Q^2 = 0$ the photon coupling is the pion charge, as it must be. The $W$ coupling carries no factor $1/\sqrt{2}$: the one that accompanies the $W$ in the Standard Model is cancelled by the $\sqrt{2}$ Clebsch--Gordan coefficient of the charged-pion transition. The $Z$ coupling is $(g/2\cos\theta_W)\,c_V$ with the same $c_V = T^3 - 2Q\sin^2\theta_W$ as for the nucleon, now with $T^3 = \pm1$, and has no axial part, because the axial current has no matrix element between two pions by parity. Finally, no current connects two neutral pions. The two-pion coupling of a gauge boson comes only from the pion kinetic term and is proportional to $\varepsilon^{abc}\phi^b\partial^\mu\phi^c$, which vanishes when both pions are the same isospin component.

\begin{figure}[t]
 \centering
 \includegraphics[width=0.45\textwidth]{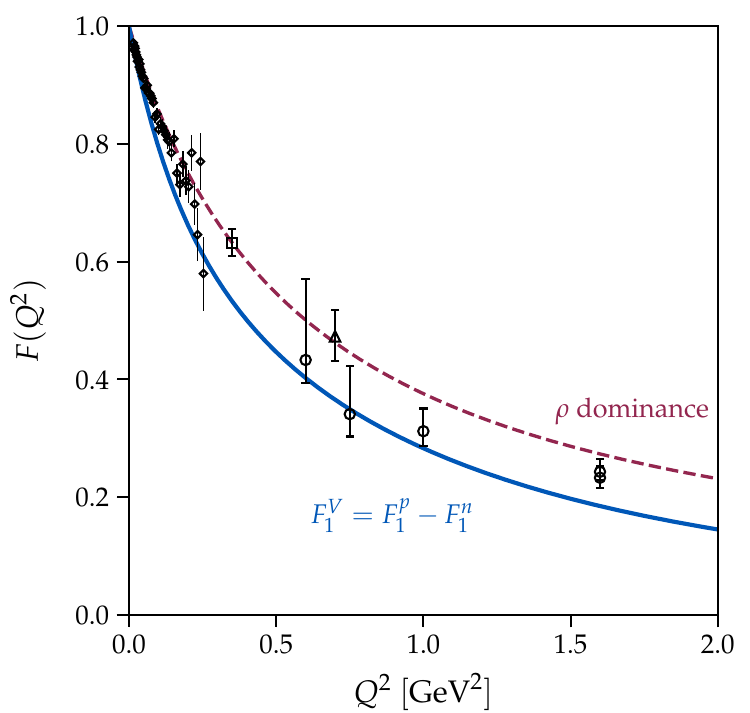}
 \caption{The form factor $F_1^V$ carried by the $\pi\pi$ vertices. The solid blue curve shows the form used in this work, and the dashed curve the $\rho$-dominance form~\cite{Sakurai:1960ju}. The data points show the measured pion form factor, from NA7~\cite{NA7:1986vav} (small diamonds) and the JLab $F_\pi$ collaboration~\cite{JeffersonLab:2008jve} (circles); the square and triangle are their reanalysis of DESY data~\cite{Ackermann:1977rp,Brauel:1979zk}.}
 \label{fig:pionFF}
\end{figure}

Gauge invariance fixes the vector form factor on these vertices. The $\pi\pi$ vertex, the contact term, and the nucleon-pole diagrams together form a gauge-invariant set, and the Ward identity holds only if the same $F_1^V = F_1^p - F_1^n$ multiplies all of them. We therefore put the nucleon isovector form factor on the $W\pi\pi$ vertex and on the vector part of the contact term, as in Ref.~\cite{Hernandez:2007qq}, rather than the measured pion form factor. The two are not the same. The measured pion form factor is well described by $\rho$-dominance~\cite{Sakurai:1960ju}, in which the photon converts to a $\rho$ meson before coupling to the pions, so that the vertex carries the $\rho$ propagator $m_\rho^2/(m_\rho^2 + Q^2)$; $F_1^V$ falls faster, as Fig.~\ref{fig:pionFF} shows. The difference is a limitation of the truncated model: the Ward identity constrains the sum of diagrams, not each vertex, and with only nucleons and pions in the set it forces the nucleon form factor onto the pion vertex. A model with an explicit $\rho$ would carry the measured form factor on the pion vertex and remain gauge invariant diagram by diagram. We leave this to future work. The $\rho$-dominance form does enter the model, but only on the axial half of the contact term, discussed next.

\subsection{Nucleon contact term}

\begin{center}
\begin{tikzpicture}
 \begin{feynman}
 \vertex (i1) at (-1.5, 0.9) {$\gamma\!/W\!/Z\ (q)$};
 \vertex (i2) at (1.5, 0.9) {$\pi^\pm(k_\pi)$};
 \vertex (v1) at (0,0);
 \vertex (o1) at (-1.5, -0.9) {$N(k_1)$};
 \vertex (o2) at (1.5, -0.9) {$N(k_2)$};
 \diagram* {
 (i1) -- [photon] (v1) , 
 (v1) -- [scalar] (i2),
 (o1) -- [fermion] (v1) -- [fermion] (o2),
 };
 \end{feynman}
 \fill (v1) circle (1.5pt);
\end{tikzpicture}
\end{center}

Gauging Eq.~\eqref{eq:pion-nucleon-gauge-lagrangian}\footnote{Gauging only the $\pi NN$ vertex, as is sometimes done, gives only the first of the three terms below, the Kroll--Ruderman term, and misses the axial structures of the full contact term. For the photon no axial pieces are needed, so there gauging only the $\pi NN$ vertex suffices.} produces four-point contact-term (CT) vertices, in which the gauge boson and the pion attach to the nucleon line at a single point. Three operators arise~\cite{Hernandez:2007qq}. The charged current, given in the physical pion states, is
\begin{multline}
    \bra{N(k_2)\,\pi(k_\pi)} J^\mu_{W,\mathrm{CT}}(0) \ket{N(k_1)} = \qquad \qquad \\
   -i\, \hat c\, \frac{g_{NNW}}{f_\pi} \bigg( g_A F_1^V(Q^2)\, \gamma^\mu\gamma^5 \qquad \qquad  \qquad \\ 
    \qquad  - F_\rho(Q^2)\, \Big( \gamma^\mu  + \frac{q^\mu\, (\slashed{q}+ \slashed{k}_\pi)}{2(Q^2 + m_\pi^2)} \Big) 
    \bigg)\,, 
    \label{eq:WCT}
\end{multline}
where $m_\pi$ is the charged pion mass.

In Eq.~\eqref{eq:pion-nucleon-gauge-lagrangian}, the gauge bosons enter through two combinations of the external fields: the vector combination $\frac12(\tilde r_\mu + \tilde l_\mu)$ and the axial combination $\frac12(\tilde r_\mu - \tilde l_\mu)$. Whether a term contributes to the vector or to the axial current is determined by which of the two it multiplies, not by its Dirac structure. Each of the three contact terms multiplies one of them. The Kroll--Ruderman term comes from the axial coupling $g_A\overline{N}\gamma^\mu\gamma^5 u_\mu N$: expanded to one pion and one gauge field, the vielbein $u_\mu$ of Eq.~\eqref{eq:chiral-vielbein} contains the commutator of the vector combination with the pion field, so the term has Dirac structure $\gamma^\mu\gamma^5$ but multiplies the vector field and belongs to the vector current. The $\gamma^\mu$ term comes from the chiral connection $\Gamma_\mu$ of Eq.~\eqref{eq:chiral-connection} in the nucleon kinetic term, which at the same order contains the commutator of the axial combination with the pion field: it multiplies the axial field, so despite its vector Dirac structure it belongs to the axial current. (The Dirac structures are interchanged relative to the nucleon current because the pion is pseudoscalar.) The third term is the pion pole: the axial current converts to a pion, which propagates and is absorbed by the $\pi\pi NN$ vertex\footnote{We do not include the $\pi\pi NN$ vertex in this model since it only contributes starting at two-pion production}. Since the UFO format has no two-point vertices, the $W$--$\pi$ mixing and the pion propagator are folded into one four-point operator. The isospin factors for $J^\mu_{W^+}$ are $\hat c = 1$ for $n \to p\,\pi^0$, $\hat c = 1/\sqrt{2}$ for $n \to n\,\pi^+$, and $\hat c = -1/\sqrt{2}$ for $p \to p\,\pi^+$, and the $W^-$ channels follow by hermitian conjugation.

The form factors follow from this assignment. The Kroll--Ruderman term is part of the vector current, so gauge invariance requires on it the same $F_1^V$ that dresses the nucleon poles and the $W\pi\pi$ vertex. The two axial terms must share one form factor if the axial current is to be conserved in the chiral limit. This can be seen by assigning $F_X$ to the $\gamma^\mu$ term and $F_{\rm PP}$ to the pion-pole term and contracting the contact term with $q_\mu$. Between on-shell nucleons, the identity $\bar u(k_2)(\slashed q+\slashed k_\pi)u(k_1) = 2\,\bar u(k_2)\slashed q\,u(k_1)$ is implied by momentum conservation ($q - k_\pi = k_2 - k_1$) and the Dirac equation. The contraction is then given by
\begin{align}
    & q_\mu\Big[F_X\gamma^\mu + F_{\rm PP}\frac{q^\mu(\slashed q+\slashed k_\pi)}{2(Q^2+m_\pi^2)}\Big] \nonumber\\
    &\quad = \slashed q\Big[(F_X-F_{\rm PP}) + F_{\rm PP}\frac{m_\pi^2}{Q^2+m_\pi^2}\Big] \nonumber\\
    &\quad \underset{m_\pi^2 \to 0}{\longrightarrow} \slashed q\,(F_X-F_{\rm PP}) \overset{!}{=} 0\,,
\end{align}
so $F_X = F_{\rm PP}$. We choose this form factor to be the $\rho$-dominance form following Ref.~\cite{Hernandez:2007qq},
\begin{equation}
    F_\rho(Q^2) = \frac{m_\rho^2}{m_\rho^2 + Q^2}\,,
\end{equation}
with $m_\rho = 775.8$~MeV. Note that Ref.~\cite{Hernandez:2007qq} evaluates it at the momentum transferred to the nucleon line, $t = (q - k_\pi)^2$, whereas we use $Q^2$. The neutral current follows the same pattern, with $\hat c = 1/\sqrt{2}$ for $p \to n\pi^+$ and $-1/\sqrt{2}$ for $n \to p\pi^-$, with $F_1^V$ replaced by the neutral-current combination carried by the $Z\pi\pi$ vertex, and with the neutral pion mass in the pole. No contact term exists for neutral-pion emission, since the isospin factor vanishes when both the current and the pion carry the third isospin component, and the same holds for the photon, for which only the Kroll-Ruderman term survives, dressed with the same $F_1^V$.


\subsection{\texorpdfstring{$N \to \Delta$}{Nucleon to Delta} Electroweak transition currents}

\begin{center}
\begin{tikzpicture}
 \begin{feynman}
 \vertex (i1) at (0, 1) {$\gamma\!/W\!/Z\ (q)$}; 
 \vertex (v1) at (0,0); 
 \vertex (o1) at (-1.5, -0.9) {$N(p)$}; 
 \vertex (o2) at (1.5, -0.9) {$\Delta(p')$};
 \diagram* {
 (i1) -- [photon] (v1) , 
 (o1) -- [fermion] (v1) -- [double_arrow] (o2), 
 };
 \end{feynman}
 \fill (v1) circle (1.5pt);
\end{tikzpicture}
\end{center}

The electroweak transition vertices follow from the operators of Sec.~\ref{sec:delta-ew}, Eq.~\eqref{eq:VNDelta} for the vector part and, through the gauging, the $\pi N\Delta$ operator of Eq.~\eqref{eq:LpiNDelta-GI} for the axial part. We write the vector part directly in the form the Lagrangian gives, with the photon couplings $G_1$ and $G_2$, and the axial part in the Adler form factors $C^A_i$ that experimental analyses use. Following the conventions of Ref.~\cite{Hernandez:2007qq}, with the coupling $g_{NNW}$ included as in the nucleon current of Eq.~\eqref{eq:nucleonW},  the form factors are defined through the matrix element 
\begin{multline}
    \langle \Delta^+(p')|J^\mu_{W^+}(0)|n(p)\rangle
    = g_{NNW}\,\bar u_\alpha(p')\left[\Gamma_V^{\alpha\mu}+\Gamma_A^{\alpha\mu}\right]u(p) \, .
    \label{eq:Delta-EW-vertex-definition}
\end{multline}
with $u_\alpha$ the Rarita-Schwinger spinor and the remaining charged channels fixed by isospin (the $p \to \Delta^{++}$ amplitude is larger by $\sqrt{3}$). 

The vector part follows from Eq.~\eqref{eq:VNDelta},
\begin{multline}
    \Gamma_V^{\alpha\mu} = \sqrt{\tfrac23}\,\frac{3\,F_V^{N\Delta}(Q^2)}{2m_N(m_N+m_\Delta)} \\
    \Big[\, i G_1\,\epsilon^{\beta\alpha\rho\mu}p'_\beta q_\rho + G_2\big(g^{\alpha\mu}\,q\!\cdot\! p' - q^\alpha p'^\mu\big)\gamma_5\Big] \, .
    \label{eq:Delta-N-Vector-vertex}
\end{multline}
The photon and the weak bosons share one CVC dipole, which multiplies the photon-point couplings $G_1$ and $G_2$,
\begin{equation}
    F_V^{N\Delta}(Q^2) = \frac{1}{\left(1 + Q^2/M_V^2\right)^2}\, \frac{1}{1 + Q^2/4M_V^2}\,,
    \label{eq:FVND}
\end{equation}
with $M_V^2 = 0.71~\mathrm{GeV}^2$, normalized to unity at the photon point so that the photoproduction couplings are unchanged. The dipole mass is the standard value describing the nucleon elastic form factors, and both it and the extra factor relative to a dipole follow Ref.~\cite{Lalakulich:2005cs}, where the latter accounts for the electro-production data falling faster than the dipole form. 
On shell, Eq.~\eqref{eq:Delta-N-Vector-vertex} reduces to the standard form of Ref.~\cite{Adler:1968tw, Hernandez:2007qq},
\begin{multline}
    \Gamma_V^{\alpha\mu}=\Big[
    \tfrac{C_3^V(Q^2)}{m_N}t^{\mu \alpha \beta}\gamma_\beta
    +\tfrac{C_4^V(Q^2)}{m_N^2}t^{\mu \alpha \beta}p'_\beta
    \\ +\tfrac{C_5^V(Q^2)}{m_N^2}t^{\mu \alpha \beta}p_\beta
    +C_6^V(Q^2) g^{\alpha\mu}\Big]\gamma_5
    \label{eq: vector em delta vertex C form}
\end{multline}
\begin{equation}
    t^{\mu \alpha \beta} = g^{\mu \alpha}q^\beta - g^{\mu \beta} q^\alpha\,.
    \label{eq:tten}
\end{equation}
with
\begin{align}
C_3^V &=\sqrt{\tfrac23}\,\tfrac32\,\frac{m_\Delta}{m_N+m_\Delta}\,G_1F_V^{N\Delta}\, , \nonumber \\
C_4^V&=-\sqrt{\tfrac23}\,\tfrac32\,\frac{m_N}{m_N+m_\Delta}\,(G_1-G_2)\,F_V^{N\Delta}\, , \nonumber \\
C_5^V&=C_6^V=0 \, .
\end{align}
Isospin gives the neutral currents. The electromagnetic one is
\[\langle\Delta^+|J^\mu_{\rm EM}|p\rangle=\langle\Delta^0|J^\mu_{\rm EM}|n\rangle
 = -e\,\bar u_\alpha\Gamma_V^{\alpha\mu}u \, .\]
Because the isoscalar current does not contribute to the $Z$ current, from $J_{NC}$ in Appendix~\ref{app:cvc,pcac,ppd} follows 
\begin{multline*}
\langle\Delta^+|J^\mu_{Z}|p\rangle=\langle\Delta^0|J^\mu_{Z}|n\rangle \\
 = -\frac{g}{2\cos\theta_W}\,\bar u_\alpha\Big[\big(1-2\sin^2\theta_W\big)\Gamma_V^{\alpha\mu}+\Gamma_A^{\alpha\mu}\Big]u \, .
\end{multline*}

The axial part of Eq.~\eqref{eq:Delta-EW-vertex-definition}, follows from Eq.~\eqref{eq:LAND}
\begin{equation}
    \Gamma_A^{\alpha\mu} = -i\sqrt{\tfrac23}\,\frac{h_D}{m_\Delta}\,
    \epsilon^{\lambda\mu\rho\alpha}p'_\rho\gamma_\lambda\gamma_5\; F_{A}^{N\Delta}(Q^2) \, ,
    \label{eq:ourvertex piNdelta}
\end{equation}
where on-shell, via $\bar u_\alpha(p')\,\epsilon^{\lambda\mu\rho\alpha}p'_\rho\gamma_\lambda\gamma_5 = i\,m_\Delta\,\bar u^\mu(p')$, this reduces to
\begin{equation}
    \Gamma_{A,\text{ on-shell}}^{\alpha\mu} = \sqrt{\tfrac23}\,h_D\, F_{A}^{N\Delta}(Q^2)\,g^{\alpha\mu} \, ,
    \label{eq:onshellvertex piNdelta}
\end{equation}
We adopt the Adler basis of form factors~\cite{Adler:1968tw}, which most experimental analyses use,
\begin{multline}
 \Gamma_A^{\alpha \mu} = \Big[ \tfrac{C^A_3(Q^2)}{m_N} t^{\mu \alpha \beta}\gamma_\beta + \tfrac{C^A_4(Q^2)}{m_N^2} t^{\mu \alpha \beta}p'_\beta \\
 + C^A_5(Q^2)g^{\mu \alpha} + \tfrac{C^A_6(Q^2)}{m_N^2} q^\mu q^\alpha \Big] \,,
 \label{eq:adlerA}
\end{multline}
with $t^{\mu \alpha \beta}$ as defined in Eq.~\eqref{eq:tten}.
Note that no $\gamma^5$ appears,  unlike in the axial nucleon current of Eq.~\eqref{eq:nucleonW}. The $\Delta$ field carries a Lorentz index, which flips the parity of the bilinear, in the same way as the pseudoscalar pion flips it for the contact terms of the previous subsection.

Matching our derived vertex in Eq.~\eqref{eq:ourvertex piNdelta} onto Eq.~\eqref{eq:adlerA} gives $C^A_3 = C^A_4 = 0$ and
\begin{align}
 C^A_5(Q^2) & = \sqrt{\tfrac{2}{3}}\, h_D\, F_A^{N\Delta}(Q^2)\,,\\
 C^A_5(0) & = \sqrt{\tfrac{2}{3}}\, h_D = 1.271\,,
 \label{eq:C5A0}
\end{align}
the off-diagonal Goldberger-Treiman value. The leading-order Lagrangian fixes $C^A_5(0)$ alone. We stress that the approximate equality between $g_A$ and $C^A_5(0)$ is completely coincidental. For the $Q^2$ dependence, we adopt the Paschos–Lalakulich form factor shape, which is an improvement of the Adler fit,
\begin{equation}
    C^A_5(Q^2) = \frac{C^A_5(0)}{\left(1 + Q^2/M_{A\Delta}^2\right)^2 \left(1 + Q^2/3M_{A\Delta}^2\right)}\,,
    \label{eq:C5Ashape}
\end{equation}
with $M_{A\Delta} = 1.05$~GeV, which we take from~\cite{Lalakulich:2005cs}. Only the
shape is taken from the fit. Its normalization $\tilde C^A_5(0)$ drops out, and $C^A_5(0)$ remains the value of Eq.~\eqref{eq:C5A0}, in the same way as for $F_A$ in Eq.~\eqref{eq:FA form}.

PCAC fixes $C^A_6$, which for an on-shell $\Delta$ is
\begin{equation}
    C^A_6(Q^2) = C^A_5(Q^2)\, \frac{m_N^2}{Q^2 + m_\pi^2}\,.
    \label{eq:C6A}
\end{equation}
with the charged pion mass in the pole for the charged current. Note that its contribution enters the amplitude proportional to the lepton mass, as for $F_P$, and we explicitly include it in our analysis to preserve PCAC.

In single-pion production, however, the $\Delta$ is always off shell, and there the structure $q^\alpha q^\mu$ of Eq.~\eqref{eq:adlerA} no longer cancels the divergence of the vertex of Eq.~\eqref{eq:ourvertex piNdelta}. We therefore attach the pion pole to the full GI vertex, replacing
\begin{equation}
    \Gamma_A^{\alpha\mu} \;\to\; \Gamma_A^{\alpha\lambda}
    \Big[ g_\lambda{}^{\mu} + \frac{q_\lambda q^\mu}{Q^2 + m_\pi^2} \Big] ,
    \label{eq:C6Apole}
\end{equation}
with the same coupling $h_D$ and form factor $F_A^{N\Delta}$. This reproduces
Eq.~\eqref{eq:C6A} on shell and keeps the axial transition current conserved in the off-shell chiral limit. 

$C^A_3$ and $C^A_4$ multiply operators of higher order that are not present in our work, and chiral symmetry does not fix their coefficients. Most analyses follow Adler's model, $C^A_3=0$ and $C^A_4=-C^A_5/4$~\cite{Adler:1968tw}, which differs from ours. The difference matters little, since fits of $C^A_5$ are insensitive to $C^A_{3,4}$ at low $Q^2$~\cite{Hernandez:2010bx}.

Because $C^A_5(0)$ is tied through $h_D$ to the width $\Gamma(\Delta \to N\pi)$, it is a prediction of the model rather than a fit parameter. We compare with the fitted values of Ref.~\cite{Alvarez-Ruso:2015eva} in Sec.~\ref{sec:free} and Appendix~\ref{app:lineshape}.


\subsection{\texorpdfstring{$N \to \Delta$}{Nucleon to Delta} pion transition currents}

\begin{center}
\begin{tikzpicture}
 \begin{feynman}
 \vertex (i1) at (0, 1) {$\pi(p_\pi)$}; 
 \vertex (v1) at (0,0); 
 \vertex (o1) at (-1.5, -0.9) {$N(p)$}; 
 \vertex (o2) at (1.5, -0.9) {$\Delta(p')$};
 \diagram* {
 (i1) -- [scalar] (v1) ,  
 (o1) -- [fermion] (v1) -- [double_arrow] (o2), 
 };
 \end{feynman}
 \fill (v1) circle (1.5pt);
\end{tikzpicture}
\end{center}

The strong $\pi N\Delta$ vertex follows from the GI Lagrangian of Eq.~\eqref{eq:LpiNDelta-GI}.
Expanding the vielbein to one pion, $u^a_\nu = -\partial_\nu\phi^a/f_\pi + \mathcal{O}(\phi^3)$, as for the nucleon in Eq.~\eqref{eq:LpiNN}, gives
\begin{equation}
    \mathcal{L}_{\pi N\Delta} \supset \frac{h_D}{m_\Delta f_\pi}\,
    \varepsilon^{\mu\nu\rho\sigma}\big(\partial_\rho\bar\Psi^a_\sigma\big)\gamma_\mu\gamma^5 N\,
    \partial_\nu\phi^a + \text{h.c.} \, ,
\end{equation}
with $h_D = 1.557$ derived from the pole width $\Gamma(\Delta \to N\pi) = 101.4$~MeV as in Sec.~\ref{sec:delta-couplings}. The current for $\Delta^+$ formation from $p\,\pi^0$ is
\begin{multline}
    \langle\Delta^+(p')|\,J_{\pi N \Delta}(0)\,|p(p)\,\rangle= \\
    \sqrt{\tfrac{2}{3}}\,\frac{h_D}{f_\pi m_\Delta}\,F_{\pi N\Delta}(Q^2)\,
    \bar u_\alpha(p')\,\epsilon^{\alpha}{}_{\rho\sigma\delta}
    \big(p_\pi^\rho\gamma^\sigma p'^\delta\big)\gamma^5u(p)\,,
     \label{eq:VpiNDelta}
\end{multline}
and the other charge channels follow by Clebsch--Gordan coefficients. The form factor $F_{\pi N\Delta}$ depends on the pion virtuality, which PCAC and PPD tie to the axial transition form factor, as for the nucleon. Since the pion is on shell in every diagram (except in the pion-in-flight diagram), $Q^2 = -m_\pi^2$ is fixed, and because $h_D$ is defined from the decay width at this same point, the vertex is point-like with $F_{\pi N\Delta} = 1$. The nucleon vertex of Eq.~\eqref{eq:GT-function} is also point-like, but the Goldberger--Treiman relation is used in the opposite direction: for the nucleon it fixes the pion coupling from $g_A$, while for the $\Delta$ it fixes the axial coupling $C^A_5(0)$ of Eq.~\eqref{eq:C5A0} from the pion coupling $h_D$. The few-percent Goldberger--Treiman discrepancy therefore sits on the pion vertex for the nucleon and on the axial vertex for the $\Delta$.

The on-shell form factor above constrains the pion leg only. The $\Delta$ leg of the strong vertex is off shell in every diagram, and the vertex of Eq.~\eqref{eq:VpiNDelta} grows with the $\Delta$ virtuality, so the resonance term does not fall off above the peak. We therefore multiply the strong $\pi N\Delta$ vertex, and the contact vertices that gauging Eq.~\eqref{eq:LpiNDelta-GI} generates with it, by a regulating function of the invariant mass of the $\Delta$ leg,
\begin{equation}
    F_\Delta(p_\Delta^2)=\frac{\Lambda^4}{\Lambda^4+\left(p_\Delta^2-m_\Delta^2\right)^2}\,,
    \label{eq:FDeltaLam}
\end{equation}
with $\Lambda = 1.121$~GeV, the cutoff that Ref.~\cite{Fernandez-Ramirez:2005nuq} uses for its Born terms. The function is one for an on-shell $\Delta$, so neither $h_D$ nor the PCAC and PPD relations above are affected, and it does not vanish below the $\pi N$ threshold, so the crossed $u$-channel $\Delta$ is damped rather than removed. The gauge-boson $N \to \Delta$ transition vertices of Sec.~\ref{sec:delta-ew} carry no such factor, so each $\Delta$ diagram is multiplied by $F_\Delta$ once.


\section{CVC, PCAC and PPD relations}
\label{app:cvc,pcac,ppd}

In this appendix we derive the three relations used to constrain the form factors in Sec.~\ref{sec:formfactors} and Appendix~\ref{app:FFs}: the conserved vector current (CVC), the partially conserved axial current (PCAC), and pion-pole dominance (PPD). We write them for the nucleon; the same arguments apply to the $N \to \Delta$ transition as long as the $\Delta$ is close to its mass shell, with different Lorentz structures. 

The currents involved are the isovector vector currents $J_V^{\mu,a}$, $a = 1,2,3$, the isoscalar vector current $J_S^{\mu}$, and the axial currents $J_A^{\mu,a}$, whose charges generate the chiral group $SU(2)_L\otimes SU(2)_R\otimes U(1)_B$ of Sec.~\ref{sec:chipt}. The vector currents are conserved and the axial currents are not. CVC and PCAC are statements about the divergences of these currents, and PPD is the assumption that the pion pole which PCAC implies for the pseudoscalar form factor is the dominant contribution to it.


\subsection{Conserved Vector Current}

We first write the most general form the nucleon matrix elements of the vector currents can take, and then see what current conservation removes. Lorentz invariance limits the Dirac structures between the spinors to $\gamma^\mu$, $i\sigma^{\mu\nu}q_\nu$, and $q^\mu = \left(p' - p\right)^\mu$. Isospin symmetry, $H = SU(2)_V\otimes U(1)_B$ of the strong Hamiltonian, fixes the isospin dependence: the matrix element must be an isospin singlet, and there is exactly one way to build a singlet from two nucleon doublets and an isoscalar current ($\mathbf{2}\otimes\mathbf{1}\otimes\mathbf{2} = \mathbf{1}\oplus\mathbf{3}$) or an isovector one ($\mathbf{2}\otimes\mathbf{3}\otimes\mathbf{2} = \mathbf{1}\oplus\mathbf{3}\oplus\mathbf{3}\oplus\mathbf{5}$), namely $\delta_{ij}$ and $\tau^a_{ij}$. The general forms are therefore
\begin{align}
\braket{N_j|J_S^\mu(0)|N_i} & = \overline{u}(p')\left[F_1^S\gamma^\mu + F_2^S\frac{i\sigma^{\mu\nu}q_\nu}{2m_N}\right.\nonumber\\
&\qquad\qquad\left. + F_3^S \frac{q^\mu}{m_N}\right]u(p)\delta_{ij}\,,\label{eq:chiral-basis-form-factors-S}\\
\braket{N_j|J_V^{\mu, a}(0)|N_i} & = \overline{u}(p')\left[F_1^V\gamma^\mu + F_2^V\frac{i\sigma^{\mu\nu}q_\nu}{2m_N}\right.\nonumber\\
&\qquad\qquad\left. + F_3^V\frac{q^\mu}{m_N}\right]u(p)\frac{\tau^a_{ij}}{2}\,,\label{eq:chiral-basis-form-factors-V}
\end{align}
with separate form factors for the isoscalar ($S$) and isovector ($V$) currents, since isospin symmetry does not mix the two. Conservation of each of these currents then removes its $q^\mu$ term. Between momentum eigenstates, where $\braket{N_j|\mathcal{O}(x)|N_i} = e^{iqx}\braket{N_j|\mathcal{O}(0)|N_i}$ for any local operator $\mathcal{O}$, $\partial_\mu J^\mu = 0$ becomes
\begin{align}
    \braket{N_j|\partial_\mu J^\mu(x)|N_i} = 0\implies q_\mu\braket{N_j|J^\mu(0)|N_i} = 0\,.
\end{align}
The $\gamma^\mu$ term drops out because $\overline{u}(p')\slashed{q}u(p) = 0$ by the Dirac equation, the $\sigma^{\mu\nu}$ term because it is antisymmetric, and what is left is $q^2 F_3/m_N$, so $F_3^S = F_3^V = 0$. This is the statement $\mathcal{F}_S = 0$ of Sec.~\ref{sec:formfactors}.

The electromagnetic current is the combination of the isovector current $J_V^{3,\mu}$ and the isoscalar current $J_S^\mu$ fixed by the charge $Q = I_3 + \frac{1}{2}B$,
\begin{align}
    J_{EM}^\mu & = J_V^{3, \mu} +\frac{1}{2}J^\mu_S\,.
\end{align}
Its proton and neutron matrix elements give $F_i^{p,n} = \frac{1}{2}\left(F_i^S\pm F_i^V\right)$, or, since $F_i^{p,n}$ are the measured quantities,
\begin{align}
    F^{S,V}_i = F^p_i \pm F^n_i\,.
\end{align}
On the SM side, the weak hypercharge current follows from $Q = T_3 + Y$,
\begin{align}
    J_Y^{\mu} & = J_{EM}^\mu - J_L^{3,\mu}\,.
\end{align}
To compute the $Z$ and $W$ form factors we express the SM currents in terms of the $L$ and $R$ currents,
\begin{align}
J_{R/L}^{\mu,a} & = \frac{1}{2}\left(J_V^{\mu,a} \pm J_A^{\mu,a}\right)
\end{align}
\begin{align}
    \mathcal{L}_{int}^{SM} & = g_1 J_Y^\mu B_\mu + g_2 J_L^{\mu, a}W_\mu^a\nonumber\\
    & = g_1 (J_{EM}^\mu - J_L^{3,\mu}) \left(\cos\theta_W A_\mu - \sin\theta_W Z_\mu\right)\nonumber\\
    & +\sum_{a=1,2} g_2 J_L^{\mu, a}W_\mu^a +g_2 J_L^{\mu, 3}\left(\cos\theta_W Z_\mu + \sin\theta_W A_\mu\right)
\end{align}
Because the photon only couples to the electromagnetic current we identify the coefficients in front of relevant terms with
\begin{align}
    J_{EM}^\mu A_\mu\,:\,\hspace{6em} g_1\cos\theta_W & = e\,,\\
    J_L^{\mu, 3} A_\mu\,:\,\hspace{1em}-g_1\cos\theta_W + g_2\sin\theta_W & = 0\,.
\end{align}
Finally, the SM interaction Lagrangian is reproduced 
\begin{align}
    \mathcal{L}_{int}^{SM} & = e J_{EM}^\mu A_\mu + \frac{g_2}{2\cos\theta_W}J_{NC}^\mu Z_\mu\nonumber\\
    & \hspace{5em}+ \frac{g_2}{2\sqrt{2}}J_{CC}^{+,\mu}W^+_\mu + \text{h.c.}
\end{align}
with
\begin{align}
    J_{NC}^\mu & = \left(1 - 2\sin^2\theta_W\right)J_V^{3,\mu} - \sin^2\theta_W\, J_S^\mu - J_A^{3,\mu}\,,\nonumber\\
    J_{CC}^{\pm,\mu} & = V_{ud}\left[\left(J_V^{1,\mu}\pm iJ_V^{2,\mu}\right) - \left(J_A^{1,\mu}\pm iJ_A^{2,\mu}\right)\right]\,.
\end{align}
expressed in the basis of $J_V, J_A$ and $J_S$ adapted to the chiral symmetry. The matrix elements of $J_{CC}, J_{NC}$ and $J_{EM}$ with nucleons can now be expressed in terms of $F^S$ and $F^V$ using Eqs.~\eqref{eq:chiral-basis-form-factors-S} and~\eqref{eq:chiral-basis-form-factors-V}, or, equivalently, in terms of the measured $F^p$ and $F^n$ vector form factors. In particular, the vector form factors of the charged current are the isovector electromagnetic ones, $F_i^V = F_i^p - F_i^n$, which is the CVC statement. The axial current is not conserved, so its matrix elements are instead constrained by PCAC, which becomes exact in the chiral limit.


\subsection{Partially Conserved Axial Current}

PCAC relates the axial form factors $F_A$ and $F_P$ to the pion--nucleon coupling. It states that the divergence of the axial current is proportional to the pion field,
\begin{align}
    \partial_\mu A^{a,\mu}(x) = f_\pi m_\pi^2\,\pi^a(x)\,,
    \label{eq:pcac-operator}
\end{align}
so that the axial current is conserved only in the chiral limit $m_\pi^2 \to 0$. To turn this operator identity into a relation between form factors, we sandwich both sides between nucleon states. The left-hand side is then expressed through $F_A$ and $F_P$, the right-hand side through the pion--nucleon coupling, and equating the two gives the relation. 

We start with the right-hand side. The pion field obeys a Klein--Gordon equation whose source $J^a_\pi$ is whatever the pion couples to,
\begin{align}
    \left(\Box + m_\pi^2\right)\pi^a(x) = J^a_\pi(x)\,.
\end{align}
The nucleon matrix element of the equation of motion at $x = 0$ is, with $q = p' - p$,
\begin{align}
    \braket{N_j|\left(\Box + m_\pi^2\right)\pi^a(0)|N_i}
    &= \left(m_\pi^2 - q^2\right)\braket{N_j|\pi^a(0)|N_i} \nonumber\\
    &= \braket{N_j|J^a_\pi(0)|N_i}\,.
\end{align}
The matrix element of the source, in turn, by definition, is the pion--nucleon vertex, with coupling $g_{\pi NN}(q^2)$,
\begin{align}
    \braket{N_j(p')|J_\pi^a(0)|N_i(p)} = i g_{\pi NN}(q^2)\,\overline{u}(p')\gamma_5 u(p)\,\tau^a_{ij}\,,
\end{align}
and therefore
\begin{align}
    \braket{N_j|\pi^a(0)|N_i} & = \frac{ig_{\pi NN}(q^2)}{m_\pi^2-q^2}\,\overline{u}(p')\gamma_5 u(p)\,\tau^a_{ij}\,.
    \label{eq:pion-matrix-element}
\end{align}
That is, the nucleon matrix element of the pion field is the pion--nucleon vertex times a pion propagator.

Now the left-hand side. The axial current matrix element is parametrized as
\begin{align}
    \braket{N_j|A^a_\mu(0)|N_i} & = \overline{u}(p')\left[\gamma_\mu F_A(q^2)\right.\nonumber\\
    & \!\!\qquad\left.+ \frac{q_\mu}{2m_N}F_P(q^2)\right]\gamma_5\,\frac{\tau^a_{ij}}{2}\,u(p)\,.
\end{align}
(A third structure, $\sigma_{\mu\nu}q^\nu\gamma_5$, is allowed by Lorentz invariance but forbidden by G-parity.) Its divergence at $x = 0$ is $iq_\mu$ times this, and since the Dirac equation gives
\begin{align}
    \overline{u}(p')\slashed{q}\gamma_5 u(p) & = \overline{u}(p')\left(\slashed{p}'\gamma_5 + \gamma_5\slashed{p}\right)u(p)\nonumber\\
    & = 2m_N\,\overline{u}(p')\gamma_5 u(p)\,,
\end{align}
we obtain
\begin{multline}
    \braket{N_j|\partial_\mu A^{a,\mu}(0)|N_i} =\\ i\left[m_N F_A(q^2) + \frac{q^2}{4m_N}F_P(q^2)\right]\overline{u}(p')\gamma_5 u(p)\,\tau^a_{ij}\,.
\end{multline}
On the other hand, the nucleon matrix element of the PCAC relation, Eq.~\eqref{eq:pcac-operator}, with Eq.~\eqref{eq:pion-matrix-element} for the pion field, gives
\begin{align}
    \braket{N_j|\partial_\mu A^{a,\mu}(0)|N_i} & = f_\pi m_\pi^2\braket{N_j|\pi^a(0)|N_i}\nonumber\\
    & = i\,\frac{f_\pi m_\pi^2\, g_{\pi NN}(q^2)}{m_\pi^2 - q^2}\,\overline{u}(p')\gamma_5 u(p)\,\tau^a_{ij}\,.
\end{align}
Equating the two gives the PCAC relation between the form factors,
\begin{align}
    \label{eq:pcac}
    m_N F_A(q^2) + \frac{q^2}{4m_N}F_P(q^2) = \frac{f_\pi m_\pi^2\, g_{\pi NN}(q^2)}{m_\pi^2 - q^2}\,.
\end{align}

Its best-known consequence follows at $q^2 = 0$. Since $F_P$ is regular there (its pole is at the pion mass, as shown below), the $F_P$ term drops out and $m_N F_A(0) = f_\pi\, g_{\pi NN}(0)$. The physical pion--nucleon coupling is $g_{\pi NN}$ at the pion pole, $q^2 = m_\pi^2$; if $g_{\pi NN}(q^2)$ varies little between $q^2 = 0$ and there, this is the Goldberger--Treiman relation,
\begin{align}
    g_{\pi NN} \simeq \frac{g_A \, m_N}{f_\pi}\,,\qquad g_A \equiv F_A(0)\,.
\end{align}
%


\subsection{Pion Pole Dominance}

PCAC also fixes $F_P$ itself, up to one assumption. Multiplying Eq.~\eqref{eq:pcac} by $(m_\pi^2-q^2)$ and taking $q^2\to m_\pi^2$ shows that $F_P$ has a pole at the pion mass with residue
\begin{align}
    \lim_{q^2\to m_\pi^2}(m_\pi^2-q^2)\,F_P(q^2) = 4m_N f_\pi\, g_{\pi NN}(m_\pi^2)\,.
\end{align}
The pole is the process in which the axial current creates a pion that the nucleon then absorbs. Pion-pole dominance is the assumption that this process accounts for the full $q^2$ dependence of $F_P$, which amounts to using the Goldberger--Treiman relation at all $q^2$, $f_\pi g_{\pi NN}(q^2) \simeq m_N F_A(q^2)$. Inserting this into Eq.~\eqref{eq:pcac} gives
\begin{align}
    \label{eq:ppd}
    F_P(q^2) \simeq \frac{4m_N^2\,F_A(q^2)}{m_\pi^2-q^2}\,,
\end{align}
which with $q^2 = -Q^2$ is the form used in Eq.~\eqref{eq:F_P}.


\section{The \texorpdfstring{$\Delta$}{Delta} Line Shape and the Resonance Region}
\label{app:lineshape}

Both free-nucleon comparisons in Sec.~\ref{sec:free} show the same excess, a $\Delta$ that is too strong above its pole. This appendix explains where the excess comes from and shows which changes remove it. Every curve in this appendix is computed with a semi-analytic evaluation of the same UFO vertices that \textsc{Sherpa} uses, which reproduces the \textsc{Sherpa} scan to within one percent when run with \textsc{Sherpa}'s prescriptions.


\subsection{Where the excess comes from}

\begin{figure}
    \centering
    \includegraphics[width=\columnwidth]{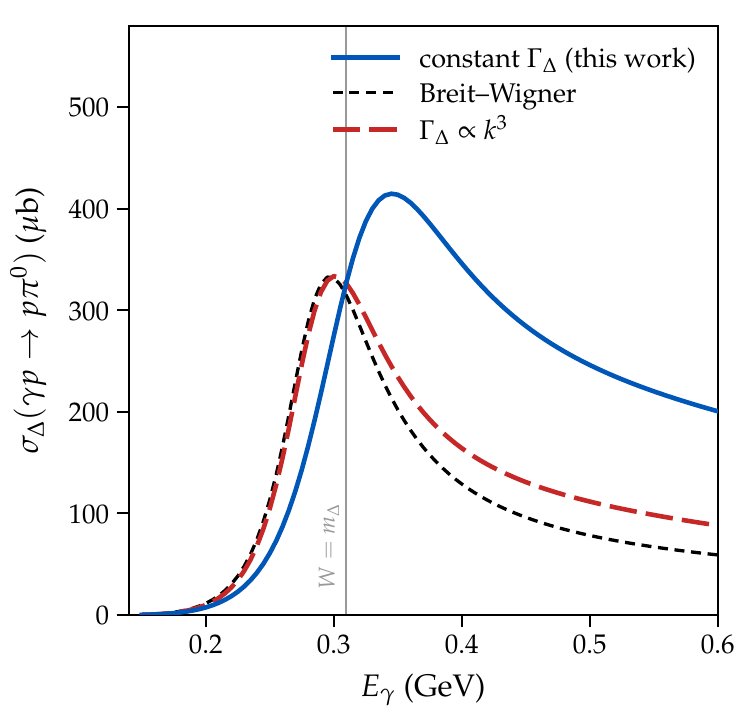}
    \caption{Line shape of the $\Delta$ term alone in $\gamma p \to p\pi^0$. Solid blue: this work, a constant width $\Gamma_\Delta = 102$~MeV at the pole mass $m_\Delta = 1.209$~GeV. Dashed black: the Breit--Wigner cross section, Eq.~(\ref{eq:bw}), with the same pole and widths, $\Gamma_{N\pi} \propto k^3$ for the $P$-wave decay and $\Gamma_{N\gamma} \propto q^3$ for the magnetic dipole decay. Dashed red: this work with that energy-dependent width in the $\Delta$ propagator and nothing else changed. The vertical line marks $W = m_\Delta$.}
    \label{fig:lineshape}
\end{figure}

We first diagnose the problem, namely why the peak is too high when the couplings reproduce the $\Delta$ widths. The Breit--Wigner cross section of a resonance~\cite{Peskin:1995ev,ParticleDataGroup:2026mpi} is fixed at $W = m_\Delta$ by the branching ratios and spins alone,
\begin{equation}
\label{eq:bwpeak}
\sigma(W = m_\Delta) = \frac{4\pi}{q^2}\,\frac{2J+1}{(2s_\gamma+1)(2s_N+1)}\,B_{N\gamma}\,B_{N\pi}\,,
\end{equation}
times the isospin factor $2/3$ for $\gamma p \to p\pi^0$. Here $q$ is the photon momentum in the $\pi N$ center of mass, $J = 3/2$ is the spin of the $\Delta$, the factor $(2J+1)/[(2s_\gamma+1)(2s_N+1)]$ counts spin states ($2s_\gamma + 1 = 2$ for the two photon polarizations and $2s_N + 1 = 2$ for the nucleon), and $B_{N\gamma} = \Gamma_{N\gamma}/\Gamma$ and $B_{N\pi} = \Gamma_{N\pi}/\Gamma$ are the branching ratios of the $\Delta$ into the initial and final states. Because it contains no coupling constant, it tests whether the $\Delta$ amplitude has the strength its widths imply. With our widths, $\Gamma_{N\pi} = 101.4$~MeV and $\Gamma_{N\gamma} = 0.57$~MeV, it gives $315\,\mu$b, and our calculation with the $\Delta$ alone gives $327\,\mu$b at the same energy ($E_\gamma = 0.31$~GeV in Fig.~\ref{fig:pi0p-variants}). The agreement at the pole indicates that the discrepancy does not arise from the coupling normalizations.

Away from the pole the shape is wrong, as Fig.~\ref{fig:lineshape} shows by comparing three line shapes for the $\Delta$ term alone. The blue curve is our tree-level calculation with the constant width. The black curve is the Breit--Wigner cross section, which, as we argue below, should be accurate in the region near the $\Delta$ mass. In a Breit--Wigner the widths are the decay rates of the resonance, and a decay rate grows with the phase space of the decay, so the widths depend on the energy $W$. For the $P$-wave decay $\Delta \to N\pi$ it is $\Gamma_{N\pi}(W) \propto k^3$, and for the magnetic dipole decay $\Delta \to N\gamma$ it is $\Gamma_{N\gamma}(W) \propto q^3$, where $k$ and $q$ are the pion and photon momenta in the $\pi N$ center of mass. These energy-dependent widths are what enter the Breit--Wigner cross section. 

The cross section through a resonance factorizes into its production, its propagation, and its decay. The production rate from $\gamma p$ is the same quantity as the decay rate to $N\gamma$, and the decay rate to $\pi N$ is $\Gamma_{N\pi}$, so the cross section is

\begin{equation}
\label{eq:bw}
\sigma_\Delta(W) = \frac{4\pi}{q^2}\,\frac{2J+1}{(2s_\gamma+1)(2s_N+1)}\;\frac{\Gamma_{N\gamma}(W)\,\Gamma_{N\pi}(W)/4}{(W - m_\Delta)^2 + \Gamma(W)^2/4}\,,
\end{equation}

times the same isospin factor. The prefactor is the largest cross section a single partial wave can carry (the unitarity limit), and Eq.~(\ref{eq:bwpeak}) is this expression evaluated at $W = m_\Delta$. The widths enter twice, in the numerator as the rates of the two steps and in the denominator as the total decay rate that broadens the peak, so their momentum dependence fixes the whole line shape. We note that even though Eq.~\eqref{eq:bw} is an approximation, it should hold within a few widths of the resonance mass. 

For $\gamma p \to p\pi^0$ near the $\Delta$ peak, a single partial wave dominates with little non-resonant background, and the momenta are small compared with the inverse size of the $\Delta$, so the $k^3$ and $q^3$ threshold forms of the widths hold. The $\Delta$ is also isolated, with the next resonance being the $N(1440)$. The black curve is Eq.~\eqref{eq:bw}, and it falls off on both sides of the resonance mass. Our constant-width curve instead keeps rising above the pole. This arises from the structure of the amplitude. The $\gamma N\Delta$ and $\pi N\Delta$ vertices each bring one power of momentum, so the numerator grows with $W$ in the same way as the product of the two partial widths, while the propagator denominator keeps the constant width. In a Breit--Wigner the two grow together and the resonance turns over above its pole. With a constant width, the cross section instead continues to increase above the pole. 

\begin{figure*}[t]
    \centering
    \includegraphics[width=0.9\linewidth]{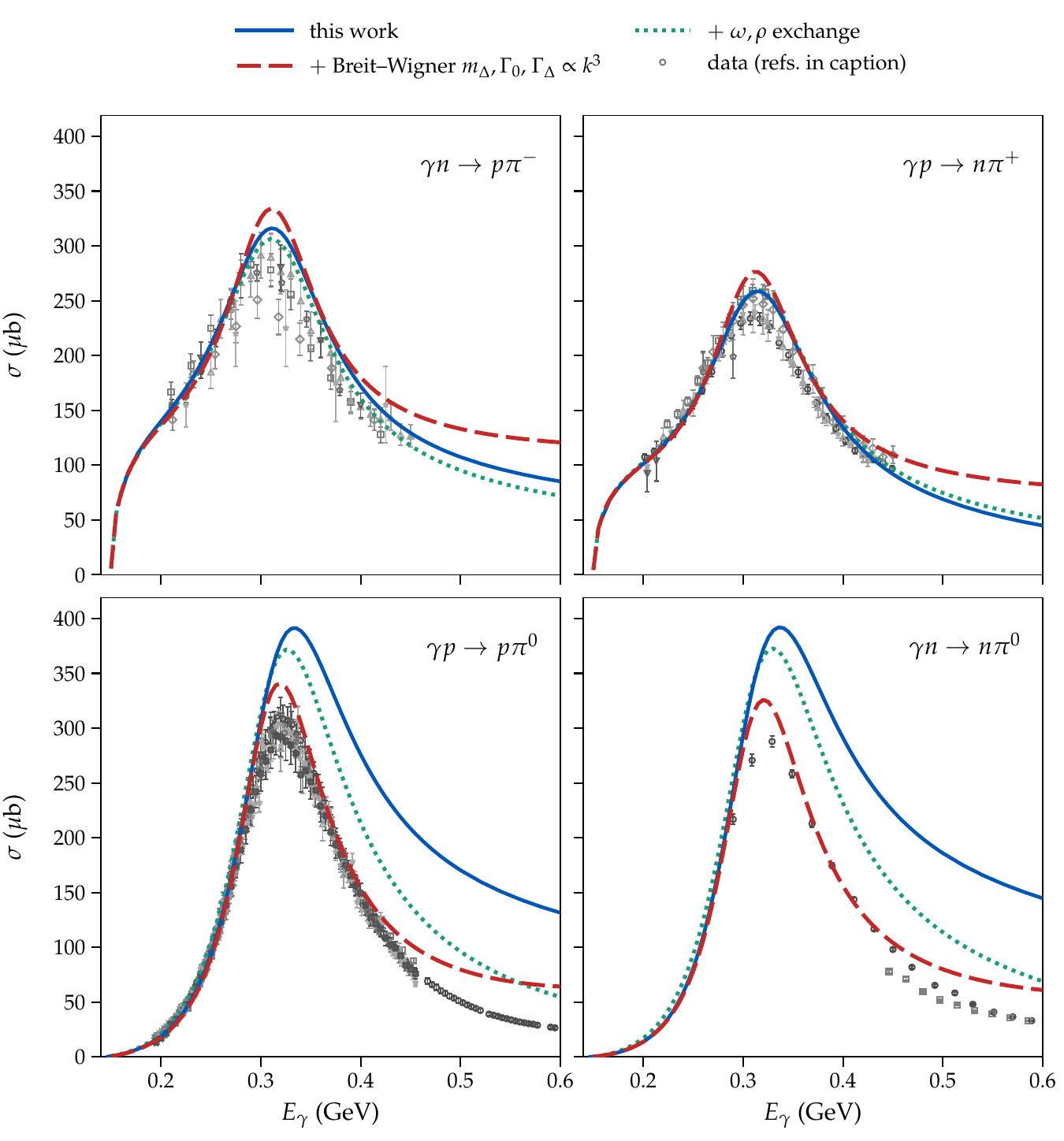}
    \caption{Total pion photoproduction cross sections, as in Fig.~\ref{fig:photoproduction}, with the successive modifications discussed in the text. Solid blue: this work (constant width, pole parameters, vertex-level form factors). Dotted green: with $t$-channel $\omega$ and $\rho$ exchange added with the couplings of Ref.~\cite{Fernandez-Ramirez:2005nuq}. Dashed red: with, in addition, the energy-dependent width $\Gamma(W) = \Gamma_0\,(k/k_0)^3\,(m_\Delta/W)$ and the Breit--Wigner parameters $m_\Delta = 1.232$~GeV, $\Gamma_0 = 117.7$~MeV. The couplings $h_D$, $G_1$, and $G_2$ are re-derived from the Breit--Wigner widths $\Gamma(\Delta \to N\pi) = 117$~MeV and $\Gamma(\Delta \to N\gamma) = 0.66$~MeV. Data for $\gamma n \to p\pi^-$ are from Refs.~\cite{White:1960ukk,Aachen-Bonn-Hamburg-Heidelberg-Muenchen:1973kaj,Rossi:1973wf,VonHoltey:1974br,Fujii:1976jg,Bagheri:1987kf,Wang:1992thesis}; for $\gamma p \to n\pi^+$ from Refs.~\cite{Knapp:1959zz,Fischer:1972mt,Fujii:1976jg,Fissum:1996fi,MacCormick:1996jz,Blanpied:2001ae,GDH:2004ydy}; and for $\gamma p \to p\pi^0$ from Refs.~\cite{Vasilkov:1960jetp,Govorkov:1968sjnp,Hitzeroth:1969du,Fischer:1971zp245,Fuchs:1996ja,Haerter:1996thesis,MacCormick:1996jz,GDH:2004ydy,CB-ELSA:2004sqg,Schumann:2010js,A2:2012lnr,A2:2015mhs}; and for $\gamma n \to n\pi^0$ from Refs.~\cite{A2:2018jcd,A2:2019yud}.}
    \label{fig:photoprod-appendix}
\end{figure*}

The excess is a consequence of this mismatch. The red curve shows that the energy-dependent width, put in the propagator with nothing else changed, restores the Breit--Wigner shape at the peak, while above it the amplitude falls more slowly than Eq.~\eqref{eq:bw}, by about 25\% at $E_\gamma = 0.4$ GeV.

\vspace{-1.2em}
\subsection{Width, parameters, and vector mesons}

Figure~\ref{fig:photoprod-appendix} shows the modifications of Fig.~\ref{fig:pi0p-variants} for all four channels. Two of them were discussed in Sec.~\ref{sec:free}. $\omega$ and $\rho$ exchange changes the neutral-pion tails and little else. The $\omega$ couples to the neutral pion only, so it is absent from the charged channels, where only the weaker $\rho^\pm$ exchange enters, at the $\sim$10\% level. 
The vector-meson couplings are the fit values of Ref.~\cite{Fernandez-Ramirez:2005nuq}, not parameters of this model, and the vector mesons are not part of the exported UFO. The dotted curve shows what the framework gives once they are added, which the follow-up to this work will do at the Lagrangian level.

Fixing the peak requires the energy-dependent width of Fig.~\ref{fig:lineshape}, $\Gamma(W) = \Gamma_0\,(k/k_0)^3\,(m_\Delta/W)$ with $k_0$ the pion momentum at $W = m_\Delta$~\cite{Hernandez:2007qq}. With it in the propagator the resonance turns over above its pole. With the pole parameters, however, the peak sits at $E_\gamma = 0.30$~GeV, below the data, and is still $\sim$$20\%$ too high. The reason is that the energy-dependent width and the mass that goes with it are not independent. The width is the imaginary part of the $\Delta$ self-energy, the effect of the $\Delta$ turning into $\pi N$ and back, and the same process has a real part that shifts the mass. Pole and Breit--Wigner parameters differ by precisely that shift, $1.209$ against $1.232$~GeV, so a width that runs belongs with the Breit--Wigner mass and width, and a constant width belongs with the pole values. With $m_\Delta = 1.232$~GeV, $\Gamma_0 = 117.7$~MeV, and the energy-dependent width, and with $h_D$, $G_1$, and $G_2$ re-derived from the Breit--Wigner widths $\Gamma(\Delta \to N\pi) = 117$~MeV and $\Gamma(\Delta \to N\gamma) = 0.66$~MeV (the red curves in Fig.~\ref{fig:photoprod-appendix}), all four peaks come to within about ten percent of the data. The mixed choices do not work. Breit--Wigner parameters with a constant width shift the resonance by $35$~MeV to higher energy and describe the data worse than the pole values do. We also tried the $X_\pi(W)$ form of the width used in Ref.~\cite{Fernandez-Ramirez:2005nuq}, which grows like $k^3$ near threshold but saturates at the pole and falls slowly above it, and it changes the peak by only a few percent. Reference~\cite{Fernandez-Ramirez:2005nuq} describes the data with this width because its electromagnetic $\Delta$ couplings are fitted to the photoproduction data rather than fixed by the radiative width.

\begin{figure}[t]
    \centering
    \includegraphics[width=\columnwidth]{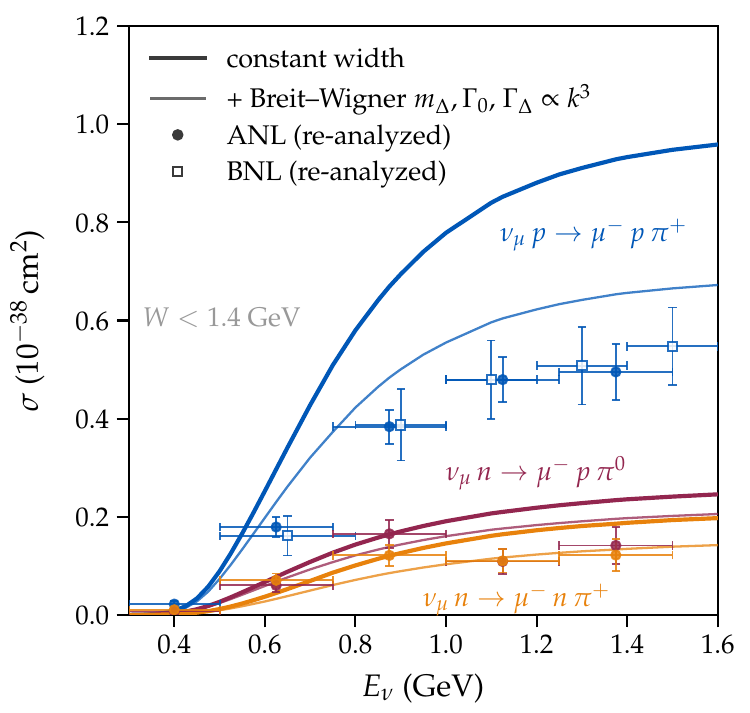}
    \caption{Charged-current single-pion production, as in Fig.~\ref{fig:nuN}. Thick lines: this work. Thin lines: with the energy-dependent width $\Gamma(W) = \Gamma_0 (k/k_0)^3 (m_\Delta/W)$ and the Breit--Wigner parameters $m_\Delta = 1.232$~GeV, $\Gamma_0 = 117.7$~MeV, with $h_D$, $G_1$, and $G_2$ re-derived from the Breit--Wigner widths as in Fig.~\ref{fig:photoprod-appendix} and all other inputs unchanged. A cut $W < 1.4$~GeV is applied to both the calculation and the data. Data are from the ANL~\cite{Radecky:1981fn} and BNL~\cite{Kitagaki:1986ct} deuterium bubble-chamber experiments in the normalization of the re-analysis of Refs.~\cite{Wilkinson:2014yfa,Rodrigues:2016xjj}.}
    \label{fig:cc1pi-appendix}
\end{figure}

The same change has the same effect for neutrinos. Figure~\ref{fig:cc1pi-appendix} shows the three charged-current channels with the energy-dependent width and the Breit--Wigner parameters, and nothing else changed. As we can see, the calculation is now broadly consistent with the data in all three channels. Note that the axial coupling in this comparison is again the Goldberger--Treiman value $C_5^A(0) = \sqrt{2/3}\,h_D$, which with $h_D$ re-derived from the Breit--Wigner width is $1.18$ instead of our $1.27$. Reference~\cite{Alvarez-Ruso:2015eva}, using the model of Ref.~\cite{Hernandez:2007qq} with the same energy-dependent width and the Breit--Wigner mass $m_\Delta = 1.232$~GeV, fitted $C_5^A(0)$ to the re-analyzed data and found $1.14 \pm 0.07$, consistent with the Goldberger--Treiman value. Their fit also includes deuteron effects, the Adler form factor $C_4^A=-C_5^A/4$, and the Olsson phases~\cite{Olsson:1974sw} discussed below. When the fit is performed without including the phases, the fitted $C_5^A(0)$ is found to be $1.05 \pm 0.07$.

The main text keeps the constant width at the pole parameters because it is the leading-order result. It follows from the Lagrangian with no function inserted by hand, it is equivalent to a complex $\Delta$ mass, which keeps the amplitude gauge invariant, and it is what the generator evaluates. The energy-dependent width is the imaginary part of the one-loop $\pi N$ correction to the $\Delta$ propagator. To include it consistently, the real part of the same loop, which shifts the mass from the pole value toward the Breit--Wigner value, has to be included with it, and the gauge invariance of the amplitude has to be restored. This is the next order of the calculation. Until then, the comparisons of this appendix quantify the leading-order uncertainty in the resonance region, about 20--30\% at the $\Delta$ peak in neutral-pion photoproduction and 20--40\% in the integrated charged-current rates.

Furthermore, Watson's theorem~\cite{Watson:1952ji} fixes the phase of the pion-production amplitude.
Each multipole is required to carry the phase of $\pi N$ scattering in the same partial wave.
This theorem follows from unitarity and time-reversal invariance and should hold where $\pi N$ scattering is elastic, which is a good approximation in the $\Delta$ region.
In our model, Watson's theorem is not satisfied since the non-resonant terms are real at tree level and the $\Delta$ amplitude only carries a phase from its propagator.
This results in the wrong phase in the $P_{33}$ partial wave, and the interference of the $\Delta$ with the non-resonant amplitudes is not constrained. 
There are two approaches to resolve this issue.
\begin{figure*}[t]
    \centering
    \includegraphics[width=0.9\linewidth]{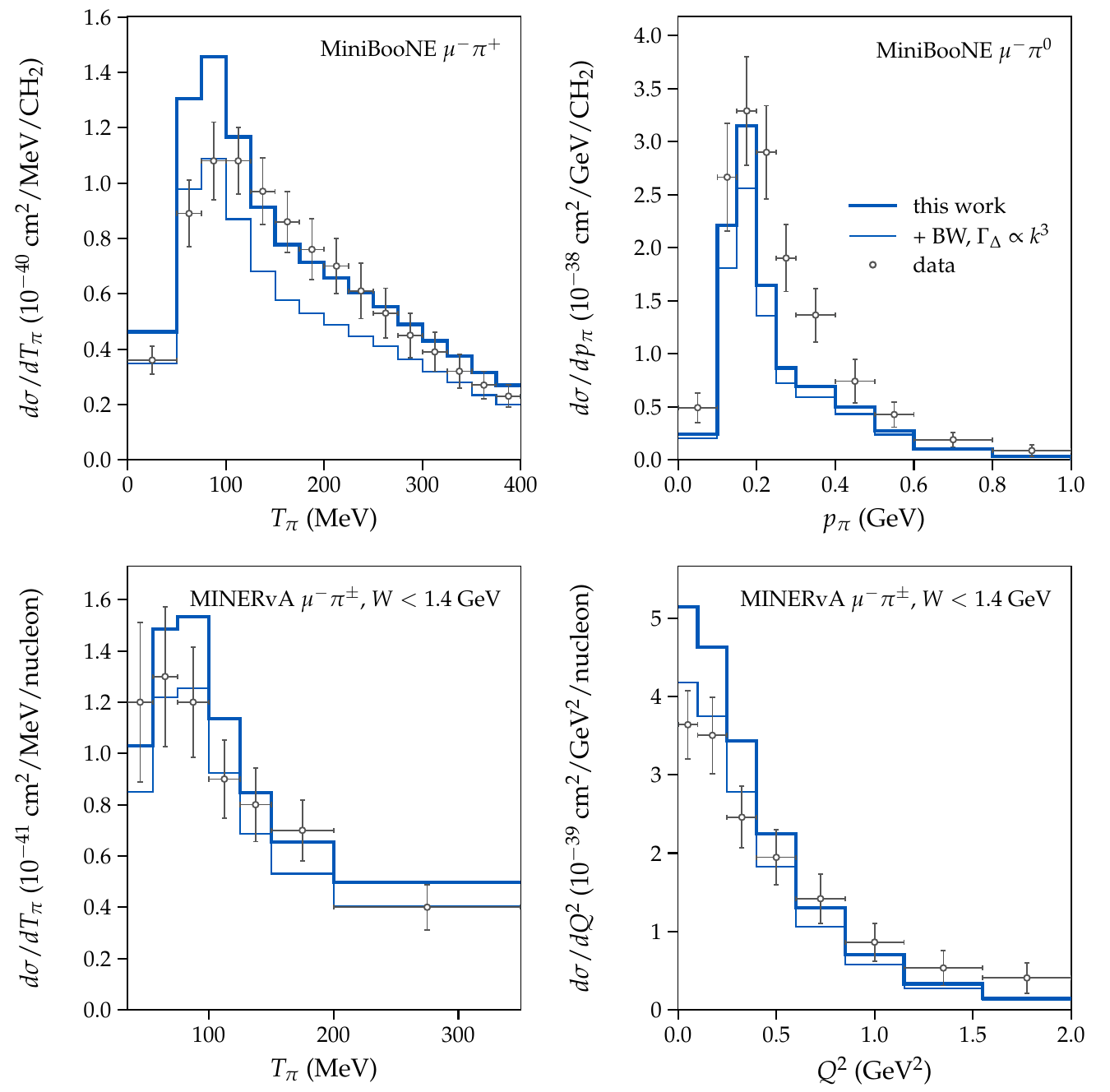}
    \caption{Charged-current single-pion production on hydrocarbon, as in Figs.~\ref{fig:minerva-pi0} and~\ref{fig:t2k-pip}. Top: the $\pi^+$ kinetic energy and the $\pi^0$ momentum measured by MiniBooNE~\cite{MiniBooNE:2010eis,MiniBooNE:2010cxl}, $0.5 < E_\nu < 2$~GeV, per CH$_2$ molecule. Bottom: the $\pi^\pm$ kinetic energy and $Q^2$ measured by MINERvA with $W < 1.4$~GeV~\cite{MINERvA:2014ogb,MINERvA:2016sfc}, $1.5 < E_\nu < 10$~GeV. Thick lines: this work. Thin lines: the Breit--Wigner variant. The pipeline reproduces the MiniBooNE and MINERvA $\pi^\pm$ data only because it sits above the free-nucleon data, and the variant that describes T2K sits about $30\%$ below them and a factor of two below the MiniBooNE $\pi^0$ data. These are the normalization tensions among the nuclear datasets discussed in Ref.~\cite{Betancourt:2018bpu}.}
    \label{fig:nuclear-appendix}
\end{figure*}
The first approach is that of Olsson~\cite{Olsson:1974sw}, in which the $\Delta$ amplitude is multiplied by phases chosen so that the dominant multipoles carry the $P_{33}$ phase, and the non-resonant amplitudes stay real. This is the approach taken in Ref.~\cite{Alvarez-Ruso:2015eva}.
The other approach is the $K$-matrix approach~\cite{Drechsel:2007if,Hooft:2026yyg}. In this approach, the tree-level amplitude is unitarized partial wave by partial wave, which gives the non-resonant amplitudes their $\pi N$ phases. The two approaches are compared in Ref.~\cite{Tsaran:2024sue}.
Neither of the two approaches above is possible to implement without significant modifications of both \textsc{Sherpa} and \textsc{Achilles}, since they require modifications on partial-wave projections of the amplitude. These improvements are left to future work.


\section{Nuclear Comparisons}
\label{app:nuclear}

The nuclear single-pion datasets are known to disagree with one another in normalization beyond their quoted uncertainties~\cite{Betancourt:2018bpu}, and no event generator describes them together. In this appendix we show where our model sits relative to the measurements that Sec.~\ref{sec:nuclear-data} left out, the MiniBooNE $\pi^+$ and $\pi^0$ measurements and the MINERvA $\pi^\pm$ measurement, and we use the model as a common ruler to quantify the tensions. Figure~\ref{fig:nuclear-appendix} shows the comparison in the same format as Figs.~\ref{fig:minerva-pi0} and~\ref{fig:t2k-pip}. The MINERvA $\pi^\pm$ data are the 2017 re-extraction of Ref.~\cite{MINERvA:2014ogb} with the flux of Ref.~\cite{MINERvA:2016sfc}, which the collaboration recommends.

The pipeline reproduces the $\pi^\pm$ spectra of both experiments in shape and in normalization, as well as the MiniBooNE $Q^2$ distribution from $0$ to $1.5$~GeV$^2$, but it does so while sitting $50$--$60\%$ above the free-nucleon data, so that the Breit--Wigner variant, which describes T2K, sits about $30\%$ below MiniBooNE and MINERvA. Relative to the same model, therefore, the MiniBooNE and T2K $\pi^+$ data differ by about $35\%$, which is the size of the tension reported in Ref.~\cite{Betancourt:2018bpu} between MiniBooNE and MINERvA. 

The tension is larger in the $\pi^0$ channel. The MiniBooNE $\pi^0$ rate is about twice what the ANL $p\pi^0$ data of Fig.~\ref{fig:nuN} imply through our model, and MiniBooNE itself reports its $\pi^0$ cross section to be $1.56 \pm 0.26$ times the prediction it was compared with~\cite{MiniBooNE:2010cxl}. Note that the ANL $p\pi^0$ channel rests on about $200$ events with systematic uncertainties that could not be assessed~\cite{Rodrigues:2016xjj}, so the factor of two is as much a statement about the free-nucleon data as about the nuclear ones. We therefore read the $\pi^0$ comparison as a statement about the datasets rather than about the model, and we do not tune to any of them.


\section{Replicating this work}
\label{app:howtouse}

\begin{table}[!t]
\centering
\footnotesize
\caption{\centering{Software used for the results of this work.}}
\label{tab:software-versions}
\renewcommand{\arraystretch}{1.2}
\setlength{\tabcolsep}{2.5pt}
\begin{tabular}{ll}
\toprule
Software & Version \\
\midrule
FeynRules & 2.3.49 with the patch in Ref.~\cite{ChiEFT:software}, Mathematica 14.3 \\
\textsc{Sherpa} & \texttt{570-update-achilles-interface}, commit \texttt{1cd6f079f} \\
\textsc{Achilles} & \texttt{sherpa3-0}, commit \texttt{ff313b48} \\
NUISANCE & NUISANCE2 \texttt{f5f79db6} with the patches in Ref.~\cite{ChiEFT:software}, \\
& via NUISANCE3 \texttt{32bcc1f8} \\
\bottomrule
\end{tabular}
\end{table}
\begin{figure}[t]
\centering
\vspace{1.2em}
\begin{tikzpicture}[
    box/.style={draw, rounded corners=2pt, thick, align=center, font=\footnotesize,
                minimum width=1.9cm, minimum height=0.85cm, fill=blue!4},
    lab/.style={font=\scriptsize, align=center, text=black!70},
    >=latex, thick]
  \node[box] (fr)  at (0,0)     {FeynRules\\[-1pt]{\scriptsize (patched)}};
  \node[box] (ufo) at (2.9,0)  {UFO 2.0\\[-1pt]{\scriptsize + param card}};
  \node[box] (sh)  at (5.8,0)   {\textsc{Sherpa}\\[-1pt]{\scriptsize (\textsc{Comix})}};
  \node[box] (ach) at (5.8,-2.1) {\textsc{Achilles}};
  \node[box] (nu)  at (2.9,-2.1) {NUISANCE};
  \node[box, fill=none, dashed] (dat) at (0,-2.1) {Data};
  \draw[->] (fr) -- (ufo);
  \draw[->] (ufo) -- (sh);
  \draw[->] ([xshift=-4pt]sh.south) -- node[lab, left=1pt] {matrix\\element} ([xshift=-4pt]ach.north);
  \draw[->] ([xshift=4pt]ach.north) -- node[lab, right=1pt] {phase-space\\point} ([xshift=4pt]sh.south);
  \draw[->] (ach) -- (nu);
  \draw[->, dashed] (dat) -- (nu);
\end{tikzpicture}
\caption{The chain used to produce the results of this work, from the Lagrangian to the comparison with data.
\textsc{Achilles} supplies the beam, the nucleus, the nuclear model, and the phase-space point; \textsc{Sherpa}
returns the matrix element. The events are written in the NuHepMC format and compared to data with NUISANCE.}
\vspace{-1.0em}
\label{fig:chain}
\end{figure}
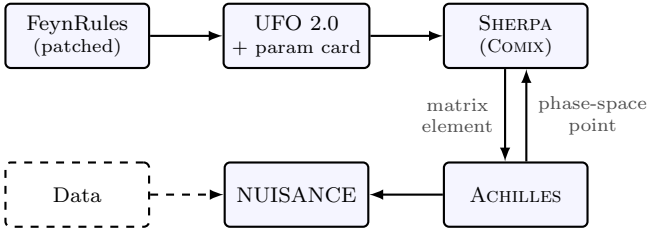

All files needed to reproduce this work can be found in the \texttt{ChiEFT} repository on GitHub (\url{https://github.com/AchillesGen/ChiEFT}), and archived on Zenodo~\cite{ChiEFT:software}.
The repository contains the FeynRules~\cite{Alloul:2013bka} model and export notebook, the UFO~2.0~\cite{Degrande:2011ua,Darme:2023jdn} model with its parameter card, a \textsc{Sherpa}~\cite{Sherpa:2024mfk} plugin implementing the form factors, the FeynRules patch, and the run cards, scripts, and data needed to reproduce all figures.
The software versions are listed in Table~\ref{tab:software-versions}.

Figure~\ref{fig:chain} shows the chain of the calculation. FeynRules does not provide a method to write form factors into the Lorentz structures as dictated by the UFO~2.0 format, nor does it handle spin-$3/2$ fields correctly. The patch within the repository fixes both. However, the patch is only required to regenerate the UFO files if desired. The generated UFO files and parameter card are included for direct use. The \textsc{Sherpa} model library is built from the UFO together with the form-factor plugin. \textsc{Achilles}~\cite{Isaacson:2022cwh,Isaacson:2025cnk} calls into \textsc{Sherpa} as a matrix element backend, extending the interface developed in Ref.~\cite{Isaacson:2021xty}. In this interface, \textsc{Achilles} provides the beam, the nucleus, the nuclear model, and the phase-space point, while \textsc{Sherpa} loads the model and calculates the full matrix element with \textsc{COMIX}~\cite{Gleisberg:2008fv,Hoppe2026} using the provided information. Events are written in the NuHepMC format~\cite{Gardiner:2025NuHepMC} and binned against data with NUISANCE~\cite{Stowell:2016jfr}.

The figures are reproduced by running \texttt{make} in the \texttt{paper/} directory of the repository. Its steps produce the semi-analytic tables (running $\Delta$ width, vector-meson exchange, $\Delta$ line shape), the free-nucleon scans with \textsc{Sherpa}, the \textsc{Achilles} events for the nuclear and electron samples, the NUISANCE analysis in a docker container, and the figures. The \texttt{README} files in the repository provide additional details for each step.


\bibliographystyle{apsrev4-2}
\bibliography{refs}

@article{ParticleDataGroup:2026mpi,
    author = "Takahashi, F. and others",
    collaboration = "Particle Data Group",
    title = "{Review of Particle Physics}",
    doi = "10.1142/s0217751x26300115",
    journal = "Int. J. Mod. Phys. A",
    volume = "41",
    number = "22",
    pages = "2630011",
    year = "2026"
}

@article{Carlson:2014vla,
    author = "Carlson, J. and Gandolfi, S. and Pederiva, F. and Pieper, Steven C. and Schiavilla, R. and Schmidt, K. E. and Wiringa, R. B.",
    title = "{Quantum Monte Carlo methods for nuclear physics}",
    eprint = "1412.3081",
    archivePrefix = "arXiv",
    primaryClass = "nucl-th",
    reportNumber = "LA-UR-14-29251, JLAB-THY-14-1989, LA-UR-14-29251",
    doi = "10.1103/RevModPhys.87.1067",
    journal = "Rev. Mod. Phys.",
    volume = "87",
    pages = "1067",
    year = "2015"
}

@misc{MicroBooNE:2015bmn,
    author = "Acciarri, R. and others",
    collaboration = "MicroBooNE, LAr1-ND, ICARUS-WA104",
    title = "{A Proposal for a Three Detector Short-Baseline Neutrino Oscillation Program in the Fermilab Booster Neutrino Beam}",
    eprint = "1503.01520",
    archivePrefix = "arXiv",
    primaryClass = "physics.ins-det",
    reportNumber = "FERMILAB-PUB-15-0726-AD-PPD",
    month = "3",
    year = "2015"
}

@misc{Hyper-Kamiokande:2018ofw,
    author = "Abe, K. and others",
    collaboration = "Hyper-Kamiokande",
    title = "{Hyper-Kamiokande Design Report}",
    eprint = "1805.04163",
    archivePrefix = "arXiv",
    primaryClass = "physics.ins-det",
    month = "5",
    year = "2018"
}

@misc{DUNE:2020ypp,
    author = "Abi, Babak and others",
    collaboration = "DUNE",
    title = "{Deep Underground Neutrino Experiment (DUNE), Far Detector Technical Design Report, Volume II: DUNE Physics}",
    eprint = "2002.03005",
    archivePrefix = "arXiv",
    primaryClass = "hep-ex",
    reportNumber = "FERMILAB-PUB-20-025-ND, FERMILAB-DESIGN-2020-02",
    doi = "10.2172/1599307",
    month = "2",
    year = "2020"
}

@article{Meyer:2016oeg,
    author = "Meyer, Aaron S. and Betancourt, Minerba and Gran, Richard and Hill, Richard J.",
    title = "{Deuterium target data for precision neutrino-nucleus cross sections}",
    eprint = "1603.03048",
    archivePrefix = "arXiv",
    primaryClass = "hep-ph",
    reportNumber = "FERMILAB-PUB-16-185-ND-T",
    doi = "10.1103/PhysRevD.93.113015",
    journal = "Phys. Rev. D",
    volume = "93",
    number = "11",
    pages = "113015",
    year = "2016"
}

@article{GENIE:2021npt,
    author = "Alvarez-Ruso, Luis and others",
    collaboration = "GENIE",
    title = "{Recent highlights from GENIE v3}",
    eprint = "2106.09381",
    archivePrefix = "arXiv",
    primaryClass = "hep-ph",
    reportNumber = "FERMILAB-PUB-21-266-SCD-T",
    doi = "10.1140/epjs/s11734-021-00295-7",
    journal = "Eur. Phys. J. ST",
    volume = "230",
    number = "24",
    pages = "4449--4467",
    year = "2021"
}

@article{Gonzalez-Jimenez:2016qqq,
    author = "Gonz\'alez-Jim\'enez, R. and Jachowicz, N. and Niewczas, K. and Nys, J. and Pandey, V. and Van Cuyck, T. and Van Dessel, N.",
    title = "{Electroweak single-pion production off the nucleon: from threshold to high invariant masses}",
    eprint = "1612.05511",
    archivePrefix = "arXiv",
    primaryClass = "nucl-th",
    doi = "10.1103/PhysRevD.95.113007",
    journal = "Phys. Rev. D",
    volume = "95",
    number = "11",
    pages = "113007",
    year = "2017"
}

@article{Abe:2024avs,
    author = "Abe, Seisho",
    title = "{Implementation and investigation of electron-nucleus scattering in the neut neutrino event generator}",
    eprint = "2412.07466",
    archivePrefix = "arXiv",
    primaryClass = "hep-ph",
    doi = "10.1103/PhysRevD.111.033006",
    journal = "Phys. Rev. D",
    volume = "111",
    number = "3",
    pages = "033006",
    year = "2025"
}

@article{Hayato:2009zz,
    author = "Hayato, Yoshinari",
    editor = "Ankowski, Arthur and Sobczyk, Jan",
    title = "{A neutrino interaction simulation program library NEUT}",
    journal = "Acta Phys. Polon. B",
    volume = "40",
    pages = "2477--2489",
    year = "2009"
}

@article{Golan:2012wx,
    author = "Golan, Tomasz and Juszczak, Cezary and Sobczyk, Jan T.",
    title = "{Final State Interactions Effects in Neutrino-Nucleus Interactions}",
    eprint = "1202.4197",
    archivePrefix = "arXiv",
    primaryClass = "nucl-th",
    reportNumber = "FERMILAB-PUB-12-280-E",
    doi = "10.1103/PhysRevC.86.015505",
    journal = "Phys. Rev. C",
    volume = "86",
    pages = "015505",
    year = "2012"
}

@article{Andreopoulos:2009rq,
    author = "Andreopoulos, C. and others",
    title = "{The GENIE Neutrino Monte Carlo Generator}",
    eprint = "0905.2517",
    archivePrefix = "arXiv",
    primaryClass = "hep-ph",
    reportNumber = "FERMILAB-PUB-09-418-CD",
    doi = "10.1016/j.nima.2009.12.009",
    journal = "Nucl. Instrum. Meth. A",
    volume = "614",
    pages = "87--104",
    year = "2010"
}

@article{Isaacson:2020wlx,
    author = "Isaacson, Joshua and Jay, William I. and Lovato, Alessandro and Machado, Pedro A. N. and Rocco, Noemi",
    title = "{New approach to intranuclear cascades with quantum Monte Carlo configurations}",
    eprint = "2007.15570",
    archivePrefix = "arXiv",
    primaryClass = "hep-ph",
    reportNumber = "FERMILAB-PUB-20-368-T",
    doi = "10.1103/PhysRevC.103.015502",
    journal = "Phys. Rev. C",
    volume = "103",
    number = "1",
    pages = "015502",
    year = "2021"
}

@article{Buss:2011mx,
    author = "Buss, O. and Gaitanos, T. and Gallmeister, K. and van Hees, H. and Kaskulov, M. and Lalakulich, O. and Larionov, A. B. and Leitner, T. and Weil, J. and Mosel, U.",
    title = "{Transport-theoretical Description of Nuclear Reactions}",
    eprint = "1106.1344",
    archivePrefix = "arXiv",
    primaryClass = "hep-ph",
    doi = "10.1016/j.physrep.2011.12.001",
    journal = "Phys. Rept.",
    volume = "512",
    pages = "1--124",
    year = "2012"
}

@article{Mosel:2019vhx,
    author = "Mosel, Ulrich",
    title = "{Neutrino event generators: foundation, status and future}",
    eprint = "1904.11506",
    archivePrefix = "arXiv",
    primaryClass = "hep-ex",
    doi = "10.1088/1361-6471/ab3830",
    journal = "J. Phys. G",
    volume = "46",
    number = "11",
    pages = "113001",
    year = "2019"
}

@article{DeForest:1983ahx,
    author = "De Forest, T.",
    title = "{Off-Shell electron Nucleon Cross-Sections. The Impulse Approximation}",
    doi = "10.1016/0375-9474(83)90124-0",
    journal = "Nucl. Phys. A",
    volume = "392",
    pages = "232--248",
    year = "1983"
}

@article{Golan:2012rfa,
    author = "Golan, T. and Sobczyk, J. T. and Zmuda, J.",
    editor = "Tzanakos, George S.",
    title = "{NuWro: the Wroclaw Monte Carlo Generator of Neutrino Interactions}",
    doi = "10.1016/j.nuclphysbps.2012.09.136",
    journal = "Nucl. Phys. B Proc. Suppl.",
    volume = "229-232",
    pages = "499--499",
    year = "2012"
}

@article{Hayato:2021heg,
    author = "Hayato, Yoshinari and Pickering, Luke",
    title = "{The NEUT neutrino interaction simulation program library}",
    eprint = "2106.15809",
    archivePrefix = "arXiv",
    primaryClass = "hep-ph",
    doi = "10.1140/epjs/s11734-021-00287-7",
    journal = "Eur. Phys. J. ST",
    volume = "230",
    number = "24",
    pages = "4469--4481",
    year = "2021"
}

@article{Isaacson:2022cwh,
    author = "Isaacson, Joshua and Jay, William I. and Lovato, Alessandro and Machado, Pedro A. N. and Rocco, Noemi",
    title = "{Introducing a novel event generator for electron-nucleus and neutrino-nucleus scattering}",
    eprint = "2205.06378",
    archivePrefix = "arXiv",
    primaryClass = "hep-ph",
    reportNumber = "FERMILAB-PUB-22-411-T, MIT-CTP/5428",
    doi = "10.1103/PhysRevD.107.033007",
    journal = "Phys. Rev. D",
    volume = "107",
    number = "3",
    pages = "033007",
    year = "2023"
}

@article{Isaacson:2021xty,
    author = {Isaacson, Joshua and H\"oche, Stefan and Lopez Gutierrez, Diego and Rocco, Noemi},
    title = "{Novel event generator for the automated simulation of neutrino scattering}",
    eprint = "2110.15319",
    archivePrefix = "arXiv",
    primaryClass = "hep-ph",
    reportNumber = "FERMILAB-PUB-21-537-T, MCNET-21-31",
    doi = "10.1103/PhysRevD.105.096006",
    journal = "Phys. Rev. D",
    volume = "105",
    number = "9",
    pages = "096006",
    year = "2022"
}

@article{Coleman:1969sm,
    author = "Coleman, Sidney R. and Wess, J. and Zumino, Bruno",
    title = "{Structure of phenomenological Lagrangians. 1.}",
    doi = "10.1103/PhysRev.177.2239",
    journal = "Phys. Rev.",
    volume = "177",
    pages = "2239--2247",
    year = "1969"
}

@article{Callan:1969sn,
    author = "Callan, Jr., Curtis G. and Coleman, Sidney R. and Wess, J. and Zumino, Bruno",
    title = "{Structure of phenomenological Lagrangians. 2.}",
    doi = "10.1103/PhysRev.177.2247",
    journal = "Phys. Rev.",
    volume = "177",
    pages = "2247--2250",
    year = "1969"
}

@article{Goldstone:1962es,
    author = "Goldstone, Jeffrey and Salam, Abdus and Weinberg, Steven",
    title = "{Broken Symmetries}",
    doi = "10.1103/PhysRev.127.965",
    journal = "Phys. Rev.",
    volume = "127",
    pages = "965--970",
    year = "1962"
}

@inproceedings{Mukhopadhyay:1998iw,
    author = "Mukhopadhyay, Nimai C.",
    title = "{Weak form-factors of the nucleon}",
    booktitle = "{5th International WEIN Symposium: A Conference on Physics Beyond the Standard Model (WEIN 98)}",
    eprint = "nucl-th/9810039",
    archivePrefix = "arXiv",
    reportNumber = "RPI-N124-98",
    pages = "222--238",
    month = "6",
    year = "1998"
}

@incollection{scherer2003introduction,
  title={Introduction to chiral perturbation theory},
  author={Scherer, Stefan},
  booktitle={Advances in Nuclear Physics, Volume 27},
  pages={277--538},
  year={2003},
  publisher={Springer}
}

@article{MINERvA:2019kfr,
    author = "Stowell, P. and others",
    collaboration = "MINERvA",
    title = "{Tuning the GENIE Pion Production Model with MINER$\nu$A Data}",
    eprint = "1903.01558",
    archivePrefix = "arXiv",
    primaryClass = "hep-ex",
    reportNumber = "FERMILAB-PUB-19-093-ND",
    doi = "10.1103/PhysRevD.100.072005",
    journal = "Phys. Rev. D",
    volume = "100",
    number = "7",
    pages = "072005",
    year = "2019"
}

@misc{Isaacson:2025lyx,
    author = "Isaacson, Joshua",
    title = "{Revisiting Gauge Ambiguities for DUNE Precision}",
    eprint = "2502.08727",
    archivePrefix = "arXiv",
    primaryClass = "hep-ph",
    reportNumber = "FERMILAB-PUB-25-0059-T",
    month = "2",
    year = "2025"
}

@article{Kelly:1997zza,
    author = "Kelly, James J.",
    title = "{Gauge ambiguities in (e --{\ensuremath{>}}, e-prime N --{\ensuremath{>}}) reactions}",
    doi = "10.1103/PhysRevC.56.2672",
    journal = "Phys. Rev. C",
    volume = "56",
    pages = "2672--2687",
    year = "1997"
}

@article{Barreau:1983ht,
      author = "Barreau, P. and others",
      title = "{Deep Inelastic electron Scattering from Carbon}",
      doi = "10.1016/0375-9474(83)90217-8",
      journal = "Nucl. Phys. A",
      volume = "402",
      pages = "515--540",
      year = "1983"
  }

@article{Coyle:2025xjk,
    author = "Coyle, Nina M. and Li, Shirley Weishi and Machado, Pedro A. N.",
    title = "{Neutrino-nucleus cross section impacts on neutrino oscillation measurements}",
    eprint = "2502.19467",
    archivePrefix = "arXiv",
    primaryClass = "hep-ph",
    reportNumber = "FERMILAB-PUB-25-0001-T, UCI-HEP-TR-2025-01",
    doi = "10.1103/PhysRevD.111.093010",
    journal = "Phys. Rev. D",
    volume = "111",
    number = "9",
    pages = "093010",
    year = "2025"
}

@article{Coyle:2022bwa,
    author = "Coyle, Nina M. and Li, Shirley Weishi and Machado, Pedro A. N.",
    title = "{The impact of neutrino-nucleus interaction modeling on new physics searches}",
    eprint = "2210.03753",
    archivePrefix = "arXiv",
    primaryClass = "hep-ph",
    reportNumber = "FERMILAB-PUB-22-726-T",
    doi = "10.1007/JHEP12(2022)166",
    journal = "JHEP",
    volume = "2022",
    number = "12",
    pages = "166",
    month = "10",
    year = "2022"
}

@article{NOvA:2021nfi,
    author = "Acero, M. A. and others",
    collaboration = "NOvA",
    title = "{Improved measurement of neutrino oscillation parameters by the NOvA experiment}",
    eprint = "2108.08219",
    archivePrefix = "arXiv",
    primaryClass = "hep-ex",
    reportNumber = "FERMILAB-PUB-21-373-ND",
    doi = "10.1103/PhysRevD.106.032004",
    journal = "Phys. Rev. D",
    volume = "106",
    number = "3",
    pages = "032004",
    year = "2022"
}

@article{Ankowski:2015kya,
    author = "Ankowski, Artur M. and Coloma, Pilar and Huber, Patrick and Mariani, Camillo and Vagnoni, Erica",
    title = "{Missing energy and the measurement of the CP-violating phase in neutrino oscillations}",
    eprint = "1507.08561",
    archivePrefix = "arXiv",
    primaryClass = "hep-ph",
    reportNumber = "FERMILAB-PUB-15-320-T",
    doi = "10.1103/PhysRevD.92.091301",
    journal = "Phys. Rev. D",
    volume = "92",
    number = "9",
    pages = "091301",
    year = "2015"
}

@article{T2K:2023mcm,
    author = "Abe, K. and others",
    collaboration = "T2K",
    title = "{Updated T2K measurements of muon neutrino and antineutrino disappearance using 3.6\texttimes{}1021 protons on target}",
    eprint = "2305.09916",
    archivePrefix = "arXiv",
    primaryClass = "hep-ex",
    doi = "10.1103/PhysRevD.108.072011",
    journal = "Phys. Rev. D",
    volume = "108",
    number = "7",
    pages = "072011",
    year = "2023"
}

@article{Kelly:2004hm,
    author = "Kelly, J. J.",
    title = "{Simple parametrization of nucleon form factors}",
    doi = "10.1103/PhysRevC.70.068202",
    journal = "Phys. Rev. C",
    volume = "70",
    pages = "068202",
    year = "2004"
}

@article{Hernandez:2013jka,
    author = "Hern\'andez, E. and Nieves, J. and Vicente Vacas, M. J.",
    title = "{Single \ensuremath{\pi} production in neutrino-nucleus scattering}",
    eprint = "1304.1320",
    archivePrefix = "arXiv",
    primaryClass = "hep-ph",
    doi = "10.1103/PhysRevD.87.113009",
    journal = "Phys. Rev. D",
    volume = "87",
    number = "11",
    pages = "113009",
    year = "2013"
}

@article{Alvarez-Ruso:2015eva,
    author = "Alvarez-Ruso, L. and Hern\'andez, E. and Nieves, J. and Vicente Vacas, M. J.",
    title = "{Watson's theorem and the $N\Delta(1232)$ axial transition}",
    eprint = "1510.06266",
    archivePrefix = "arXiv",
    primaryClass = "hep-ph",
    doi = "10.1103/PhysRevD.93.014016",
    journal = "Phys. Rev. D",
    volume = "93",
    number = "1",
    pages = "014016",
    year = "2016"
}

@article{Hernandez:2007qq,
    author = "Hernandez, E. and Nieves, J. and Valverde, M.",
    title = "{Weak Pion Production off the Nucleon}",
    eprint = "hep-ph/0701149",
    archivePrefix = "arXiv",
    doi = "10.1103/PhysRevD.76.033005",
    journal = "Phys. Rev. D",
    volume = "76",
    pages = "033005",
    year = "2007"
}

@article{Yao:2018pzc,
    author = "Yao, De-Liang and Alvarez-Ruso, Luis and Hiller Blin, Astrid N. and Vicente Vacas, M. J.",
    title = "{Weak pion production off the nucleon in covariant chiral perturbation theory}",
    eprint = "1806.09364",
    archivePrefix = "arXiv",
    primaryClass = "hep-ph",
    doi = "10.1103/PhysRevD.98.076004",
    journal = "Phys. Rev. D",
    volume = "98",
    number = "7",
    pages = "076004",
    year = "2018"
}

@article{Rarita:1941mf,
    author = "Rarita, William and Schwinger, Julian",
    title = "{On a theory of particles with half integral spin}",
    doi = "10.1103/PhysRev.60.61",
    journal = "Phys. Rev.",
    volume = "60",
    pages = "61",
    year = "1941"
}

@article{Moldauer:1956zz,
    author = "Moldauer, P. A. and Case, K. M.",
    title = "{Properties of Half-Integral Spin Dirac-Fierz-Pauli Particles}",
    doi = "10.1103/PhysRev.102.279",
    journal = "Phys. Rev.",
    volume = "102",
    pages = "279--285",
    year = "1956"
}

@article{Weinberg:1964cn,
    author = "Weinberg, Steven",
    title = "{Feynman Rules for Any Spin}",
    doi = "10.1103/PhysRev.133.B1318",
    journal = "Phys. Rev.",
    volume = "133",
    pages = "B1318--B1332",
    year = "1964"
}

@misc{Criado:2021gcb,
    author = {Criado, Juan Carlos and Djouadi, Abdelhak and Koivunen, Niko and M{\"u}{\"u}rsepp, Kristjan and Raidal, Martti and Veerm{\"a}e, Hardi},
    title = "{An effective field theory of the Delta-resonance}",
    eprint = "2106.09031",
    archivePrefix = "arXiv",
    primaryClass = "hep-ph",
    month = "6",
    year = "2021"
}

@article{Criado:2020jkp,
    author = {Criado, Juan Carlos and Koivunen, Niko and Raidal, Martti and Veerm{\"a}e, Hardi},
    title = "{Dark matter of any spin -- an effective field theory and applications}",
    eprint = "2010.02224",
    archivePrefix = "arXiv",
    primaryClass = "hep-ph",
    doi = "10.1103/PhysRevD.102.125031",
    journal = "Phys. Rev. D",
    volume = "102",
    number = "12",
    pages = "125031",
    year = "2020"
}

@article{Pascalutsa:1999zz,
    author = "Pascalutsa, Vladimir and Timmermans, Rob",
    title = "{Field theory of nucleon to higher spin baryon transitions}",
    eprint = "nucl-th/9905065",
    archivePrefix = "arXiv",
    reportNumber = "NIKHEF-99-017, KVI-1466",
    doi = "10.1103/PhysRevC.60.042201",
    journal = "Phys. Rev. C",
    volume = "60",
    pages = "042201",
    year = "1999",
    url = {https://arxiv.org/abs/nucl-th/9905065}
}

@article{Johnson:1960vt,
    author = "Johnson, Kenneth and Sudarshan, E. C. G.",
    title = "{Inconsistency of the local field theory of charged spin 3/2 particles}",
    doi = "10.1016/0003-4916(61)90030-6",
    journal = "Annals Phys.",
    volume = "13",
    pages = "126--145",
    year = "1961"
}

@article{Velo:1969bt,
    author = "Velo, Giorgio and Zwanziger, Daniel",
    title = "{Propagation and quantization of Rarita-Schwinger waves in an external electromagnetic potential}",
    doi = "10.1103/PhysRev.186.1337",
    journal = "Phys. Rev.",
    volume = "186",
    pages = "1337--1341",
    year = "1969"
}

@article{Pascalutsa:2006up,
    author = "Pascalutsa, Vladimir and Vanderhaeghen, Marc and Yang, Shin Nan",
    title = "{Electromagnetic excitation of the Delta(1232)-resonance}",
    eprint = "hep-ph/0609004",
    archivePrefix = "arXiv",
    reportNumber = "JLAB-THY-06-537",
    doi = "10.1016/j.physrep.2006.09.006",
    journal = "Phys. Rept.",
    volume = "437",
    pages = "125--232",
    year = "2007",
    url ={https://arxiv.org/pdf/hep-ph/0609004}
}

@article{Pascalutsa:2000kd,
    author = "Pascalutsa, V.",
    title = "{Correspondence of consistent and inconsistent spin - 3/2 couplings via the equivalence theorem}",
    eprint = "hep-ph/0008026",
    archivePrefix = "arXiv",
    doi = "10.1016/S0370-2693(01)00140-X",
    journal = "Phys. Lett. B",
    volume = "503",
    pages = "85--90",
    year = "2001"
}

@article{Mart:2019jtb,
    author = "Mart, T. and Kristiano, J. and Clymton, S.",
    title = "{Pure spin-3/2 representation with consistent interactions}",
    eprint = "1909.04282",
    archivePrefix = "arXiv",
    primaryClass = "nucl-th",
    doi = "10.1103/PhysRevC.100.035207",
    journal = "Phys. Rev. C",
    volume = "100",
    number = "3",
    pages = "035207",
    year = "2019",
    url = {https://arxiv.org/pdf/1909.04282}
}

@article{Pascalutsa:1998pw,
    author = "Pascalutsa, V.",
    title = "{Quantization of an interacting spin - 3 / 2 field and the Delta isobar}",
    eprint = "hep-ph/9802288",
    archivePrefix = "arXiv",
    reportNumber = "THU-98-07",
    doi = "10.1103/PhysRevD.58.096002",
    journal = "Phys. Rev. D",
    volume = "58",
    pages = "096002",
    year = "1998",
    url = {https://arxiv.org/abs/hep-ph/9802288v2}
}

@article{Praet:2008yn,
    author = "Praet, C. and Lalakulich, O. and Jachowicz, N. and Ryckebusch, J.",
    title = "{Delta-mediated pion production in nuclei}",
    eprint = "0804.2750",
    archivePrefix = "arXiv",
    primaryClass = "nucl-th",
    doi = "10.1103/PhysRevC.79.044603",
    journal = "Phys. Rev. C",
    volume = "79",
    pages = "044603",
    year = "2009"
}

@article{A2:2015mhs,
    author = "Adlarson, P. and others",
    collaboration = "A2",
    title = "{Measurement of $\pi^0$ photoproduction on the proton at MAMI C}",
    eprint = "1506.08849",
    archivePrefix = "arXiv",
    primaryClass = "hep-ex",
    doi = "10.1103/PhysRevC.92.024617",
    journal = "Phys. Rev. C",
    volume = "92",
    number = "2",
    pages = "024617",
    year = "2015"
}

@article{A2:2012lnr,
    author = "Hornidge, D. and others",
    collaboration = "A2, CB-TAPS",
    title = "{Accurate Test of Chiral Dynamics in the $\vec{\gamma} p \to \pi^0 p$ Reaction}",
    eprint = "1211.5495",
    archivePrefix = "arXiv",
    primaryClass = "nucl-ex",
    doi = "10.1103/PhysRevLett.111.062004",
    journal = "Phys. Rev. Lett.",
    volume = "111",
    number = "6",
    pages = "062004",
    year = "2013"
}

@article{Schumann:2010js,
    author = "Schumann, S. and others",
    title = "{Radiative pi0 photoproduction on protons in the Delta+(1232) region}",
    eprint = "1001.3626",
    archivePrefix = "arXiv",
    primaryClass = "hep-ex",
    doi = "10.1140/epja/i2010-10925-y",
    journal = "Eur. Phys. J. A",
    volume = "43",
    pages = "269--282",
    year = "2010"
}

@article{Fuchs:1996ja,
    author = "Fuchs, M and others",
    title = "{Neutral pion photoproduction from the proton near threshold}",
    doi = "10.1016/0370-2693(95)01488-8",
    journal = "Phys. Lett. B",
    volume = "368",
    pages = "20--25",
    year = "1996"
}

@article{MacCormick:1996jz,
    author = "MacCormick, M. and others",
    title = "{Total photoabsorption cross-sections for H-1, H-2 and He-3 from 200-MeV to 800-MeV}",
    reportNumber = "DAPNIA-SPHN-96-03",
    doi = "10.1103/PhysRevC.53.41",
    journal = "Phys. Rev. C",
    volume = "53",
    pages = "41--49",
    year = "1996"
}

@article{GDH:2004ydy,
    author = "Ahrens, J. and others",
    collaboration = "GDH, A2",
    title = "{Helicity dependence of the $\gamma p\to  N \pi$ channels and multipole analysis in the $\Delta$ region}",
    doi = "10.1140/epja/i2003-10216-x",
    journal = "Eur. Phys. J. A",
    volume = "21",
    number = "2",
    pages = "323--333",
    year = "2004"
}

@article{CB-ELSA:2004sqg,
    author = "Bartholomy, O. and others",
    collaboration = "CB-ELSA",
    title = "{Neutral pion photoproduction off protons in the energy range 0.3-GeV $< E_\gamma <$ 3-GeV}",
    eprint = "hep-ex/0407022",
    archivePrefix = "arXiv",
    doi = "10.1103/PhysRevLett.94.012003",
    journal = "Phys. Rev. Lett.",
    volume = "94",
    pages = "012003",
    year = "2005"
}

@article{Hitzeroth:1969du,
    author = "Hitzeroth, W.",
    title = "{Photoproduction of neutral pions on protons near threshold}",
    doi = "10.1007/BF02757017",
    journal = "Nuovo Cim. A",
    volume = "60",
    pages = "467--480",
    year = "1969"
}

@article{Benhar:1994hw,
    author = "Benhar, O. and Fabrocini, A. and Fantoni, S. and Sick, I.",
    title = "{Spectral function of finite nuclei and scattering of GeV electrons}",
    doi = "10.1016/0375-9474(94)90920-2",
    journal = "Nucl. Phys. A",
    volume = "579",
    pages = "493--517",
    year = "1994"
}

@article{Nikolakopoulos:2024mjj,
    author = "Nikolakopoulos, A. and Ershova, A. and Gonz{\'a}lez-Jim{\'e}nez, R. and Isaacson, J. and Kelly, A. M. and Niewczas, K. and Rocco, N. and S{\'a}nchez, F.",
    title = "{Final-state interactions in neutrino-induced proton knockout from argon in MicroBooNE}",
    eprint = "2406.09244",
    archivePrefix = "arXiv",
    primaryClass = "nucl-th",
    reportNumber = "FERMILAB-PUB-24-0327-T",
    doi = "10.1103/PhysRevC.110.054611",
    journal = "Phys. Rev. C",
    volume = "110",
    number = "5",
    pages = "054611",
    year = "2024"
}

@article{Murphy:2019wed,
    author = "Murphy, M. and others",
    title = "{Measurement of the cross sections for inclusive electron scattering in the E12-14-012 experiment at Jefferson Lab}",
    eprint = "1908.01802",
    archivePrefix = "arXiv",
    primaryClass = "hep-ex",
    reportNumber = "JLAB-PHY-19-3013",
    doi = "10.1103/PhysRevC.100.054606",
    journal = "Phys. Rev. C",
    volume = "100",
    number = "5",
    pages = "054606",
    year = "2019"
}

@article{JeffersonLabHallA:2022cit,
    author = "Jiang, L. and others",
    collaboration = "Jefferson Lab Hall A",
    title = "{Determination of the argon spectral function from (e,e'p) data}",
    eprint = "2203.01748",
    archivePrefix = "arXiv",
    primaryClass = "nucl-ex",
    reportNumber = "JLAB-PHY-22-3575, SLAC-PUB-17650",
    doi = "10.1103/PhysRevD.105.112002",
    journal = "Phys. Rev. D",
    volume = "105",
    number = "11",
    pages = "112002",
    year = "2022"
}

@article{JeffersonLabHallA:2022ljj,
    author = "Jiang, L. and others",
    collaboration = "Jefferson Lab Hall A",
    title = "{Determination of the titanium spectral function from (e,{\,}e'p) data}",
    eprint = "2209.14108",
    archivePrefix = "arXiv",
    primaryClass = "nucl-ex",
    doi = "10.1103/PhysRevD.107.012005",
    journal = "Phys. Rev. D",
    volume = "107",
    number = "1",
    pages = "012005",
    year = "2023"
}

@article{Blanpied:2001ae,
    author = "Blanpied, G. and others",
    title = "{$N \to \Delta$ transition and proton polarizabilities from measurements of $p(\vec{\gamma},\gamma)$, $p(\vec{\gamma},\pi^0)$, and $p(\vec{\gamma},\pi^+)$}",
    doi = "10.1103/PhysRevC.64.025203",
    journal = "Phys. Rev. C",
    volume = "64",
    pages = "025203",
    year = "2001"
}

@article{Fissum:1996fi,
    author = "Fissum, K. G. and Caplan, H. S. and Hallin, E. L. and Skopik, D. M. and Vogt, J. M. and Frodyma, M. and Rosenzweig, D. P. and Storm, D. W. and O'Rielly, G. V. and Garrow, K. R.",
    title = "{Inclusive positive pion photoproduction}",
    doi = "10.1103/PhysRevC.53.1278",
    journal = "Phys. Rev. C",
    volume = "53",
    pages = "1278--1289",
    year = "1996"
}

@article{Fischer:1972mt,
    author = "Fischer, G. and Von Holtey, G. and Knop, G. and Stuempfig, J.",
    title = "{Angular distributions of pions from $\gamma p \to \pi^+ n$ at photon energies between 220 and 425 MeV}",
    doi = "10.1007/BF01387168",
    journal = "Z. Phys.",
    volume = "253",
    pages = "38--52",
    year = "1972"
}

@article{Fujii:1976jg,
    author = "Fujii, T. and Kondo, T. and Takasaki, F. and Yamada, S. and Homma, S. and Huke, K. and Kato, S. and Okuno, H. and Endo, I. and Fujii, H.",
    title = "{Photoproduction of Charged pi Mesons from Hydrogen and Deuterium in the Energy Range Between 250-MeV and 790-MeV}",
    reportNumber = "INS-256",
    doi = "10.1016/0550-3213(77)90084-0",
    journal = "Nucl. Phys. B",
    volume = "120",
    pages = "395--422",
    year = "1977"
}

@article{Knapp:1959zz,
    author = "Knapp, Edward A. and Kenney, Robert W. and Perez-Mendez, Victor",
    title = "{Angular Distributions of Photopions from Hydrogen}",
    doi = "10.1103/PhysRev.114.605",
    journal = "Phys. Rev.",
    volume = "114",
    pages = "605--611",
    year = "1959"
}

@article{Bagheri:1987kf,
    author = "Bagheri, A. and Aniol, K. A. and Entezami, F. and Hasinoff, M. D. and Measday, David F. and Poutissou, J. M. and Salomon, M. and Robertson, B. C.",
    title = "{The Reaction $\pi^- p \to \gamma n$ Below the $\Delta$ Resonance}",
    reportNumber = "TRI-PP-87-89",
    doi = "10.1103/PhysRevC.38.875",
    journal = "Phys. Rev. C",
    volume = "38",
    pages = "875--884",
    year = "1988"
}

@article{Aachen-Bonn-Hamburg-Heidelberg-Muenchen:1973kaj,
    author = "Benz, P. and others",
    collaboration = "Aachen-Bonn-Hamburg-Heidelberg-Muenchen",
    title = "{Measurement of the reaction $\gamma d \to \pi^- pp$, and determination of cross-sections for the reaction $\gamma n \to \pi^- p$, at photon energies between 0.2-GeV and 2.0-GeV}",
    reportNumber = "DESY-73-19",
    doi = "10.1016/0550-3213(73)90280-0",
    journal = "Nucl. Phys. B",
    volume = "65",
    pages = "158--209",
    year = "1973"
}

@article{Rossi:1973wf,
    author = "Rossi, V. and others",
    title = "{Analysis of the reaction $\gamma n \to p \pi^-$ in the first and second resonance regions}",
    doi = "10.1007/BF02788907",
    journal = "Nuovo Cim. A",
    volume = "13",
    pages = "59--89",
    year = "1973"
}

@article{VonHoltey:1974br,
    author = "Von Holtey, G. and Knop, G. and Stein, H. and Stuempfig, J. and Wahlen, H.",
    title = "{Photoproduction of charged pions on deuterium in the first resonance region}",
    doi = "10.1016/0550-3213(74)90436-2",
    journal = "Nucl. Phys. B",
    volume = "70",
    pages = "379--389",
    year = "1974"
}

@article{White:1960ukk,
    author = "White, D. H. and Schectman, R. M. and Chasan, B. M.",
    title = "{Photoproduction of Negative Mesons in Deuterium}",
    doi = "10.1103/PhysRev.120.614",
    journal = "Phys. Rev.",
    volume = "120",
    number = "2",
    pages = "614",
    year = "1960"
}

@article{Fernandez-Ramirez:2005nuq,
    author = "Fernandez-Ramirez, C. and Moya de Guerra, E. and Udias, J. M.",
    title = "{Effective Lagrangian approach to pion photoproduction from the nucleon}",
    eprint = "nucl-th/0509020",
    archivePrefix = "arXiv",
    doi = "10.1016/j.aop.2006.02.009",
    journal = "Annals Phys.",
    volume = "321",
    pages = "1408--1456",
    year = "2006"
}

@article{Fischer:1971zp245,
    author = "Fischer, G. and Fischer, H. and von Holtey, G. and K{\"a}mpgen, H. and Knop, G. and Schulz, P. and Wessels, H.",
    title = "{Photoproduction of neutral pions on hydrogen at photon energies between 200 and 440 MeV}",
    doi = "10.1007/BF01395234",
    journal = "Z. Phys.",
    volume = "245",
    pages = "225--252",
    year = "1971"
}

@phdthesis{Haerter:1996thesis,
    author = "H{\"a}rter, F.",
    title = "{Photoproduktion neutraler Pionen am Proton im ersten und zweiten Resonanzgebiet}",
    school = "Johannes Gutenberg-Universit{\"a}t Mainz",
    year = "1996"
}

@phdthesis{Wang:1992thesis,
    author = "Wang, Maozhi",
    title = "{Radiative Decay of the $\Delta$ Resonance}",
    school = "University of Kentucky",
    year = "1992",
    note = "UMI-92-19658"
}

@article{Radecky:1981fn,
    author = "Radecky, G. M. and others",
    title = "{Study of Single Pion Production by Weak Charged Currents in Low-energy Neutrino $d$ Interactions}",
    reportNumber = "PU-81-495, ANL-HEP-PR-81-07, COO-3066-175",
    doi = "10.1103/PhysRevD.25.1161",
    journal = "Phys. Rev. D",
    volume = "25",
    pages = "1161--1173",
    year = "1982",
    note = "[Erratum: Phys. Rev. D 26, 3297 (1982)]"
}

@article{Kitagaki:1986ct,
    author = "Kitagaki, T. and others",
    title = "{Charged Current Exclusive Pion Production in Neutrino Deuterium Interactions}",
    doi = "10.1103/PhysRevD.34.2554",
    journal = "Phys. Rev. D",
    volume = "34",
    pages = "2554--2565",
    year = "1986"
}

@article{Wilkinson:2014yfa,
    author = "Wilkinson, Callum and Rodrigues, Philip and Cartwright, Susan and Thompson, Lee and McFarland, Kevin",
    title = "{Reanalysis of bubble chamber measurements of muon-neutrino induced single pion production}",
    eprint = "1411.4482",
    archivePrefix = "arXiv",
    primaryClass = "hep-ex",
    doi = "10.1103/PhysRevD.90.112017",
    journal = "Phys. Rev. D",
    volume = "90",
    number = "11",
    pages = "112017",
    year = "2014"
}

@article{Alvarez-Ruso:2014bla,
    author = "Alvarez-Ruso, L. and Hayato, Y. and Nieves, J.",
    title = "{Progress and open questions in the physics of neutrino cross sections at intermediate energies}",
    eprint = "1403.2673",
    archivePrefix = "arXiv",
    primaryClass = "hep-ph",
    doi = "10.1088/1367-2630/16/7/075015",
    journal = "New J. Phys.",
    volume = "16",
    pages = "075015",
    year = "2014"
}

@article{Ankowski:2020qbe,
    author = "Ankowski, Artur M. and Friedland, Alexander",
    title = "{Assessing the accuracy of the GENIE event generator with electron-scattering data}",
    eprint = "2006.11944",
    archivePrefix = "arXiv",
    primaryClass = "hep-ph",
    reportNumber = "SLAC-PUB-17541",
    doi = "10.1103/PhysRevD.102.053001",
    journal = "Phys. Rev. D",
    volume = "102",
    number = "5",
    pages = "053001",
    year = "2020"
}

@article{MiniBooNE:2009dxl,
    author = "Aguilar-Arevalo, Alexis A. and others",
    collaboration = "MiniBooNE",
    title = "{Measurement of $\nu_\mu$ and $\bar{\nu}_\mu$ induced neutral current single $\pi^0$ production cross sections on mineral oil at $E_\nu \sim {\cal O}(1 {\rm GeV})$}",
    eprint = "0911.2063",
    archivePrefix = "arXiv",
    primaryClass = "hep-ex",
    reportNumber = "FERMILAB-PUB-09-583-E",
    doi = "10.1103/PhysRevD.81.013005",
    journal = "Phys. Rev. D",
    volume = "81",
    pages = "013005",
    year = "2010"
}

@article{MicroBooNE:2022zhr,
    author = "Abratenko, P. and others",
    collaboration = "MicroBooNE",
    title = "{Measurement of neutral current single {\ensuremath{\pi}}0 production on argon with the MicroBooNE detector}",
    eprint = "2205.07943",
    archivePrefix = "arXiv",
    primaryClass = "hep-ex",
    reportNumber = "FERMILAB-PUB-22-412-ND-SCD",
    doi = "10.1103/PhysRevD.107.012004",
    journal = "Phys. Rev. D",
    volume = "107",
    number = "1",
    pages = "012004",
    year = "2023"
}

@article{Alexandrou:2006mc,
    author = "Alexandrou, C. and Leontiou, Th. and Negele, John W. and Tsapalis, A.",
    title = "{The Axial N to Delta transition form factors from Lattice QCD}",
    eprint = "hep-lat/0607030",
    archivePrefix = "arXiv",
    doi = "10.1103/PhysRevLett.98.052003",
    journal = "Phys. Rev. Lett.",
    volume = "98",
    pages = "052003",
    year = "2007"
}

@article{Alexandrou:2010uk,
    author = "Alexandrou, C. and Koutsou, G. and Negele, J. W. and Proestos, Y. and Tsapalis, A.",
    title = "{Nucleon to Delta transition form factors with $N_F=2+1$ domain wall fermions}",
    eprint = "1011.3233",
    archivePrefix = "arXiv",
    primaryClass = "hep-lat",
    doi = "10.1103/PhysRevD.83.014501",
    journal = "Phys. Rev. D",
    volume = "83",
    pages = "014501",
    year = "2011"
}

@article{Kronfeld:2019nfb,
    author = "Kronfeld, Andreas S. and Richards, David G. and Detmold, William and Gupta, Rajan and Lin, Huey-Wen and Liu, Keh-Fei and Meyer, Aaron S. and Sufian, Raza and Syritsyn, Sergey",
    collaboration = "USQCD",
    title = "{Lattice QCD and Neutrino-Nucleus Scattering}",
    eprint = "1904.09931",
    archivePrefix = "arXiv",
    primaryClass = "hep-lat",
    reportNumber = "FERMILAB-PUB-19-172-T",
    doi = "10.1140/epja/i2019-12916-x",
    journal = "Eur. Phys. J. A",
    volume = "55",
    number = "11",
    pages = "196",
    year = "2019"
}

@article{Rein:1980wg,
    author = "Rein, Dieter and Sehgal, Lalit M.",
    title = "{Neutrino Excitation of Baryon Resonances and Single Pion Production}",
    reportNumber = "PITHA-80-10",
    doi = "10.1016/0003-4916(81)90242-6",
    journal = "Annals Phys.",
    volume = "133",
    pages = "79--153",
    year = "1981"
}

@article{Berger:2007rq,
    author = "Berger, Ch. and Sehgal, L. M.",
    title = "{Lepton mass effects in single pion production by neutrinos}",
    eprint = "0709.4378",
    archivePrefix = "arXiv",
    primaryClass = "hep-ph",
    doi = "10.1103/PhysRevD.76.113004",
    journal = "Phys. Rev. D",
    volume = "76",
    pages = "113004",
    year = "2007"
}

@article{Yan:2024kkg,
    author = "Yan, Qiyu and Niewczas, Kajetan and Nikolakopoulos, Alexis and Gonz{\'a}lez-Jim{\'e}nez, Ra{\'u}l and Jachowicz, Natalie and Lu, Xianguo and Sobczyk, Jan and Zheng, Yangheng",
    title = "{The Ghent Hybrid model in NuWro: a new neutrino single-pion production model in the GeV regime}",
    eprint = "2405.05212",
    archivePrefix = "arXiv",
    primaryClass = "hep-ph",
    reportNumber = "FERMILAB-PUB-24-0235-T",
    doi = "10.1007/JHEP12(2024)141",
    journal = "JHEP",
    volume = "2024",
    number = "12",
    pages = "141",
    year = "2024"
}

@article{Kamano:2013iva,
    author = "Kamano, H. and Nakamura, S. X. and Lee, T. -S. H. and Sato, T.",
    title = "{Nucleon resonances within a dynamical coupled-channels model of $\pi N$ and $\gamma N$ reactions}",
    eprint = "1305.4351",
    archivePrefix = "arXiv",
    primaryClass = "nucl-th",
    doi = "10.1103/PhysRevC.88.035209",
    journal = "Phys. Rev. C",
    volume = "88",
    number = "3",
    pages = "035209",
    year = "2013"
}

@article{Nakamura:2015rta,
    author = "Nakamura, S. X. and Kamano, H. and Sato, T.",
    title = "{Dynamical coupled-channels model for neutrino-induced meson productions in resonance region}",
    eprint = "1506.03403",
    archivePrefix = "arXiv",
    primaryClass = "hep-ph",
    doi = "10.1103/PhysRevD.92.074024",
    journal = "Phys. Rev. D",
    volume = "92",
    number = "7",
    pages = "074024",
    year = "2015"
}

@article{Alloul:2013bka,
    author = "Alloul, Adam and Christensen, Neil D. and Degrande, C{\'e}line and Duhr, Claude and Fuks, Benjamin",
    title = "{FeynRules  2.0 - A complete toolbox for tree-level phenomenology}",
    eprint = "1310.1921",
    archivePrefix = "arXiv",
    primaryClass = "hep-ph",
    reportNumber = "CERN-PH-TH-2013-239, MCNET-13-14, IPPP-13-71, DCPT-13-142, PITT-PACC-1308",
    doi = "10.1016/j.cpc.2014.04.012",
    journal = "Comput. Phys. Commun.",
    volume = "185",
    pages = "2250--2300",
    year = "2014"
}

@article{Degrande:2011ua,
    author = "Degrande, Celine and Duhr, Claude and Fuks, Benjamin and Grellscheid, David and Mattelaer, Olivier and Reiter, Thomas",
    title = "{UFO - The Universal FeynRules Output}",
    eprint = "1108.2040",
    archivePrefix = "arXiv",
    primaryClass = "hep-ph",
    reportNumber = "CP3-11-25, IPHC-PHENO-11-04, IPPP-11-39, DCPT-11-78, MPP-2011-68",
    doi = "10.1016/j.cpc.2012.01.022",
    journal = "Comput. Phys. Commun.",
    volume = "183",
    pages = "1201--1214",
    year = "2012"
}

@article{Darme:2023jdn,
    author = "Darm{\'e}, Luc and others",
    title = "{UFO 2.0: the {\textquoteleft}Universal Feynman Output{\textquoteright} format}",
    eprint = "2304.09883",
    archivePrefix = "arXiv",
    primaryClass = "hep-ph",
    reportNumber = "BONN-TH-2023-03, DESY-23-051, FERMILAB-PUB-23-138-T, KA-TP-06-2023,
  MCNET-23-06, P3H-23-023, TIF-UNIMI-2023-11",
    doi = "10.1140/epjc/s10052-023-11780-9",
    journal = "Eur. Phys. J. C",
    volume = "83",
    number = "7",
    pages = "631",
    year = "2023"
}

@article{Sherpa:2024mfk,
    author = "Bothmann, Enrico and others",
    collaboration = "Sherpa",
    title = "{Event generation with Sherpa 3}",
    eprint = "2410.22148",
    archivePrefix = "arXiv",
    primaryClass = "hep-ph",
    reportNumber = "IPPP/24/67, LTH-1385, FERMILAB-PUB-24-0748-T, ZU-TH 51/24, MCNET-24-17, CERN-TH-2024-171",
    doi = "10.1007/JHEP12(2024)156",
    journal = "JHEP",
    volume = "2024",
    number = "12",
    pages = "156",
    year = "2024"
}

@article{Gleisberg:2008fv,
    author = "Gleisberg, Tanju and Hoeche, Stefan",
    title = "{Comix, a new matrix element generator}",
    eprint = "0808.3674",
    archivePrefix = "arXiv",
    primaryClass = "hep-ph",
    reportNumber = "SLAC-PUB-13232, IPPP-08-31, DCPT-08-62, MCNET-08-08",
    doi = "10.1088/1126-6708/2008/12/039",
    journal = "JHEP",
    volume = "2008",
    number = "12",
    pages = "039",
    year = "2008"
}

@article{Alwall:2014hca,
    author = "Alwall, J. and Frederix, R. and Frixione, S. and Hirschi, V. and Maltoni, F. and Mattelaer, O. and Shao, H. -S. and Stelzer, T. and Torrielli, P. and Zaro, M.",
    title = "{The automated computation of tree-level and next-to-leading order differential cross sections, and their matching to parton shower simulations}",
    eprint = "1405.0301",
    archivePrefix = "arXiv",
    primaryClass = "hep-ph",
    reportNumber = "CERN-PH-TH-2014-064, CP3-14-18, LPN14-066, MCNET-14-09, ZU-TH-14-14",
    doi = "10.1007/JHEP07(2014)079",
    journal = "JHEP",
    volume = "2014",
    number = "07",
    pages = "079",
    year = "2014"
}

@article{Kilian:2007gr,
    author = "Kilian, Wolfgang and Ohl, Thorsten and Reuter, Jurgen",
    title = "{WHIZARD: Simulating Multi-Particle Processes at LHC and ILC}",
    eprint = "0708.4233",
    archivePrefix = "arXiv",
    primaryClass = "hep-ph",
    reportNumber = "DESY-11-126, EDINBURGH-2010-36, FR-PHENO-2010-037, SI-HEP-2010-18",
    doi = "10.1140/epjc/s10052-011-1742-y",
    journal = "Eur. Phys. J. C",
    volume = "71",
    pages = "1742",
    year = "2011"
}

@article{Adler:1969gk,
    author = "Adler, Stephen L.",
    title = "{Axial vector vertex in spinor electrodynamics}",
    doi = "10.1103/PhysRev.177.2426",
    journal = "Phys. Rev.",
    volume = "177",
    pages = "2426--2438",
    year = "1969"
}

@article{Bell:1969ts,
    author = "Bell, J. S. and Jackiw, R.",
    title = "{A PCAC puzzle: $\pi^0 \to \gamma \gamma$ in the $\sigma$ model}",
    doi = "10.1007/BF02823296",
    journal = "Nuovo Cim. A",
    volume = "60",
    pages = "47--61",
    year = "1969"
}

@article{Nath:1971wp,
    author = "Nath, L. M. and Etemadi, B. and Kimel, J. D.",
    title = "{Uniqueness of the interaction involving spin 3/2 particles}",
    doi = "10.1103/PhysRevD.3.2153",
    journal = "Phys. Rev. D",
    volume = "3",
    pages = "2153--2161",
    year = "1971"
}

@article{Goldberger:1958tr,
    author = "Goldberger, M. L. and Treiman, S. B.",
    title = "{Decay of the pi meson}",
    doi = "10.1103/PhysRev.110.1178",
    journal = "Phys. Rev.",
    volume = "110",
    pages = "1178--1184",
    year = "1958"
}

@article{Feynman:1958ty,
    author = "Feynman, R. P. and Gell-Mann, Murray",
    editor = "Brown, L. M.",
    title = "{Theory of Fermi interaction}",
    doi = "10.1103/PhysRev.109.193",
    journal = "Phys. Rev.",
    volume = "109",
    pages = "193--198",
    year = "1958"
}

@article{Burkert:2025coj,
    author = "Burkert, Volker and Eichmann, Gernot and Klempt, Eberhard",
    title = "{The impact of {\ensuremath{\gamma}}N and {\ensuremath{\gamma}}{\ensuremath{*}}N interactions on our understanding of nucleon excitations}",
    eprint = "2506.16482",
    archivePrefix = "arXiv",
    primaryClass = "hep-ph",
    reportNumber = "JLAB-PHY-25-4380",
    doi = "10.1016/j.ppnp.2025.104214",
    journal = "Prog. Part. Nucl. Phys.",
    volume = "146",
    pages = "104214",
    year = "2026"
}

@article{Benmerrouche:1989uc,
    author = "Benmerrouche, M. and Davidson, R. M. and Mukhopadhyay, N. C.",
    title = "{Problems of Describing Spin 3/2 Baryon Resonances in the Effective Lagrangian Theory}",
    doi = "10.1103/PhysRevC.39.2339",
    journal = "Phys. Rev. C",
    volume = "39",
    pages = "2339--2348",
    year = "1989"
}

@article{JeffersonLab:2008jve,
    author = "Huber, G. M. and others",
    collaboration = "Jefferson Lab",
    title = "{Charged pion form-factor between Q**2 = 0.60-GeV**2 and 2.45-GeV**2. II. Determination of, and results for, the pion form-factor}",
    eprint = "0809.3052",
    archivePrefix = "arXiv",
    primaryClass = "nucl-ex",
    reportNumber = "JLAB-PHY-08-864",
    doi = "10.1103/PhysRevC.78.045203",
    journal = "Phys. Rev. C",
    volume = "78",
    pages = "045203",
    year = "2008"
}

@article{NA7:1986vav,
    author = "Amendolia, S. R. and others",
    editor = "Loken, S. C.",
    collaboration = "NA7",
    title = "{A Measurement of the Space - Like Pion Electromagnetic Form-Factor}",
    reportNumber = "CERN-EP-86-34",
    doi = "10.1016/0550-3213(86)90437-2",
    journal = "Nucl. Phys. B",
    volume = "277",
    pages = "168",
    year = "1986"
}

@article{Ackermann:1977rp,
    author = "Ackermann, H. and Azemoon, T. and Gabriel, W. and Mertiens, H. D. and Reich, H. D. and Specht, G. and Janata, F. and Schmidt, D.",
    title = "{Determination of the Longitudinal and the Transverse Part in pi+ Electroproduction}",
    reportNumber = "DESY-77-64",
    doi = "10.1016/0550-3213(78)90523-0",
    journal = "Nucl. Phys. B",
    volume = "137",
    pages = "294--300",
    year = "1978"
}

@article{Brauel:1979zk,
    author = "Brauel, P. and Canzler, T. and Cords, D. and Felst, R. and Grindhammer, Guenter and Helm, M. and Kollmann, W. D. and Krehbiel, H. and Schadlich, M.",
    title = "{Electroproduction of $\pi^+ n$, $\pi^- p$ and $K^+ \Lambda$, $K^+ \Sigma^0$ Final States Above the Resonance Region}",
    reportNumber = "DESY-79-59",
    doi = "10.1007/BF01443698",
    journal = "Z. Phys. C",
    volume = "3",
    pages = "101",
    year = "1979"
}

@article{Sakurai:1960ju,
    author = "Sakurai, J. J.",
    title = "{Theory of strong interactions}",
    doi = "10.1016/0003-4916(60)90126-3",
    journal = "Annals Phys.",
    volume = "11",
    pages = "1--48",
    year = "1960"
}

@article{Lalakulich:2005cs,
    author = "Lalakulich, Olga and Paschos, Emmanuel A.",
    title = "{Resonance production by neutrinos. I. J = 3/2 resonances}",
    eprint = "hep-ph/0501109",
    archivePrefix = "arXiv",
    reportNumber = "DO-TH-04-12",
    doi = "10.1103/PhysRevD.71.074003",
    journal = "Phys. Rev. D",
    volume = "71",
    pages = "074003",
    year = "2005"
}

@article{Rodrigues:2016xjj,
    author = "Rodrigues, Philip and Wilkinson, Callum and McFarland, Kevin",
    title = "{Constraining the GENIE model of neutrino-induced single pion production using reanalyzed bubble chamber data}",
    eprint = "1601.01888",
    archivePrefix = "arXiv",
    primaryClass = "hep-ex",
    doi = "10.1140/epjc/s10052-016-4314-3",
    journal = "Eur. Phys. J. C",
    volume = "76",
    number = "8",
    pages = "474",
    year = "2016"
}

@book{Peskin:1995ev,
    author = "Peskin, Michael E. and Schroeder, Daniel V.",
    title = "{An Introduction to quantum field theory}",
    doi = "10.1201/9780429503559",
    isbn = "978-0-201-50397-5, 978-0-429-50355-9, 978-0-429-49417-8",
    publisher = "Addison-Wesley",
    address = "Reading, USA",
    year = "1995"
}

@article{Adler:1968tw,
    author = "Adler, Stephen L.",
    title = "{Photoproduction, electroproduction and weak single pion production in the (3,3) resonance region}",
    doi = "10.1016/0003-4916(68)90278-9",
    journal = "Annals Phys.",
    volume = "50",
    pages = "189--311",
    year = "1968"
}

@book{Scherer:2012xha,
    author = "Scherer, Stefan and Schindler, Matthias R.",
    title = "{A Primer for Chiral Perturbation Theory}",
    doi = "10.1007/978-3-642-19254-8",
    isbn = "978-3-642-19253-1",
    publisher = "Springer",
    series = "Lect. Notes Phys.",
    volume = "830",
    year = "2012"
}

@article{Isaacson:2025cnk,
    author = "Isaacson, Joshua and Jay, William and Lovato, Alessandro and Machado, Pedro and Nikolakopoulos, Alexis and Rocco, Noemi and Steinberg, Noah",
    title = "{Single pion production and pion propagation in Achilles}",
    eprint = "2508.19213",
    archivePrefix = "arXiv",
    primaryClass = "hep-ph",
    reportNumber = "FERMILAB-PUB-25-0473-T, MSUHEP-25-018",
    doi = "10.1103/13bh-22lm",
    journal = "Phys. Rev. D",
    volume = "113",
    pages = "036005",
    year = "2026"
}

@article{Vasilkov:1960jetp,
    author = "Vasil'kov, R. G. and Govorkov, B. B. and Gol'danskii, V. I.",
    title = "{Photoproduction of Neutral $\pi$ Mesons on Hydrogen by $\gamma$ Quanta with Energy between the Threshold and 240 MeV}",
    journal = "Sov. Phys. JETP",
    volume = "10",
    pages = "7",
    year = "1960"
}

@article{Govorkov:1968sjnp,
    author = "Govorkov, B. B. and Denisov, S. P. and Minarik, E. V.",
    journal = "Sov. J. Nucl. Phys.",
    volume = "6",
    pages = "370",
    year = "1968"
}

@article{A2:2019yud,
    author = "Briscoe, W. J. and others",
    collaboration = "A2",
    title = "{Cross section for $\gamma n \to \pi^0 n$ at the Mainz A2 experiment}",
    eprint = "1908.02730",
    archivePrefix = "arXiv",
    primaryClass = "nucl-ex",
    doi = "10.1103/PhysRevC.100.065205",
    journal = "Phys. Rev. C",
    volume = "100",
    pages = "065205",
    year = "2019"
}

@article{A2:2018jcd,
    author = "Dieterle, M. and others",
    collaboration = "A2",
    title = "{Photoproduction of $\pi^0$ Mesons off Protons and Neutrons in the Second and Third Nucleon Resonance Region}",
    eprint = "1805.10504",
    archivePrefix = "arXiv",
    primaryClass = "nucl-ex",
    doi = "10.1103/PhysRevC.97.065205",
    journal = "Phys. Rev. C",
    volume = "97",
    pages = "065205",
    year = "2018"
}

@book{halzenmartinQUARKSLEPTONSINTRODUCTORY1984,
  title = "Quarks and Leptons: An Introductory Course in Modern Particle Physics",
  author = "Halzen, F. and Martin, Alan D.",
  publisher = "John Wiley \& Sons",
  location = "United States of America",
  isbn = "978-0-471-88741-6",
  pagetotal = "369",
  year = "1984"
}

@article{Christensen:2013aua,
    author = "Christensen, Neil D. and de Aquino, P. and Deutschmann, N. and Duhr, C. and Fuks, B. and Garcia-Cely, C. and Mattelaer, O. and Mawatari, K. and Oexl, B. and Takaesu, Y.",
    title = "{Simulating spin-$ \frac{3}{2}$ particles at colliders}",
    eprint = "1308.1668",
    archivePrefix = "arXiv",
    primaryClass = "hep-ph",
    reportNumber = "CERN-PH-TH-2013-187, IPPP-13-56, DCPT-13-172, MCNET-13-09, PITT-PACC-1307, KIAS-P13044, TUM-HEP-900-13",
    doi = "10.1140/epjc/s10052-013-2580-x",
    journal = "Eur. Phys. J. C",
    volume = "73",
    number = "10",
    pages = "2580",
    year = "2013"
}

@article{ParticleDataGroup:2025Resonances,
    author        = "D. M. Asner and C. Hanhart and M. Mikhasenko",
    collaboration = "Particle Data Group",
    title         = "{Review of Particle Physics: Resonances}",
    journal       = "Phys. Rev. D",
    volume        = "110",
    pages         = "030001",
    year          = "2024",
    note          = "2025 update; review 50, ``Resonances''",
    url           = "https://pdg.lbl.gov/2025/reviews/rpp2025-rev-resonances.pdf"
}

@article{Gardiner:2025NuHepMC,
   title={NuHepMC: A standardized event record format for neutrino event generators},
   ISSN={2949-804X},
   url={http://dx.doi.org/10.21468/SciPostPhysCodeb.57},
   DOI={10.21468/scipostphyscodeb.57},
   journal={SciPost Physics Codebases},
   publisher={Stichting SciPost},
   author={Gardiner, Steven and Isaacson, Joshua and Pickering, Luke},
   year={2025},
   month=Aug }

@book{Srednicki:2007qs,
    author = "Srednicki, M.",
    title = "{Quantum field theory}",
    doi = "10.1017/CBO9780511813917",
    isbn = "978-0-521-86449-7, 978-0-511-26720-8",
    publisher = "Cambridge University Press",
    month = "1",
    year = "2007"
}

@article{MINERvA:2017okh,
    author = "Altinok, O. and others",
    collaboration = "MINERvA",
    title = "{Measurement of $\nu_{\mu}$ charged-current single $\pi^{0}$ production on hydrocarbon in the few-GeV region using MINERvA}",
    eprint = "1708.03723",
    archivePrefix = "arXiv",
    primaryClass = "hep-ex",
    reportNumber = "FERMILAB-PUB-17-220-ND",
    doi = "10.1103/PhysRevD.96.072003",
    journal = "Phys. Rev. D",
    volume = "96",
    number = "7",
    pages = "072003",
    year = "2017"
}

@article{MINERvA:2014ogb,
    author = "Eberly, B. and others",
    collaboration = "MINERvA",
    title = "{Charged Pion Production in $\nu_\mu$ Interactions on Hydrocarbon at $\langle E_{\nu}\rangle$= 4.0 GeV}",
    eprint = "1406.6415",
    archivePrefix = "arXiv",
    primaryClass = "hep-ex",
    reportNumber = "FERMILAB-PUB-14-193-E",
    doi = "10.1103/PhysRevD.92.092008",
    journal = "Phys. Rev. D",
    volume = "92",
    number = "9",
    pages = "092008",
    year = "2015"
}

@article{MINERvA:2016sfc,
    author = "McGivern, C. L. and others",
    collaboration = "MINERvA",
    title = "{Cross sections for $\nu_\mu$ and $\bar \nu_\mu$ induced pion production on hydrocarbon in the few-GeV region using MINERvA}",
    eprint = "1606.07127",
    archivePrefix = "arXiv",
    primaryClass = "hep-ex",
    reportNumber = "FERMILAB-PUB-16-228-ND",
    doi = "10.1103/PhysRevD.94.052005",
    journal = "Phys. Rev. D",
    volume = "94",
    number = "5",
    pages = "052005",
    year = "2016"
}

@article{MiniBooNE:2010eis,
    author = "Aguilar-Arevalo, A. A. and others",
    collaboration = "MiniBooNE",
    title = "{Measurement of Neutrino-Induced Charged-Current Charged Pion Production Cross Sections on Mineral Oil at E$_{\nu}\sim 1~\textrm{GeV}$}",
    eprint = "1011.3572",
    archivePrefix = "arXiv",
    primaryClass = "hep-ex",
    reportNumber = "FERMILAB-PUB-10-470-E",
    doi = "10.1103/PhysRevD.83.052007",
    journal = "Phys. Rev. D",
    volume = "83",
    pages = "052007",
    year = "2011"
}

@article{MiniBooNE:2010cxl,
    author = "Aguilar-Arevalo, A. A. and others",
    collaboration = "MiniBooNE",
    title = "{Measurement of $\nu_\mu$-induced charged-current neutral pion production cross sections on mineral oil at $E_\nu\in0.5-2.0$ GeV}",
    eprint = "1010.3264",
    archivePrefix = "arXiv",
    primaryClass = "hep-ex",
    reportNumber = "FERMILAB-PUB-10-426-E",
    doi = "10.1103/PhysRevD.83.052009",
    journal = "Phys. Rev. D",
    volume = "83",
    pages = "052009",
    year = "2011"
}

@article{T2K:2019yqu,
    author = "Abe, K. and others",
    collaboration = "T2K",
    title = "{Measurement of the muon neutrino charged-current single $\pi^+$ production on hydrocarbon using the T2K off-axis near detector ND280}",
    eprint = "1909.03936",
    archivePrefix = "arXiv",
    primaryClass = "hep-ex",
    doi = "10.1103/PhysRevD.101.012007",
    journal = "Phys. Rev. D",
    volume = "101",
    number = "1",
    pages = "012007",
    year = "2020"
}

@article{T2K:2021naz,
    author = "Abe, K. and others",
    collaboration = "T2K",
    title = "{First T2K measurement of transverse kinematic imbalance in the muon-neutrino charged-current single-$\pi^+$ production channel containing at least one proton}",
    eprint = "2102.03346",
    archivePrefix = "arXiv",
    primaryClass = "hep-ex",
    doi = "10.1103/PhysRevD.103.112009",
    journal = "Phys. Rev. D",
    volume = "103",
    number = "11",
    pages = "112009",
    year = "2021"
}

@article{Stowell:2016jfr,
    author = "Stowell, P. and others",
    title = "{NUISANCE: a neutrino cross-section generator tuning and comparison framework}",
    eprint = "1612.07393",
    archivePrefix = "arXiv",
    primaryClass = "hep-ex",
    doi = "10.1088/1748-0221/12/01/P01016",
    journal = "JINST",
    volume = "12",
    number = "01",
    pages = "P01016",
    year = "2017"
}

@article{Betancourt:2018bpu,
    author = "Betancourt, M. and others",
    title = "{Comparisons and Challenges of Modern Neutrino Scattering Experiments (TENSIONS2016 Report)}",
    eprint = "1805.07378",
    archivePrefix = "arXiv",
    primaryClass = "hep-ex",
    reportNumber = "FERMILAB-CONF-18-208-ND",
    doi = "10.1016/j.physrep.2018.08.003",
    journal = "Phys. Rept.",
    volume = "773-774",
    pages = "1--28",
    year = "2018"
}

@article{alexandrouetalAxialNucleonNucleon2007,
  title = {Axial Nucleon and Nucleon to {$\Delta$} Form Factors and the {Goldberger-Treiman} Relations from Lattice {QCD}},
  author = {Alexandrou, C. and Koutsou, G. and Leontiou, Th. and Negele, J. W. and Tsapalis, A.},
  journal = {Phys. Rev. D},
  volume = {76},
  number = {9},
  pages = {094511},
  year = {2007},
  doi = {10.1103/PhysRevD.76.094511}
}

@article{DelgadoAcosta:2015ypa,
    author = "Delgado Acosta, E. G. and Banda Guzm{\'a}n, V. M. and Kirchbach, M.",
    title = "{Bosonic and fermionic Weinberg-Joos (j,0) {\ensuremath{\oplus}} (0,j) states of arbitrary spins as Lorentz tensors or tensor-spinors and second-order theory}",
    eprint = "1503.07230",
    archivePrefix = "arXiv",
    primaryClass = "hep-ph",
    doi = "10.1140/epja/i2015-15035-x",
    journal = "Eur. Phys. J. A",
    volume = "51",
    pages = "35",
    year = "2015"
}

@article{Hernandez:2010bx,
    author = "Hernandez, E. and Nieves, J. and Valverde, M. and Vicente Vacas, M. J.",
    title = "{N-Delta(1232) axial form factors from weak pion production}",
    eprint = "1001.4416",
    archivePrefix = "arXiv",
    primaryClass = "hep-ph",
    doi = "10.1103/PhysRevD.81.085046",
    journal = "Phys. Rev. D",
    volume = "81",
    pages = "085046",
    year = "2010"
}

@Article{Hoppe2026,
  author  = {Mareen Hoppe},
  title   = {Simulating spin-3/2 particles with Sherpa},
  journal = {in preparation},
  year = {2026}
}

@article{Denner:1999gp,
    author = "Denner, Ansgar and Dittmaier, S. and Roth, M. and Wackeroth, D.",
    title = "{Predictions for all processes e+ e- ---{\ensuremath{>}} 4 fermions + gamma}",
    eprint = "hep-ph/9904472",
    archivePrefix = "arXiv",
    reportNumber = "BI-TP-99-10, PSI-PR-99-12",
    doi = "10.1016/S0550-3213(99)00437-X",
    journal = "Nucl. Phys. B",
    volume = "560",
    pages = "33--65",
    year = "1999"
}

@article{Davidson:2001rk,
    author = "Davidson, R. M. and Workman, Ron",
    title = "{Form-factors and photoproduction amplitudes}",
    eprint = "nucl-th/0101066",
    archivePrefix = "arXiv",
    doi = "10.1103/PhysRevC.63.025210",
    journal = "Phys. Rev. C",
    volume = "63",
    pages = "025210",
    year = "2001"
}

@article{Sealock:1989nx,
  author = {Sealock, R. M. and others},
  title = {{Electroexcitation of the $\Delta(1232)$ in Nuclei}},
  doi = {10.1103/PhysRevLett.62.1350},
  journal = {Phys. Rev. Lett.},
  volume = {62},
  pages = {1350--1353},
  year = {1989},
}

@article{Ohta:1989ji,
    author = "Ohta, K.",
    title = "{Electromagnetic Interactions of Extended Nucleons}",
    doi = "10.1103/PhysRevC.40.1335",
    journal = "Phys. Rev. C",
    volume = "40",
    pages = "1335--1346",
    year = "1989"
}

@article{Workman:1992hr,
    author = "Workman, R. L. and Naus, H. W. L. and Pollock, S. J.",
    title = "{On form-factors and gauge invariance in pion photoproduction}",
    reportNumber = "VPI-CAPS-92-2",
    doi = "10.1103/PhysRevC.45.2511",
    journal = "Phys. Rev. C",
    volume = "45",
    pages = "2511--2513",
    year = "1992"
}

@article{Haberzettl:1997jg,
    author = "Haberzettl, Helmut",
    title = "{Gauge invariant theory of pion photoproduction with dressed hadrons}",
    eprint = "nucl-th/9704057",
    archivePrefix = "arXiv",
    doi = "10.1103/PhysRevC.56.2041",
    journal = "Phys. Rev. C",
    volume = "56",
    pages = "2041--2058",
    year = "1997"
}

@article{Haberzettl:1998aqi,
    author = "Haberzettl, H. and Bennhold, C. and Mart, T. and Feuster, T.",
    title = "{Gauge-invariant tree-level photoproduction amplitudes with form factors}",
    eprint = "nucl-th/9804051",
    archivePrefix = "arXiv",
    doi = "10.1103/PhysRevC.58.R40",
    journal = "Phys. Rev. C",
    volume = "58",
    number = "1",
    pages = "R40--R44",
    year = "1998"
}

@article{Benhar:2006hh,
  author = {Benhar, Omar},
  editor = {Cavanna, F. and Morfin, J. G. and Nakaya, T.},
  title = {{Final state interactions in the electroweak nuclear response}},
  eprint = {hep-ph/0602108},
  archiveprefix = {arXiv},
  doi = {10.1016/j.nuclphysbps.2006.08.034},
  journal = {Nucl. Phys. B Proc. Suppl.},
  volume = {159},
  pages = {168--173},
  year = {2006},
}

@article{MicroBooNE:2024pdj,
  author = {Abratenko, P. and others},
  collaboration = {MicroBooNE},
  title = {{First Double-Differential Cross Section Measurement of Neutral-Current {\ensuremath{\pi}}0 Production in Neutrino-Argon Scattering in the MicroBooNE Detector}},
  eprint = {2404.10948},
  archiveprefix = {arXiv},
  primaryclass = {hep-ex},
  reportnumber = {FERMILAB-PUB-24-0125},
  doi = {10.1103/PhysRevLett.134.161802},
  journal = {Phys. Rev. Lett.},
  volume = {134},
  number = {16},
  pages = {161802},
  year = {2025},
}

@article{Rein:1982pf,
    author = "Rein, Dieter and Sehgal, Lalit M.",
    title = "{Coherent pi0 Production in Neutrino Reactions}",
    reportNumber = "PITHA-82-27",
    doi = "10.1016/0550-3213(83)90090-1",
    journal = "Nucl. Phys. B",
    volume = "223",
    pages = "29--44",
    year = "1983"
}

@article{MiniBooNE:2008mmr,
    author = "Aguilar-Arevalo, A. A. and others",
    collaboration = "MiniBooNE",
    title = "{First Observation of Coherent $\pi^0$ Production in Neutrino Nucleus Interactions with $E_{\nu}<$ 2
GeV}",
    eprint = "0803.3423",
    archivePrefix = "arXiv",
    primaryClass = "hep-ex",
    reportNumber = "FERMILAB-PUB-08-102-E",
    doi = "10.1016/j.physletb.2008.05.006",
    journal = "Phys. Lett. B",
    volume = "664",
    pages = "41--46",
    year = "2008"
}

@article{MINERvA:2014ani,
    author = "Higuera, A. and others",
    collaboration = "MINERvA",
    title = "{Measurement of Coherent Production of $\pi^\pm$ in Neutrino and Antineutrino Beams on Carbon from
$E_\nu$ of $1.5$ to $20$ GeV}",
    eprint = "1409.3835",
    archivePrefix = "arXiv",
    primaryClass = "hep-ex",
    reportNumber = "FERMILAB-PUB-14-331-E",
    doi = "10.1103/PhysRevLett.113.261802",
    journal = "Phys. Rev. Lett.",
    volume = "113",
    number = "26",
    pages = "261802",
    year = "2014"
}

@software{ChiEFT:software,
  author       = {Graham-Jones, Zara and
                  Hoppe, Mareen and
                  Isaacson, Joshua and
                  Li, Shirley Weishi and
                  Toman, Miša},
  title        = {ChiEFT: a phenomenological chiral effective field
                   theory model of pion production for neutrino event
                   generators
                  },
  month        = oct,
  year         = 2026,
  publisher    = {Zenodo},
  version      = {v1.0.0-rc1},
  doi          = {10.5281/zenodo.23767017},
  url          = {https://doi.org/10.5281/zenodo.23767017},
}

@article{Hooft:2026yyg,
    author = "Hooft, M. and Nikolakopoulos, A. and Garc{\'\i}a-Marcos, J. and De Backer, Y. and Franco-Munoz, T. and Niewczas, K. and Gonz{\'a}lez-Jim{\'e}nez, R. and Jachowicz, N.",
    title = "{Optimizing the description of the Delta region in the Ghent hybrid model for single-pion production}",
    eprint = "2603.29486",
    archivePrefix = "arXiv",
    primaryClass = "nucl-th",
    doi = "10.1103/dqgm-8sv5",
    journal = "Phys. Rev. D",
    volume = "114",
    number = "5",
    pages = "056018",
    year = "2026"
}

@article{Tsaran:2024sue,
    author = "Tsaran, Viacheslav and Vanderhaeghen, Marc",
    title = "{Nuclear coherent {\ensuremath{\pi}}0 photoproduction with charge-exchange and spin-flip rescattering}",
    eprint = "2403.08608",
    archivePrefix = "arXiv",
    primaryClass = "nucl-th",
    doi = "10.1103/PhysRevC.110.014605",
    journal = "Phys. Rev. C",
    volume = "110",
    number = "1",
    pages = "014605",
    year = "2024"
}

@article{Olsson:1974sw,
    author = "Olsson, M. G.",
    title = "{Solutions of the Multichannel Unitarity Equations Describing the Addition of a Resonance and Background: Application to a Pole Model of Photoproduction}",
    reportNumber = "CERN-TH-1846",
    doi = "10.1016/0550-3213(74)90115-1",
    journal = "Nucl. Phys. B",
    volume = "78",
    pages = "55--76",
    year = "1974"
}

@article{Drechsel:2007if,
    author = "Drechsel, D. and Kamalov, S. S. and Tiator, L.",
    title = "{Unitary Isobar Model - MAID2007}",
    eprint = "0710.0306",
    archivePrefix = "arXiv",
    primaryClass = "nucl-th",
    doi = "10.1140/epja/i2007-10490-6",
    journal = "Eur. Phys. J. A",
    volume = "34",
    pages = "69--97",
    year = "2007"
}

@article{Watson:1952ji,
    author = "Watson, Kenneth M.",
    title = "{The Effect of final state interactions on reaction cross-sections}",
    doi = "10.1103/PhysRev.88.1163",
    journal = "Phys. Rev.",
    volume = "88",
    pages = "1163--1171",
    year = "1952"
}

\end{document}